\documentclass[aps,prx,twocolumn,superscriptaddress,showpacks]{revtex4-2}

\usepackage{comment}
\usepackage[colorlinks, citecolor=red]{hyperref}
\usepackage{hyperref}
\usepackage[utf8]{inputenc}
\usepackage[T1]{fontenc}    %
\usepackage[english]{babel}
\usepackage[pdftex]{graphicx}
\usepackage{amsmath,amssymb,amsthm,bbm, mathtools} %
\usepackage{mathrsfs}
\usepackage{tikz}   %
\usepackage[normalem]{ulem}

\renewcommand{\vec}[1]{\boldsymbol{#1}}

\begin{document}


\title{Fermi sea and sky in the Bogoliubov–de Gennes equation}

\author{Xian-Peng Zhang}

\affiliation{Centre for Quantum Physics, Key Laboratory of Advanced Optoelectronic Quantum Architecture and Measurement (MOE), School of Physics, Beijing Institute of Technology, Beijing, 100081, China}

\affiliation{Department of Physics, Hong Kong University of Science and Technology, Clear Water Bay, Hong Kong, China}

\author{Yugui Yao}
\affiliation{Centre for Quantum Physics, Key Laboratory of Advanced Optoelectronic Quantum Architecture and Measurement (MOE), School of Physics, Beijing Institute of Technology, Beijing, 100081, China}

\begin{abstract}
We develop a comprehensive logical framework for effectively handling the overcomplete basis set in the Bogoliubov–de Gennes equation that contains two orthonormal basis sets conjugate with each other, such as particle and hole orthonormal basis sets. 
We highlight the significant implications of our logical framework from theoretical concepts and experimental predictions. Firstly,  we rigorously derive all many-body eigenfunctions of arbitrary nonuniform superconductors and uncover that the many-body eigenstates are full of superconducting spin clouds—the electron configuration within the Cooper-like pair of an arbitrary nonuniform superconductor. Secondly, we demonstrate a conjugate loop formed by the effective vacuum states of two orthonormal basis sets conjugate with each other, such as the Fermi sea and sky—the effective vacuum states of positive and negative orthonormal basis sets, respectively. We emphasize the conceptual importance of the Fermi sky, which serves as the starting point for the filling of all negative quasiparticles within the Fermi sea.  Next, we emphasize the significance of our logical framework  from the experimental prediction such as superconducting tunneling spectroscopy. The existing theory of the superconducting tunneling spectroscopy, based on an excitation orthonormal basis set, necessitates the breakup and formation of the Cooper pairs that nerve was directly observed in experiments. Our conjugate loop provides insights into how to populate negative quasiparticles and achieves the one-body quasiparticle representation of the many-body superconducting state. This discovery provides a solid foundation for treating a many-body superconducting state as a one-body problem and thus proves valuable for addressing superconducting tunneling spectroscopy as simply as the standard nonsuperconducting spectroscopy theory. Thirdly, we present a gate-, field-, and phase-tunable tunnel spectroscopy asymmetry arising from the imbalanced particle-hole distribution of the subgap quasiparticles in a quantum-dot Josephson junction. These findings underscore the power of our logical framework and its implications for advancing our understanding and utilization of solid-state devices based on superconductivity.
\end{abstract}

\maketitle

\section{Introduction}

The Bogoliubov–de Gennes (BdG) formalism is widely used as a fundamental tool to investigate not only standard superconducting phenomena~\cite{de2018superconductivity,tinkham2004introduction} but also to understand novel phases in topological superconductivity~\cite{bernevig2013topological,zhu2016bogoliubov}. The particle-hole symmetry in the BdG spectra provides an intuitive illustration of the various superconducting spectroscopies~\cite{lee2014spin,valentini2021nontopological} owing to the use of an overcomplete basis set that contains both particle and hole operators of  \emph{each} electron and hence consists of two orthonormal basis sets (OBSs) whose elements are independent of each other and satisfy anticommutation relations~\cite{bernevig2013topological,de2018superconductivity,tinkham2004introduction,zhu2016bogoliubov}. Moreover, the diagonalizable description of the coexistence of  superconductivity and spin-orbit coupling, crucial for describing novel phenomena like topological superconductivity~\cite{bernevig2013topological}, necessitates the use of an overcomplete basis set~\footnote{By diagonalizable form, we mean the ability of mathematically rewriting the Hamiltonian into the form $H=\Psi^{\dagger}\mathcal{H}_{BdG}\Psi$, where the Hermitian BdG Hamlitonian matrix $\mathcal{H}_{BdG}$ can be diagonalized by a unitary transformation and the Nambu vector can contain electron and hole field operators. The pair potential itself $c^{\dagger}_{n\uparrow}c^{\dagger}_{n\downarrow}$ can be well-described by an orthonormal basis set $\Psi=\bigotimes_{n}\Psi_{n}$, with $\Psi_{n}=[c^{}_{n\uparrow},
\tilde{c}_{n\uparrow}=-c^{\dagger}_{n\downarrow}]$, where we  use the hole representation of spin-down electrons $\tilde{c}_{n\uparrow}=c^{\dagger}_{n\downarrow}$ to obtain the diagonalizable description of pair potential $-c^{\dagger}_{n\uparrow}\tilde{c}^{}_{n\uparrow}$. However, the spin-flip from the spin-orbit coupling is not diagonalizable anymore  in this orthonormal basis set, i.e., $c^{\dagger}_{n\uparrow}c_{n\downarrow}=c^{\dagger}_{n\uparrow}\tilde{c}^{\dagger}_{n\uparrow}$. Therefore, the quadratic description of the coexistence of superconductivity ($c^{\dagger}_{n\uparrow}c^{\dagger}_{n\downarrow}=-c^{\dagger}_{n\uparrow}\tilde{c}^{}_{n\uparrow}$) and spin-orbit coupling ($c^{\dagger}_{n\uparrow}c_{n\downarrow}$) requires an overcomplete basis set $\Psi=\bigotimes_{n}\Psi_{n}$, with $\Psi_{n}=[c^{}_{n\uparrow},c^{}_{n\downarrow},\tilde{c}_{n\uparrow}=-c^{\dagger}_{n\downarrow},\tilde{c}_{n\downarrow}=c^{\dagger}_{n\uparrow}]$.}.  However, working with the overcomplete basis set, whose elements are not independent of each other anymore~\cite{bernevig2013topological,zhu2016bogoliubov}, \emph{seemingly} enlarges the dimension of the Hilbert space  if we disregard the mutual dependence between particle and hole operators of each electron~\footnote{The dimension of the Hilbert space is equal to the number of all distinguishable many-body eigenstates. Assuming there are $M$ single-particle eigenstates that are independent of each other, the corresponding number of many-body eigenstates is $2^{M}$ because each single particle can be either occupied or unoccupied. Let us take the low-energy Hilbert space of the Andreev system dominated by four spinful subgap quasiparticles that are not independent of each other~\cite{janvier2015coherent,bretheau2013supercurrent}. Due to this independence, there are only four distinguishable many-body eigenstates rather than $2^{4}$ eigenstates.}. 
Hitherto, a comprehensive logical framework for dealing with the overcomplete basis set is still lacking, resulting in some ambiguities, such as  the apparent disregard of half quasiparticles in superconducting scattering~\cite{datta1999can,datta1996scattering,coleman2015introduction},  the misunderstanding prefactor ($2e$) origin  in the current-phase character~\cite{beenakker2013fermion}, as well as the perplexing occupancy of negative quasiparticles in the superconducting ground state (see details below). 

Note that a quasiparticle energy of the BdG Hamiltonian, denoted as $E_m$, contributes to the supercurrent with an amount, $I_m=\frac{2e}{\hbar}\partial_{\phi} E_m$~\cite{bretheau2013supercurrent}, where $e<0$ is the charge of the electron, $\hbar$ is the reduced Plank constant, and $\phi$ denotes the phase difference between two superconducting leads. There exists a pervasive misunderstanding that the prefactor $2e$ in the above current-phase character refers to the charge of a Cooper pair~\cite{beenakker2013fermion}. Paradoxically, when calculating the ground-state supercurrent including the contributions from all negative quasiparticles, the prefactor becomes $e$, i.e., $I_G=\frac{e}{\hbar}\sum_{E_n<0}\partial_{\phi} E_n$. Because the $\phi$-dependent part of the ground-state energy is the \emph{half} summation of all negative quaisparticle energies where spin degeneracy is included. This half summation is \emph{vaguely} explained by the double counting~\cite{chtchelkatchev2003andreev}, which originates from the mean-field treatment of the attractive interaction~\cite{bardeen1969structure,beenakker2005superconducting}. Despite the half prefactor, the above summation treats the ground state as a filled Fermi sea of all negative quasiparticles from the vacuum of electrons $\vert 0 \rangle$, i.e., $\vert G\rangle=(\prod_{E_n<0}\gamma^{\dagger}_{n})\vert 0\rangle$, which can be derived from a popular formula $\vert G\rangle=(\prod_{E_m>0}\gamma_{m})\vert 0\rangle$~\cite{coleman2015introduction,alicea2011non} because each positive solution $\gamma_m$ is always paired with a negative  solution $\gamma_n$, satisfying $\gamma_{m}=\gamma^{\dagger}_n$ and $E_{m}=-E_{n}$~\cite{aguado2017majorana,martin2012introduction}. This conceptualization of the ground state is widespread in the condensed matter physics~\cite{coleman2015introduction,himeda2002stripe,kurilovich2021microwave,alicea2011non,sajith2023signatures}, such as topological superconductor~\cite{alicea2011non}. However, this description of the ground state is \textit{unsatisfactory} because the vacuum of electrons ceases to be the effective vacuum state (EVS) of these negative quasiparticles (i.e., $\gamma^{}_{n}\vert 0\rangle\neq 0$ for $E_n<0$), rendering this description problematic when i) we study quantum dots and normal metals within the BdG formalism and ii) the ground state has a fermionic odd parity (see Sec.~\ref{quasiparticleoccupancy}). Consequently, the question of where to assign these negative quasiparticles remains unclear. However, within the field of condensed matter physics, theorists and experimentalists have primarily focused on the excitation properties of superconducting materials, with little emphasis on the precise wave function or negative quasiparticle occupancy of the superconducting ground state, which, in principle, should be crucial for understanding the dramatic features observed in various spectroscopies of the superconducting condensate characterized by particle-hole symmetry~\cite{datta1996scattering,datta1999can}.

In our work, we develop a comprehensive logical framework that effectively deals with the overcomplete basis set in the BdG equation. This framework encompasses both positive and negative energy quasiparticle states while maintaining symmetry between spin-up and spin-down quasiparticle states in scattering and tunneling theory and is applicable across a wide range of condensed matter physics.
Our logical framework is crucial for both theoretical concepts and experimental predictions. We emphasize the conceptual significance of our logical framework by introducing new techniques that accurately determine all eigenfunctions of arbitrary nonuniform superconductors. Notably,  superconducting eigenstates are full of superconducting spin clouds, representing the electron configuration within Cooper-like pairs in nonuniform superconductors. We elucidate how Bogoliubov excitations reconfigure these superconducting spin clouds. Additionally, we demonstrate the existence of a conjugate loop formed by the EVSs of two OBSs conjugate with each other, such as the Fermi sea and sky, representing the EVSs of positive and negative OBSs, respectively. We highlight the conceptual significance of the Fermi sky, serving as the starting point for filling all negative quasiparticles within the Fermi sea. Shifting our focus to experimental predictions, we highlight the importance of the conjugate loop in revealing the one-body quasiparticle representation of the many-body superconducting state. This discovery provides a solid foundation for treating a many-body state as a one-body problem and proves valuable in addressing superconducting tunneling spectroscopy as simply as the standard nonsuperconducting spectroscopy theory. For example, we demonstrate a gate-, field-, and phase-tunable tunnel spectroscopy asymmetry caused by the imbalanced particle-hole distribution of the subgap quasiparticles in a quantum-dot Josephson junction. Overall, our work presents a comprehensive logical framework for addressing the challenges posed by the overcomplete basis set and for providing valuable tools for studying and understanding various aspects of condensed matter physics, particularly in the context of nonuniform superconductors.

The paper is organized as follows. In Sec.~\ref{fvanfm},  we show the conjugate relation of arbitrary superconducting system. In Sec.~\ref{logic}, we present our generic logic, including quantum dot (Sec.~\ref{quantumdot}), normal metal (Sec.~\ref{newlogic}), superconductor (Sec.~\ref{superconductor}), and general case (Sec. \ref{logic}).  Section~\ref{theory} presents our main results including conjugate loop (Sec.~\ref{conjugatlopp}), Fermi sea and sky (Sec.~\ref{quasiparticleoccupancy}), as well as tunnel spectroscopy asymmetry (Sec.~\ref{tunnelspectrocopyanomaly}). Our paper ends with conclusion and outlook. The Hamiltonian for plane-wave and tight-binding models is provided in Appendix~\ref{ModelHamiltonian} and Appendix~\ref{tightbindingmodel} respectively. To gain an intuitive understanding of our  logical framework, we provide examples starting with a simple quantum dot case (Appendix~\ref{quantumdot}),  followed by normal metal (Appendix~\ref{newlogic}) and superconductor (Appendix~\ref{superconductor}) cases. Finally, Appendix~\ref{effectivevacuumstates} and  Appendix~\ref{tunnelingcurrent} present the microscopic derivations of effective vacuum states and tunneling current, respectively.

\section{Model} \label{fvanfm}

Our methodology involves a quadratic and diagonalizable Hamiltonian that describes a non-uniform superconducting system incorporating various elements such as a Zeeman magnetic field, spin-orbit coupling, spin-dependent tunneling, and Coulomb interaction treated in a mean-field manner. This framework allows us to investigate a wide range of superconducting systems, including superconductor-quantum dot  \cite{de2010hybrid}, -semiconductor \cite{prada2020andreev}, -ferromagnet \cite{bergeret2005odd}, and -superconductor \cite{strambini2016omega} heterostructures as well as superconducting quantum interference device \cite{fagaly2006superconducting}. Our methodology is applicable to different models, including plane-wave and tight-binding models. The Hamiltonian for each specific model is provided in Appendix \ref{ModelHamiltonian} and Appendix \ref{tightbindingmodel}, respectively. Despite the artificial redundancy introduced by the BdG Hamiltonian, we can mathematically transform it into a quadratic and diagonalizable form using an overcomplete basis set, which enables exact solvability. The resulting Hamiltonian is given by
\begin{align} \label{fdbldfl}
   H&=\frac{1}{2}\sum^{N}_{l=1}\sum_{\sigma=\Uparrow/\Downarrow}\sum_{\eta=+/-} E^{}_{l\sigma\eta}\gamma_{l\sigma\eta}^{\dagger} \gamma_{l\sigma\eta }^{}+\mathcal{E}.
\end{align} 
Here, $N$ represents the system size, and $\sigma=\Uparrow/\Downarrow$ denotes pseudospin-up and -down quasiparticles in the presence of spin-flip effects from spin-orbit coupling and spin-dependent tunneling. The appearance of the half prefactor arises when we mathematically rewrite the original Hamiltonian into the quadratic and diagonalizable BdG form using the overcomplete basis set \cite{bernevig2013topological}. The above Hamiltonian involves the Nambu and spin space~\cite{de2018superconductivity}, and an additional index $\eta=+/-$ labels the high/low energy levels of each pseudospin species ($E_{l\sigma+}>E_{l\sigma-}$), satisfying $E_{l\sigma\eta}=-E_{l-\sigma-\eta}$ and $E_{l\Uparrow\eta}>E_{l\Downarrow\eta}$. The quasiparticle operator associated with energy $E_{l\sigma\eta}$ is represented as a unit vector in the spin-Nambu space, which encompasses the degrees of freedom of the hybrid system
\begin{equation} \label{fvvldl}
    \gamma_{l\sigma\eta}= \sum^{N}_{n=1}\sum_{s=\uparrow,\downarrow}\left[(u_{\eta})^{\sigma s}_{ln}c^{}_{ns}+(v_{\eta})^{\sigma s}_{ln}(-sc^{\dagger}_{n-s})\right].
\end{equation} 
The field operator $c_{ns}^{}$ annihilates an electron with spin $s$ and energy $\epsilon_n+sh_n$, where $\epsilon_{n}$ and $h_n$ are the energy spectrum and Zeeman energy. The constant energy $\mathcal{E}=\mathcal{E}_{\epsilon}+\mathcal{E}_{\Delta}$ includes the contributions from energy spectrum $\epsilon_{n}$, given by $\mathcal{E}_{\epsilon}=\sum_{n}\epsilon_{n}$,  and pair potential $\Delta_n$, given by $\mathcal{E}_{\Delta}=\sum_{n}\vert \Delta_n\vert^2/g$, where $g$ is attractive interaction, e.g., the Gorkov contact pairing interaction and $\mathcal{E}_{\Delta}$ is the energy constant originating from the mean-field treatment of the attractive interaction~\cite{beenakker2005superconducting}. The Nambu space allows us to treat particles and holes on an equal footing and enables a unified description of superconducting systems.

The quasiparticle operators $(\gamma_{l\sigma\eta})$ for all $l$, $\sigma$, and $\eta$ form an overcomplete basis set that consists of two mutually conjugate OBSs. The quasiparticle operators described by Eq. \eqref{fvvldl} satisfy the conjugate relation
\begin{equation} \label{tpfvvldl}
\gamma_{l\sigma\eta}^\dagger = \gamma_{l-\sigma-\eta}.
\end{equation}
As a consequence of this conjugate relation, the electron ($u_\eta$) and hole ($v_\eta$) distributions satisfy the following relation $(u_\eta^*)^{\sigma s}_{ln} = s(v_{-\eta})^{-\sigma -s}_{ln}$. The conjugate relation~\eqref{tpfvvldl} offers an intuitive and comprehensive illustration of the artificial redundancy in the BdG formalism from a quasiparticle operator point of view. It is important to note that the ubiquitous conjugation relation is different from the demanding requirement of the Majorana fermion, i.e., $\gamma_{ls\eta}^\dagger = \gamma_{ls\eta}$. 
The conjugate relation \eqref{tpfvvldl} reveals that the occupancy of the $E_{l\sigma+}$ quasiparticle is entirely determined by the occupancy of the $E_{l\sigma-}$ quasiparticle. An occupied $E_{l\sigma\eta}$ quasiparticle, i.e., $\langle \gamma_{l\sigma\eta}^\dagger \gamma_{l\sigma\eta} \rangle = 1$, is equivalent to an unfilled $E_{l-\sigma-\eta}$ quasiparticle, i.e., $\langle \gamma_{l-\sigma-\eta}^\dagger \gamma_{l-\sigma-\eta} \rangle = \langle \gamma_{l\sigma\eta} \gamma_{l\sigma\eta}^\dagger \rangle = 0$. However, it is important to note that both $\gamma_{l\sigma\eta}$ and $\gamma_{l-\sigma-\eta}$ describe the same quasiparticle and no longer satisfy the anticommutation relation $\{\gamma_{l\sigma\eta},\gamma_{l-\sigma-\eta}\}=\{\gamma_{l\sigma\eta},\gamma_{l\sigma\eta}^\dagger\}\neq 0$. From the $4N$-element overcomplete basis set $(\gamma_{l\sigma\eta})$ for all $l$, $\sigma$, and $\eta$, we can select an OBS in which the $2N$ elements are independent of each other and satisfy the anticommutation relation. It is worth noting that, in principle, there are $4^N$ OBSs that can be constructed, as each quasiparticle has its conjugate counterpart with opposite pseudospin and energy. However, the choice of OBS is arbitrary and does not affect the behavior of superconducting materials. Thus, it is necessary to develop a general logic that is independent of the chosen OBS.
     
\section{Logic} \label{logic}

We present our logic for the usage of the overcomplete basis set including two OBSs conjugate with each other 
\begin{itemize}
    \item STEP I): divide the overcomplete basis set [$(\gamma^{}_{l\sigma\eta})$ for all $l$, $\sigma$, and $\eta$] into two OBSs conjugate with each other; 
    \item STEP II): derive the  EVSs of these two OBSs; 
    \item STEP III): ground state is a filled sea of all negative basis states in the chosen OBS from the corresponding EVS.
\end{itemize}
Then we attain the following three properties 
\begin{itemize}
    \item PROPERTY I): the  EVSs of any two OBSs conjugate with each other form a conjugate loop;
    \item PROPERTY II): there are $2^N$ possible divisions of the two OBSs conjugate with each other. When a pair of basis states is exchanged between the two OBSs, the fermionic parity (see definition in Ref. \footnote{The fermionic parity operator, denoted as $P=(-1)^{\sum_{ns}c^{\dagger}_{ns}c^{}_{ns}}$, provides information about the number of electrons in a system. It assigns an eigenvalue of $-1$ if the number of electrons is odd and $+1$ if the number is even. While the conservation of electron number is not applicable in a superconductor, the conservation of fermionic parity, denoted as $[P,H]=0$, holds true. This conservation of fermionic parity implies that if the initial state of the system has an even (odd) number of fermions, the final state will also have an even (odd) number of fermions for a fixed set of system parameters. However, it is important to note that the ground state of a superconductor can undergo changes in fermionic parity with variations in system parameters, such as Coulomb interaction, magnetic field, and phase difference in Josephson junctions.}) of the EVSs switches.
    \item PROPERTY III): The even-parity eigenstates are full of Bogoliubov-like singlets -- a superposition of vacuum state and Cooper-like pair, while the odd-parity eigenstates contain not only plenty of Bogoliubov-like singlets but also a superconducting pseudospin cloud.
\end{itemize} 
To gain an intuitive understanding of this logic, we provide examples starting with a simple quantum dot case (Appendix~\ref{quantumdot}),  followed by normal metal (Appendix~\ref{newlogic}) and superconductor (Appendix~\ref{superconductor}) cases. We here generalize the logic to arbitrary nonuniform superconductor case.

STEP I):  Here, we work on arbitrary nonuniform superconductor with the BdG  Hamiltonian \eqref{fdbldfl}. The conventional superconductor theory relies on an excitation OBS, whose EVS happens to to be the ground state. However, the ground state can evolve a changeover of fermionic parity, and hence it is more convenient to start with the OBSs whose quasiparticle operators are smooth functions of all system parameters and support even-parity EVSs that is analytically solvable~\cite{yanagisawa2003lattice,mitake2002possible,yanagisawa2002stripes}.  Here, we divide the overcomplete basis set into two OBSs dual to each other --  $(\gamma^{}_{l\sigma+})$ for all $l$ and $\sigma$ as well as $(\gamma^{}_{l\sigma-})$ for all $l$ and $\sigma$. For our purposes, we assume that the above division of OBSs guarantees the even-parity EVSs.  The corresponding Hamiltonian can be obtained from the Hamiltonian \eqref{fdbldfl} by transforming the remaining quasiparticles into the chosen OBS according to the conjugate relation \eqref{tpfvvldl}
\begin{align} \label{nvdknkdf}
    H&=\sum^{N}_{l=1}\sum_{\sigma=\Uparrow/\Downarrow} E^{}_{l\sigma+}\gamma_{l\sigma+}^{\dagger} \gamma_{l\sigma+ }^{}+\mathcal{E}_{+}\\
    &=\sum^{N}_{l=1}\sum_{\sigma=\Uparrow/\Downarrow} E^{}_{l\sigma-}\gamma_{l\sigma-}^{\dagger} \gamma_{l\sigma- }^{}+\mathcal{E}_{-},\notag 
\end{align} 
where the energy of the  EVS for OBS  $(\gamma^{}_{l\sigma\eta})$ for all $l$ and $\sigma$ is given by 
\begin{align} \label{fdvmdfklp}
    \mathcal{E}_{+}=\mathcal{E}+\sum_{l\sigma}\frac{1}{2}E^{}_{l\sigma-},
\end{align}
\begin{align} \label{fdvmdfklm}
    \mathcal{E}_{-}=\mathcal{E}+\sum_{l\sigma}\frac{1}{2}E^{}_{l\sigma+}.
\end{align}

STEP II): Then, we derive the even-parity  EVS. The EVS of the OBSs $(\gamma^{}_{l\Uparrow\eta},\gamma^{}_{l\Downarrow\eta})$  for all $l$ is defined by 
\begin{align} \label{fvndfjnv}
    \gamma_{l\sigma\eta}\vert V\rangle_{\eta}=0, \text{ for all $l$ and $\sigma$}.
\end{align}
We assume a generic ansatz, which is available for the spin-flip effects from spin-orbit coupling and spin-dependent tunneling 
\begin{equation} \label{ufvdfmmvl}
   \vert V\rangle_{\eta}\propto e^{\frac{1}{2}\sum_{bb'}c^{\dagger}_{b}\mathcal{C}^{\eta}_{bb'}c^{\dagger}_{b'}}\vert 0\rangle,
\end{equation}
where $|0\rangle$ is the vacuum of electrons defined by $c_{b}|0\rangle=0$ for all $b=(n,s)$. The superconducting cloud matrix $\mathcal{C}^{\eta}$ relies on the electron and hole distributions of all $\eta$ quasiparticles~\cite{yanagisawa2003lattice,mitake2002possible,yanagisawa2002stripes} (see derivations in Appendix \ref{effectivevacuumstates})
\begin{align} \label{mfvnkdk}
    \mathcal{C}^{\eta}=-\begin{bmatrix}
      (u_{\eta})^{\Uparrow\uparrow} & (u_{\eta})^{\Uparrow\downarrow} \\
      (u_{\eta})^{\Downarrow\uparrow} &  (u_{\eta})^{\Downarrow\downarrow}
    \end{bmatrix}^{-1} \begin{bmatrix}
      (v_{\eta})^{\Uparrow\downarrow} &-(v_{\eta})^{\Uparrow\uparrow} \\
       (v_{\eta})^{\Downarrow\downarrow} & -(v_{\eta})^{\Downarrow\uparrow}
    \end{bmatrix},
\end{align}
and show that all electrons can be entangled. Though the superconducting cloud matrix \eqref{mfvnkdk} is non-Hermitian, we can still diagonalize it using singular value decomposition, $\mathcal{C}^{\eta}=U_{\eta}\mathcal{A}^{\eta}V^{+}_{\eta}$. Here, $U_{\eta}$ and $V_{\eta}$ are two unitary matrices that are not independent and satisfy $U_{\eta}(n,2k-1)=e^{+i\phi^{\eta}_k}V^{+}_{\eta}(2k,n)$ for all $n$. $\mathcal{A}^{\eta}$ is a diagonal matrix with doubly degenerate Bogoliubov coefficients  $\mathcal{A}^{\eta}_{k\sigma,k\sigma}=\mathcal{A}^{\eta}_{kk}$, where $\sigma=\Uparrow,\Downarrow$ is a pseudospin index corresponding to this double degeneracy.  Then, EVSs \eqref{ufvdfmmvl} can be rewritten to the Cooper-like pair form  
\begin{equation} \label{trufvdfmmvl}
    \vert V\rangle_{\eta}=\frac{1}{N^{1/2}_{\eta}}\prod^{N}_{k=1}\left(1+e^{i\phi^{\eta}_k}\mathcal{A}^{\eta}_{kk}a^{\dagger}_{k\Uparrow\eta}a^{\dagger}_{k\Downarrow\eta}\right)\vert 0\rangle,
\end{equation}
where $N_{\eta}=\prod_{k}\left(1+\vert\mathcal{A}^{\eta}_{kk}\vert^2\right)$.
The superconducting pseudospin cloud operators show how the electrons in a nonuniform superconductor pair with each other 
\begin{align} \label{andreevcloud}
a^{\dagger}_{k\sigma\eta}=\sum_{ns}c^{\dagger}_{ns}(U_{\eta})^{ s\sigma }_{nk}.
\end{align}
The EVS \eqref{trufvdfmmvl} is full of  Bogoliubov-like singlet -- a superposition of vacuum state and Cooper-like pair. The latter is a pair of superconducting clouds  with opposite pseudospins. The superconducting pseudospin cloud operators  $(a_{k\sigma\eta})$ for all $k$ and $\sigma$ also form an OBS and shows how electrons pair with each other in the EVS \eqref{trufvdfmmvl}.

STEP III): We express the ground state in terms of the  EVSs \eqref{trufvdfmmvl} following our new logic. Let us calculate the superconducting ground state in term of the $\eta=+$ OBSs  $(\gamma^{}_{l\Uparrow+},\gamma^{}_{l\Downarrow+})$ for all $l$. The superconductor has $4^{N}$ eigenstates. They are $C(2N,0)$ EVS $\vert V \rangle_+$, $C(2N,1)$ one-Bogoliubon states  $\gamma^{\dagger}_{l_1\sigma_1+}\vert V \rangle_+$, $C(2N,2)$ two-Bogoliubon states $\gamma^{\dagger}_{l_1\sigma_1+}\gamma^{\dagger}_{l_2\sigma_2+}\vert V \rangle_+$, and so on, up to $C(2N,2N)$ $2N$-Bogolibon state $(\prod_{l\sigma}\gamma^{\dagger}_{l\sigma+})\vert V \rangle_+$ and the corresponding energy are, $\mathcal{E}^{L}_{+}$, $\mathcal{E}^{L}_{+}+E_{l_1\sigma_1+}$, $\mathcal{E}^{L}_{+}+E_{l_1\sigma_1+}+E_{l_2\sigma_2+}$, and so on, up to $\mathcal{E}^{L}_{+}+\sum_{l\sigma}E_{l\sigma+}$,  respectively, where $l_i,s_i$ are required to satisfy the Pauli exclusion principle for the eigen states with many Bogoliubov quasiparticles. The ground state is the lowest energy state among these $4^{N}$ states and therefore is the Fermi sea -- a filled sea of all negative Bogoliubov quasiparticles in the OBS $(\gamma_{l\Uparrow+},\gamma_{l\Downarrow+})$ from the  EVS $\vert V\rangle_{+}$. The same goes for the $\eta=-$ OBSs $(\gamma^{}_{l\Uparrow-},\gamma^{}_{l\Downarrow-})$ with the EVS \eqref{trufvdfmmvl}. Therefore,  we obtain the ground state of arbitrary nonuniform superconductor
\begin{align} \label{yfvklal}
    \vert G \rangle=\left(\prod_{E_{l\sigma+}<0}\gamma^{\dagger}_{l\sigma+}\right)\vert V\rangle_+=\left(\prod_{E_{l\sigma-}<0}\gamma^{\dagger}_{l\sigma-}\right)\vert V\rangle_-,
\end{align}
whose ground-state energy is given by  
\begin{align} \label{fdjvkdf} 
    \mathcal{E}_G=\mathcal{E}_{+}+\sum_{E_{l\sigma+}<0}E_{l\sigma+}=\mathcal{E}_{-}+\sum_{E_{l\sigma-}<0}E_{l\sigma-}.
\end{align}

PROPERTY I): Then, we try to find the relation between EVSs $\vert V\rangle_+$ and $\vert V\rangle_-$ [Eq. \eqref{trufvdfmmvl}]. In the absence of the Zeeman magnetic fields, the ground state can be expressed in terms of two OBSs $\vert G
\rangle=\vert V\rangle_{+}=\prod_{ l} \gamma_{ l \uparrow-\eta}^{\dagger}\gamma_{l \downarrow-\eta}^{\dagger} \vert V\rangle_{-}$. Thus, the EVSs $\vert V\rangle_{+}$ and $\vert V\rangle_{-}$  form a conjugate loop 
\begin{equation} \label{fvdkmk}
    \vert V\rangle_{\eta}= \prod_{l=1}^{N} \gamma^{\dagger}_{l\Uparrow-\eta}\gamma^{\dagger}_{l\Downarrow-\eta}\vert V\rangle_{-\eta}.
\end{equation}
The one-body quasiparticle representation of the many-body  EVSs (\ref{fvdkmk}) not only solves the conceptual difficulties of the disappeared states dual to the chosen OBS (see Sec. \ref{conjugatlopp}) but also intuitively explains the  asymmetry of the negative- $(\propto E_{-\sigma-})$ and positive-voltage $(\propto E_{\sigma+})$ peaks reveals the  electron-hole distribution imbalance of the Andreev bound states (see Sec. \ref{tunnelspectrocopyanomaly}).

\begin{figure}[t!]
\begin{center}
\includegraphics[width=0.8\linewidth]{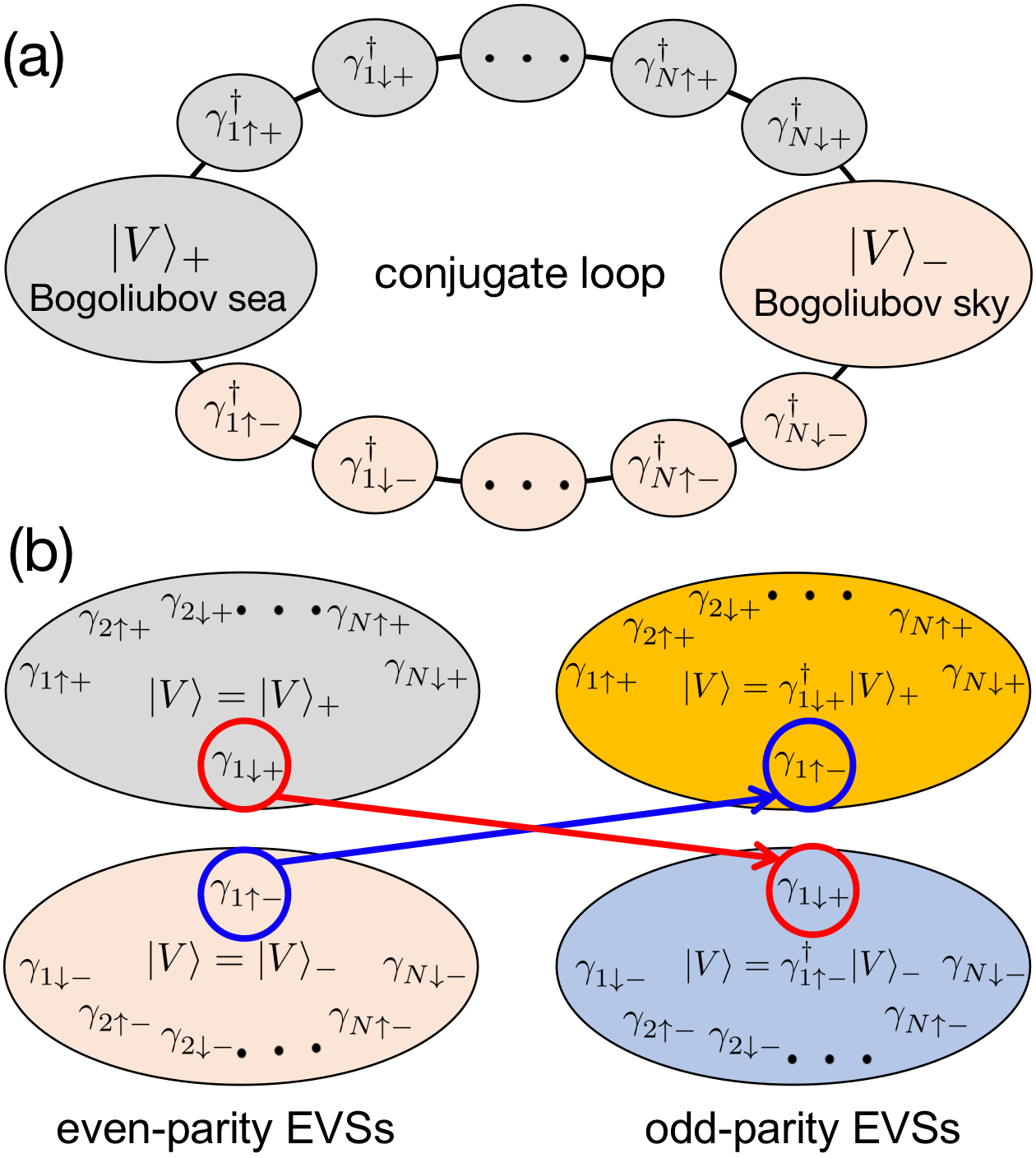}
\end{center}
\caption{Conjugate loop and parity change of effective vacuum states (EVSs). Here, we exclude spin-flip effect and assume the case of  $E_{ls+}>0$ and $E_{ls-}<0$, where Fermi sea and sky become the EVSs $\vert V \rangle_{+}$ and $\vert V \rangle_{-}$, respectively. (a) Fermi sea and sky -- the  EVSs of $\eta=+$ and $\eta=-$ orthonormal basis sets [Eq. \eqref{trufvdfmmvl}], respectively, contain all $\eta=-$ and $\eta=+$ Bogoliubons [Eq. \eqref{fvdkmk}] and therefore form a conjugate loop.  (b) Parity change of the EVSs due to the exchange of a pair of basis states conjugate with each other, e.g., Andreev operators $\gamma^{}_{1\downarrow+}$ and $\gamma^{}_{1\uparrow-}$.} 
\label{FIGSTORY} 
\end{figure}

PROPERTY II):  Next, we show that whenever a pair of basis states conjugate with each other is exchanged, the fermionic parity of  EVSs switches  [Fig.~\ref{FIGSTORY}(b)]. 
Once we have a pair of the  EVSs, for example, EVSs \eqref{trufvdfmmvl}, we can construct new EVSs by adding the exchanged quasiparticles, e.g., $\gamma^{}_{1\Downarrow+}$ and $\gamma^{}_{1\Uparrow-}$. By utilizing Eqs.~\eqref{tpfvvldl} and \eqref{fvndfjnv},  it can be proven that $\gamma^{\dagger}_{1\Downarrow+}\vert V\rangle_{+}$ and $\gamma^{\dagger}_{1\Uparrow-}\vert V\rangle_{-}$ represent the EVSs of the two OBSs after the exchange. This exchange naturally occurs when we consider the ground state with a pair of subgap levels crossing zero~\cite{zhang2022enhancement,zhang2023singlet}.

PROPERTY III):  The even-parity eigenstates $\vert e\rangle_{\imath}$ of these $4^N$ eigenstates always has Bogoliubov-like singlet form
\begin{equation} \label{trufvDdfmmvl}
    \vert e\rangle_{\imath}=\frac{1}{N^{1/2}_{\imath}}\prod^{N}_{k=1}\left(1+e^{i\phi^{\imath}_k}\mathcal{A}^{\imath}_{k}a^{\dagger}_{k\Uparrow\imath}a^{\dagger}_{k\Downarrow\imath}\right)\vert 0\rangle.
\end{equation}
Because $\vert e\rangle_{\imath}$ can be expressed by $\vert V\rangle_+$ by adding even number of $\eta=+$ Bogoliubons and hence is the EVS after exchange these Bogoliubons according to PROPERTY II), which has Cooper-like pair form. This indicates that Bogoliubov excitations reconfigure the superconducting pseudospin clouds within the superconducting condensate. The odd-parity eigen  states, $\vert o \rangle^{l\sigma}_{\imath}$ can be always expressed from one of even-parity eigen state $\vert e\rangle_{\imath}$ by adding one Bogoliubon such as $\gamma^{\dagger}_{l\sigma+}$. By means of Eqs. \eqref{andreevcloud} and \eqref{fvvldl}, we obtain odd-parity eigenstates 
\begin{align} \label{oddgroundstate}
    \vert o \rangle^{l\sigma}_{\imath} =\frac{1}{N^{1/2}_{\imath}} \sum_{k\sigma'}\mathcal{B}^{+}_{k\sigma'}a^{\dagger}_{k\sigma'\imath}\prod^{}_{k'\neq k}\left(1+e^{i\phi^{\imath}_k}\mathcal{A}^{\imath}_{k}a^{\dagger}_{k\Uparrow\imath}a^{\dagger}_{k\Downarrow\imath}\right)\vert 0\rangle,
\end{align}
with $\mathcal{B}^{l\sigma}_{k\sigma'}=\sum_{ns}[(u^{*}_{+})^{\sigma s}_{ln}(U^{*}_{\imath})^{ s\sigma'}_{nk}+(v_{+})^{\sigma s}_{ln}s\sigma'(U^{}_{\imath})^{-s-\sigma'}_{nk}e^{i\phi^{\imath}_k}\mathcal{A}^{\imath}_{k}]$. Eq. \eqref{oddgroundstate} implies that adding a Bogoliubon into Bogoliubov-like singlet results in a superconducting pseudospin cloud.   Thus, the odd-parity eigenstate \eqref{oddgroundstate} contains not only plenty of Cooper-like pairs but also a superconducting pseudospin cloud. The superconducting pseudospin clouds  $\{a_{k\sigma+}\}$ for all $k$ and $\sigma$, as an OBS, are useful to describe both even- and odd-parity eigenstates [Eqs.  \eqref{trufvDdfmmvl} and \eqref{oddgroundstate}]. Therefore, we obtain the exact expressions of all superconducting eigenstates.

\section{Results and discussions} \label{theory}

\subsection{Conjugate loop} \label{conjugatlopp}

The conceptual significance of the conjugate loop is emphasized. Note that there exists a conceptual difficulty of apparent disregard of half quasiparticles in superconducting scattering and tunneling theory~\cite{datta1999can,datta1996scattering,coleman2015introduction}. Take the superconductor case as an example (Appendix \ref{superconductor}). The Bogoliubov operators  $\gamma^{}_{\vec{k}s+}=u_{\vec{k}}c^{}_{\vec{k}s}+sv^{}_{\vec{k}}c^{\dagger}_{-\vec{k}-s}$ and $\gamma^{}_{\vec{k}s-}=-s v^{}_{\vec{k}}c^{}_{\vec{k}s}+u_{\vec{k}}c^{\dagger}_{-\vec{k}-s}$ satisfy the conjugate relation $\gamma^{\dagger}_{\vec{k}s\eta}=  \gamma^{}_{-\vec{k}-s-\eta}$, where the Bogoliubov coefficients are $u_{\vec{k}}=\frac{1}{\sqrt{2}}(1+\epsilon_{\vec{k}}/E_{\vec{k}})^{1/2}$ and $v_{\vec{k}}=\frac{1}{\sqrt{2}}(1-\epsilon_{\vec{k}}/E_{\vec{k}})^{1/2}$ with $E_{\vec{k}}=\sqrt{\Delta^2+\epsilon^2_{\vec{k}}}$.  The conventional condensed matter theory relies on an excitation OBS, which includes only positive quasiparticles $(\gamma_{\vec{k}\uparrow+},\gamma_{\vec{k}\downarrow+})$ for $\vec{k}$ in the whole Brillouin zone, and hence faces a conceptual challenge as it overlooks the existence of negative energy quasiparticle states. To address this limitation, S. Datta et al. introduced an OBS consisting of spin-up quasiparticles $(\gamma_{\vec{k}\uparrow+},\gamma_{\vec{k}\uparrow-})$ for $\vec{k}$ in the whole Brillouin zone, to include the negative quasiparticles \cite{datta1996scattering,datta1999can}. However, this OBS encounters the similar difficulty of neglecting spin-down quasiparticle states, which are crucial for understanding spin-dependent scattering and tunneling phenomena. An alternative approach, known as the semiconductor analogy \cite{coleman2015introduction}, includes both spin-up (positive) and spin-down (negative) quasiparticles by choosing the half Brillouin zone. Unfortunately, this OBS neglects the contribution of the other half of the Brillouin zone quasiparticles. All these attempts overlook the importance of the EVS.

These conceptual difficulties are completely resolved by the conjugate loop, which reveals that the EVS of one OBS contains all the remaining quasiparticles that are dual to the chosen ones [see Eqs. \eqref{pmvfnjk1}, \eqref{fdkfmvdk}, \eqref{dkfmvdk1}, and \eqref{dkfmvdk2} in Appendix \ref{superconductor}]. Thus, the disregarded half quasiparticles are contained in the EVS and also participate   in superconducting scattering and tunneling theory, as shown in Sec.~\ref{tunnelspectrocopyanomaly}. Moreover, we will show how the conjugate loop helps in understanding the correct occupancy of negative quasiparticles (Sec. \ref{quasiparticleoccupancy}). Therefore, the conjugate loop provides a comprehensive framework that incorporates both positive and negative quasiparticles, preserving the symmetry between spin-up and spin-down quasiparticles in scattering and tunneling theory. 

\subsection{Fermi sea and sky} \label{quasiparticleoccupancy}

Note that the conventional superconductor theory relies on an  excitation OBS that only considers positive quasiparticles. Then, we divide the overcomplete basis set into a positive OBS, given by  $(\gamma^{}_{l\sigma\eta})$ for all $E_{l\sigma\eta}>0$,  as well as a negative OBS, given by $(\gamma^{}_{l\sigma\eta})$ for all $E_{l\sigma\eta}<0$. The EVSs of the positive and negative OBSs, defined by $\gamma_{l\sigma\eta}\vert \text{Fermi sea}\rangle=0 \text{ for all $E_{l\sigma\eta}>0$}$ and $\gamma_{l\sigma\eta}\vert \text{Fermi sky}\rangle=0 \text{ for all $E_{l\sigma\eta}<0$}$, are dubbed as the Fermi sea and sky, respectively, because the Fermi sea and sky, in the absence of superconductivity, reduce to all negative  and positive electron states being occupied, respectively.  Thus, the Fermi sea is equal to the ground state, i.e., $ \vert G\rangle=\vert \text{Fermi sea}\rangle$.  PROPERTY I) shows that the Fermi sea and sky should form a conjugate loop as follows [see also Fig. \ref{FIGSTORY}(a)]
\begin{equation} \label{fvdkmksea}
    \vert \text{Fermi sea}\rangle= \left(\prod_{E_{l\sigma\eta<0}} \gamma^{\dagger}_{l\sigma\eta}\right)\vert \text{Fermi sky}\rangle,
\end{equation}
\begin{equation} \label{fvdkmksky}
    \vert \text{Fermi sky}\rangle= \left(\prod_{E_{l\sigma\eta>0}} \gamma^{\dagger}_{l\sigma\eta}\right)\vert \text{Fermi sea}\rangle.
\end{equation}
Notable, the Fermi sea (or ground state) is a filled sea of all negative quasiparticles from the Fermi sky which has huge energy $\mathcal{E}_{\vert \text{Fermi sky}\rangle}=\mathcal{E}+ \frac{1}{2}\sum_{E_{ls\eta}>0}E_{ls\eta}$ instead of the vacuum of electrons.  Fermi sky, as a complement of the Fermi sea,  should be a fundamental concept in the Fermion theory. In the case of  $E_{ls+}>0$ and $E_{ls-}<0$, the Fermi sea and sky become the EVSs $\vert V \rangle_{+}$ and $\vert V \rangle_{-}$, respectively. Hereafter, we exclude spin-flip effects from spin-orbit coupling and spin-dependent tunneling and thus pseudospin-up and -down $\sigma=\Uparrow/\Downarrow$ become spin-up and -down $s=\uparrow/\downarrow$.

\begin{figure}[t]
\begin{center}
\includegraphics[width=1.0\linewidth]{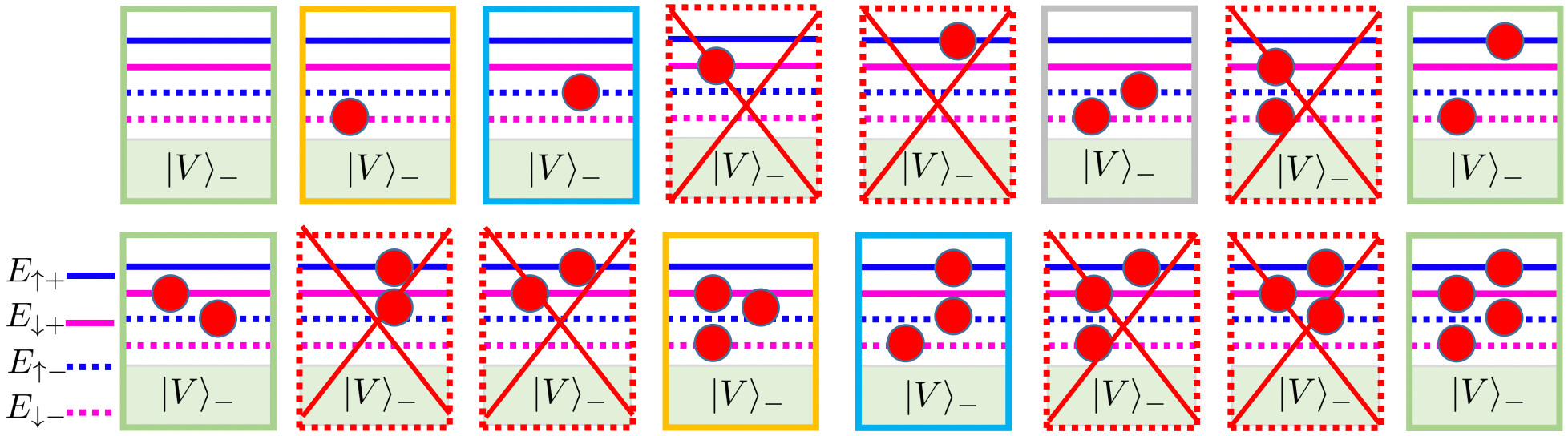}
\end{center}
\caption{The quasiparticle occupancy controversy in quantum dot.  Though there are four Andreev levels in  Nambu space, it only poses four states given in the reduced conjugate loops rather than $2^{4}$. For the rest 12 occupancies, five are identical to the four (same color boxes), while seven do not exist (red-dash boxes). }
\label{NQO}
\end{figure}

It is important to know that employing an overcomplete basis set, where the elements are not independent of each other, can lead to an increase in the dimension of the Hilbert space if we disregard their mutual dependence. This results in an artificially enlarged Hilbert space whose axis represents eigenstates of the system and necessitate careful consideration. It should be noted that certain states within this enhanced Hilbert space lack physical meaning, while some of these states are indistinguishable. To simplify matters, let's first focus on the quantum dot case (Appendix \ref{quantumdot}). Despite having four energy levels in the spin-Nambu space, only four distinct many-body states exist in the conjugate loop [Fig.~\ref{FIGQD}(e) and (f)], rather than the $2^4$ states that might be expected. The remaining 12 occupancy configurations shown in Fig. \ref{NQO} can be categorized as follows. Five of them are identical to the four states (indicated by the same color boxes), for example, $\gamma^{\dagger}_{\downarrow-}\vert V\rangle_{-}=\gamma^{\dagger}_{\downarrow+}\gamma^{\dagger}_{\uparrow-}\gamma^{\dagger}_{\downarrow-}\vert V\rangle_{-}$ (orange boxes). Additionally, seven occupancy configurations do not exist (highlighted in dashed red boxes) due to the conjugate relation, for example, $\gamma^{\dagger}_{\downarrow+}\vert V\rangle_{-}=\gamma^{}_{\uparrow-}\vert V\rangle_{-}=0$. Our results can be directly applied to low-energy subspace of the quantum-dot Josephson junction.   Furthermore, the conjugate loop implies that the EVS $\vert V\rangle_{+}$, representing the ground state at $h_D=0$ ($\vert G\rangle=\vert V\rangle_{+}$), is a filled Fermi sea consisting of all negative-energy quasiparticles in the OBS $(\gamma_{\uparrow-},\gamma_{\downarrow-})$, starting from the EVS $\vert V \rangle_-$ rather than the vacuum of electrons $\vert 0\rangle$ without electrons. Incorrectly using the EVS could lead to meaningless system states, such as $\gamma^{\dagger}_{\uparrow-}\gamma^{\dagger}_{\downarrow-}\vert 0\rangle=0$, $\gamma^{\dagger}_{\uparrow-}\vert 0\rangle=0$, $\gamma^{\dagger}_{\downarrow-}\vert 0\rangle=0$, where we consider $\epsilon_D>0$ case with $\gamma_{s+}=d_{s}$ and $\gamma_{s-}=d^{\dagger}_{-s}$. 

Next, we move on to the superconductor case (Appendix \ref{superconductor}). The zero magnetic field ground state, i.e., the Fermi sea, is a filled  sea of all negative Bogoliubons with respect to their EVS, i.e., Fermi sky, rather than the vacuum of electrons, as shown by the conjugate loop \eqref{fvdkmk}. A filled Fermi sea of all negative Bogoliubons from the vacuum of electrons is the ground state of the superconductor, i.e., $(\prod_{\vec{k}}\gamma^{\dagger}_{\vec{k}\downarrow-}\gamma^{\dagger}_{\vec{k}\uparrow-})\vert 0\rangle =\left(\prod_{\vec{k}}v^{}_{\vec{k}}\right) \vert G\rangle$ \cite{coleman2015introduction} in spite of the normalization prefactor, which is equal to creating all spin-up negative Bogoliubons and annihilating all spin-up positive Bogoliubons on 
the vacuum of electrons according to the conjugate relation, i.e., $(\prod_{\vec{k}}\gamma^{}_{\vec{k}\uparrow+}\gamma^{\dagger}_{\vec{k}\uparrow-})\vert 0\rangle=\left(\prod_{\vec{k}}v^{}_{\vec{k}}\right) \vert G\rangle$~\cite{himeda2002stripe}. The  normalization will not be a problem only when the prefactor $(\prod_{\vec{k}}v^{}_{\vec{k}})$ is nonzero.   However,  the normalization prefactor  becomes zero, i.e., $(\prod_{\vec{k}}v^{}_{\vec{k}})=0$ in the absence of superconductivity ($\Delta=0$) [see Eq. \eqref{fdvmkdfmk} in Appendix \ref{newlogic}]  and hence  $\prod_{\vec{k}}\gamma^{\dagger}_{\vec{k}\uparrow-}\gamma^{\dagger}_{\vec{k}\downarrow-}\vert 0\rangle=0$ becomes meaningless. Incorrectly using the EVS can once again result in meaningless system states and hence the conceptualization of the ground state, based on the vacuum of electrons, is unsatisfactory.  According to our logical framework, the occupancy of these negative quasiparticles should, in principle, refer to their EVS, i.e., the Fermi sky $\vert \text{Fermi sky}\rangle$, given by $\vert \text{Fermi sky}\rangle=\vert V\rangle_-=\prod_{\vec{k}}(v^{}_{\vec{k}}+u_{\vec{k}} c_{\vec{k} \uparrow}^{\dagger} c_{-\vec{k} \downarrow}^{\dagger})|0\rangle$. Note that we can express the vacuum of electrons in terms of the negative OBS and the EVS, i.e., $\vert 0\rangle=C_0\vert \text{Fermi sky}\rangle+C_{1}\vert \psi\rangle$ with $C_0=\langle \text{Fermi sky}\vert 0\rangle=\prod_{\vec{k}}v^{}_{\vec{k}} $, where $\vert \psi\rangle$ contains at least one negative quasiparticle from the EVS $\vert V\rangle_{-}$ and hence satisfies $(\prod_{\vec{k}}\gamma^{\dagger}_{\vec{k}\uparrow-}\gamma^{\dagger}_{\vec{k}\downarrow-})\vert \psi\rangle=0$ due to the Pauli exclusion principle.  Therefore, a filled Fermi sea of all negative quasiparticles from the vacuum of electrons is the ground state only when the $\vert \text{Fermi sky}\rangle$ component of the vacuum of electrons is not zero~\footnote{In this sense, ground state can be a filled Fermi sea of all negative quasiparticles from any two electrons with opposite momentum and spin  $c_{\vec{k} \uparrow}^{\dagger} c_{-\vec{k} \downarrow}^{\dagger}\vert 0\rangle $ because $\langle \text{Fermi sky}\vert c_{\vec{k} \uparrow}^{\dagger} c_{-\vec{k} \downarrow}^{\dagger}\vert 0\rangle\neq 0$. This description of the ground state is unsatisfactory because no matter what we choose as a starting point for filling all negative quasiparticles within the Fermi sea only its Fermi sky component survives.}. This fails when i) we study quantum dots and normal metals (see above) and ii) the ground state has a fermionic odd parity (as below).

Then, we consider the arbitrary nonuniform superconductors. Note that the fermionic parity of the ground state (or Fermi sea) switches when the pair of Andreev levels dual to each other cross zero [see Eq.~\eqref{yfvklal}]. Here, we focus on the odd-parity ground state, which can not be a filled Fermi sea of all negative quasiparticles from the vacuum of electrons, i.e., $\vert \text{Fermi sea}\rangle\neq  (\prod_{E_{ls\eta<0}} \gamma^{\dagger}_{ls\eta})\vert 0\rangle$ because Fermi sky, like the Fermi sea, has ferimionic odd parity and thus the $\vert \text{Fermi sky}\rangle$ component of the vacuum of electrons is zero, i.e., $C_0=0$, resulting a meaningless state, i.e., $(\prod_{E_{ls\eta<0}} \gamma^{\dagger}_{ls\eta})\vert 0\rangle=C_0 \vert \text{Fermi sea}\rangle=0$. Our conceptualization of the ground state [or Fermi sea \eqref{fvdkmksea}], applicable over the whole
field of condensed matter physics, not only naturally guarantees correct normalization conditions but also intuitively shows the ground-state energy $ \mathcal{E}_G=\mathcal{E}_{\vert \text{Fermi sky}\rangle}+\sum_{E_{ls\eta}<0}E_{ls\eta}$ according to the occupancy of the negative quasiparticles, which clarifies the half prefactor in the ground-state energy (as below).

Finally, we resolve the half prefactor in the ground-state energy. As a result of particle-hole symmetry $E_{ls\eta}=-E_{l-s-\eta}$, the ground-state energy \eqref{fdjvkdf} is equivalent to $\mathcal{E}_G=\mathcal{E}+ \frac{1}{2}\sum_{E_{ls\eta}<0}E_{ls\eta}$. Thus, the ground-state energy contains the half summation of all negative Bogoliubon energies and the energy constant of the BdG Hamiltonian \eqref{fdbldfl}. The calculation of the ground-state energy  is straightforward from the excitation OBS -- $(\gamma^{}_{ls\eta})$ for all $E_{ls\eta}>0$. The half summation naturally appears when we transform all negative Bogoliubons ($E_{ls\eta}<0$) of Eq.~\eqref{fdbldfl} into positive Bogoliubov  quasiparticles ($E_{ls\eta}>0$) using the conjugate relation \eqref{tpfvvldl}.  Alternatively, the half prefactor in Ref.~\cite{chtchelkatchev2003andreev} can be clearly understood as the fact that the zero-magnetic-field ground state $\vert G\rangle =\vert V\rangle_+$  is a filled  Fermi sea of all negative Bogoliubons with respect to their EVS $\vert V \rangle_-$, as shown by Eqs.~\eqref{yfvklal} and \eqref{fvdkmk}. Although the half prefactor is not explicitly present before $E_{l\sigma-}$ in the second expression of the ground-state energy \eqref{fdjvkdf}, it naturally emerges after the substitution of $\mathcal{E}_{-}=\mathcal{E}+\sum_{ls}\frac{1}{2}E^{}_{ls+}$. The ground-state supercurrent, quantified by the $\phi$-derivative of the ground state energy~\cite{anderson1964many,bagwell1992suppression,riedel1998low,beenakker1991universal,beenakker2013fermion}, can be expressed in different forms due to the particle-hole symmetry $E_{ls\eta}=-E_{l-s-\eta}$ 
\begin{align} \label{fvmakfvmkad}
\frac{I_G}{I_0}&=\partial_{\phi}\mathcal{E}_{+}+\sum_{E_{ls+}<0}\partial_{\phi}E_{ls+}=\partial_{\phi}\mathcal{E}_{-}+\sum_{E_{ls-}<0}\partial_{\phi}E_{ls-}\notag\\
&=\frac{1}{2}\sum_{E_{ls\eta}<0}\partial_{\phi}E_{ls\eta}=\sum_{E_{l\downarrow\eta}<0}\partial_{\phi}E_{l\downarrow\eta}=\sum_{E_{l\uparrow\eta}<0}\partial_{\phi}E_{l\uparrow\eta},
\end{align}
with $I_0=2e/\hbar$, where we assumed the $\phi$-independent magnitude of order parameter inside $\mathcal{E}$ for simplicity. 
These five expressions correspond to OBSs $(\gamma^{}_{ls+})$ for all $l$ and $s$,  $(\gamma^{}_{ls-})$ for all $l$ and $s$, $(\gamma^{}_{ls\eta})$ for all $E_{ls\eta}>0$, $(\gamma^{}_{l\downarrow\eta})$ for all $l$ and $\eta$, and $(\gamma^{}_{l\uparrow\eta})$ for all $l$ and $\eta$, respectively. The last one is originally proposed by S. Datta et al., whose  EVS, involving phase-independent energy, is full of all spin-down electrons~\cite{datta1996scattering,datta1999can}. The derivation of the ground-state supercurrent from the $\phi$ derivative of the ground-state energy naturally introduces a prefactor of $2e$ in $I_0$, which seems to be eliminated by the half prefactor in the third expression of Eq.~\eqref{fvmakfvmkad}. It is misleading that this half prefactor is due to the double counting, which adds $\mathcal{E}_{\Delta}$ into $\mathcal{E}$ to correct for a double counting of the attractive interaction energy in mean-field Hamiltonian~\cite{bardeen1969structure,beenakker2005superconducting}.  Note that the supercurrent is carried by the Cooper-like pairs rather than Bogoliubons, and each Cooper-like pair is relevant to both in-gap ($l=1$) and out-of-gap ($l>1$) quasiparticles [Eqs.~(\ref{fvndfjnv}-\ref{fdjvkdf})]. Therefore, it is necessary to properly include the out-of-gap contributions to ensure the uniqueness of the ground-state supercurrent \eqref{fvmakfvmkad}. 

In conclusion, we have clarified two widespread misunderstanding in the superconducting theory. They are: i) the ground state is a filled Fermi sea of all negative quasiparticles from the Fermi sky rather than the vacuum of electrons; ii)  the prefactor $2e$ in the above current-phase character does not refer to the charge of a Cooper pair and 
the half prefactor of the ground-state energy originates from the BdG formalism in terms of the overcomplete basis set.  Consequently, understanding the correct occupancy of negative Bogoliubons is conceptually important for a deeper understanding of the superconducting ground state.

\subsection{Tunnel spectroscopy asymmetry} \label{tunnelspectrocopyanomaly}
The existing theory of the tunneling spectroscopy of uniform superconductors, based on an excitation OBS, requires the breakup and formation of the Cooper pairs that nerve was directly demonstrated in experiments.  The tunneling Hamiltonian between superconductor and normal metal can be written as the sum of the processes $u_{\vec{k}}\gamma^{\dagger}_{\vec{k}s+}c^{}_{\vec{p}s}$ and $v^{}_{\vec{k}}c^{\dagger}_{\vec{p}s}\gamma^{\dagger}_{-\vec{k}-s+}$, together with their conjugate processes which are negligible for low enough temperature~\cite{tinkham2004introduction,anderson2006theory}. Here, field operator $c^{}_{\vec{p}s}$ annihilates an electron in the normal metal with momentum $\vec{p}$, spin $s$, and energy $E_{\vec{p}}$. Thus, the differential conductance can be separated into two terms, i.e., $\frac{dI}{dV}\propto \sum_{\vec{k}}u_{\vec{k}}^2 \delta\left(V-E_{\vec{k}}\right)+v_{\vec{k}}^2 \delta\left(V+E_{\vec{k}}\right)$~\cite{coleman2015introduction,schrieffer2018theory}. At positive bias, only the $u_{\vec{k}}$ process which transfers an electron into the superconductor contributes, while at negative bias, the $v_{\vec{k}}$ term dominates where the breakup of a Cooper pair creates a quasiparticle and ejects an electron into the normal metal.  Noting that positive $+V_0 $ and negative $-V_0 $ biases select the same Bogoliubon energy $E_{\vec{k}}=V_0$, for a fixed Bogoliubon energy $E_{\vec{k}}=V_0$, there exist two values of electron energy, $\epsilon_{\vec{k}_+}=+\sqrt{V_0^2-\Delta^2}$ and $\epsilon_{\vec{k}_{-}}=-\sqrt{V_0^2-\Delta^2}$. Thus, the differential conductance at positive and negative biases are given by $\left.\frac{dI}{dV}\right\vert_{V=+ V_0}\propto\nu_{}(\epsilon_{\vec{k}_{+}})u^2_{\vec{k}_+}+\nu_{}(\epsilon_{\vec{k}_{-}})u^2_{\vec{k}_-}$ and $\left.\frac{dI}{dV}\right\vert_{V=- V_0}\propto\nu_{}(\epsilon_{\vec{k}_{+}})u^2_{\vec{k}_-}+\nu_{}(\epsilon_{\vec{k}_{-}})u^2_{\vec{k}_+}$, respectively,  where we have used the identities $v^2_{\vec{k}_{\pm}}=u^2_{\vec{k}_{\mp}}$. Assuming $\epsilon_{\vec{k}}$-independent the density of state~\cite{coleman2015introduction}, the differential conductance exhibits symmetric tunneling conductivity between positive and negative  biases~\cite{tinkham2004introduction,coleman2015introduction,schrieffer2018theory}.

However, the breakup and formation of the Cooper pair is a mysterious process, even some experiments claimed they directly observed Cooper pairs from the pair correlations~\cite{holten2022observation}. Our conjugate loop not only tells us that the ground state is full of  $\gamma^{}_{\vec{k}s-}$ quasiparticle but also achieves the one-body quasiparticle representation of the many-body superconducting state, thus providing a solid foundation for treating a many-body superconducting state as a one-body problem.  Knowing that the $v^{}_{\vec{k}}c^{\dagger}_{\vec{p}s}\gamma^{\dagger}_{-\vec{k}-s+}(=v^{}_{\vec{k}}c^{\dagger}_{\vec{p}s}\gamma^{}_{\vec{k}s-})$ process can be intuitively reexplained by the annihilation of $\gamma^{}_{\vec{k}s-}$ quasiparticle by ejecting  an  electron into the tip. This description,  without the need for the Cooper pair, allows us to intuitively explain the tunneling spectroscopy of superconductors akin to the conventional nonsuperconducting tunnel spectroscopy.

\begin{figure}[t]
\begin{center}
\includegraphics[width=0.98\linewidth]{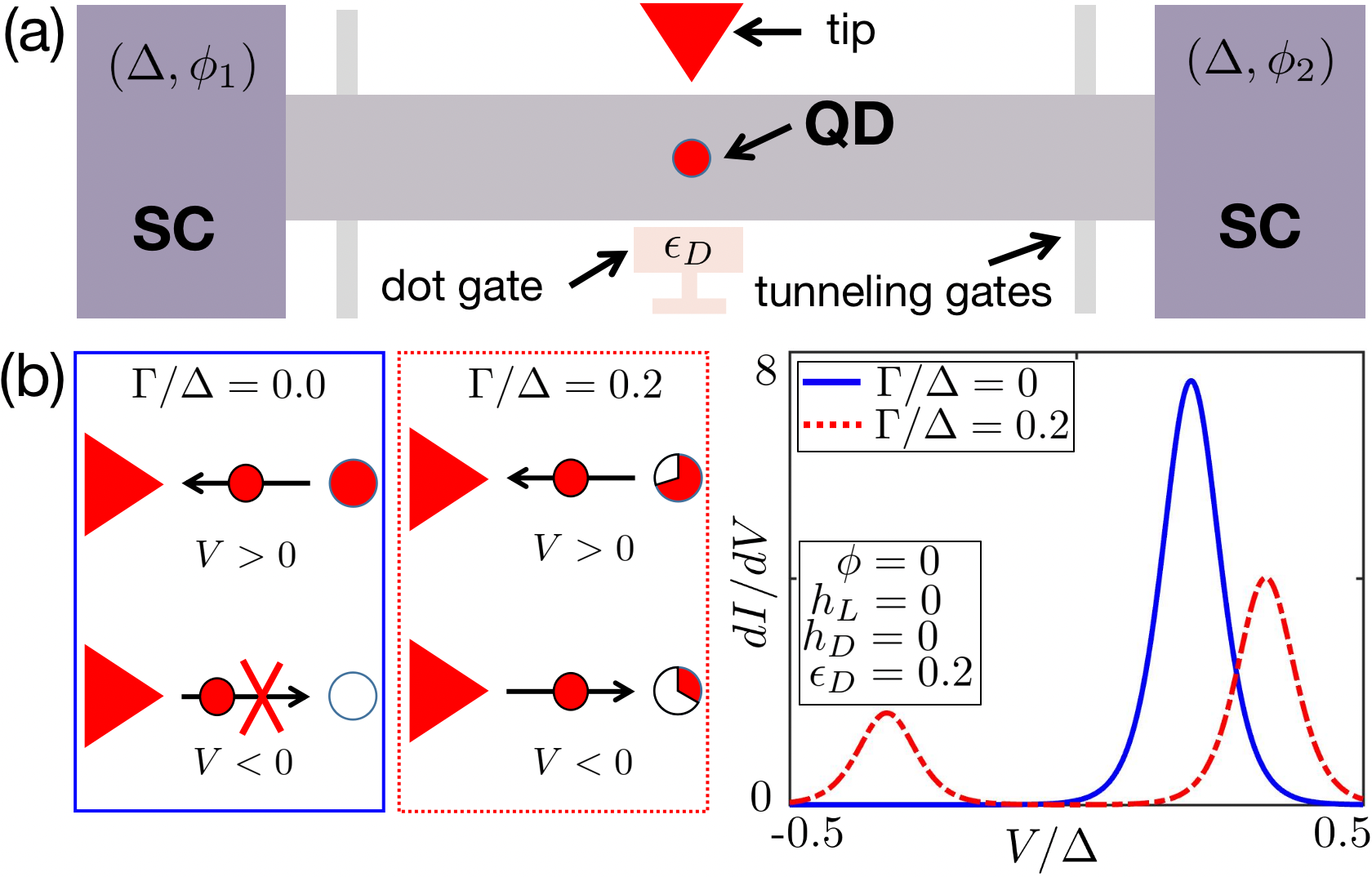}
\includegraphics[width=1\linewidth]{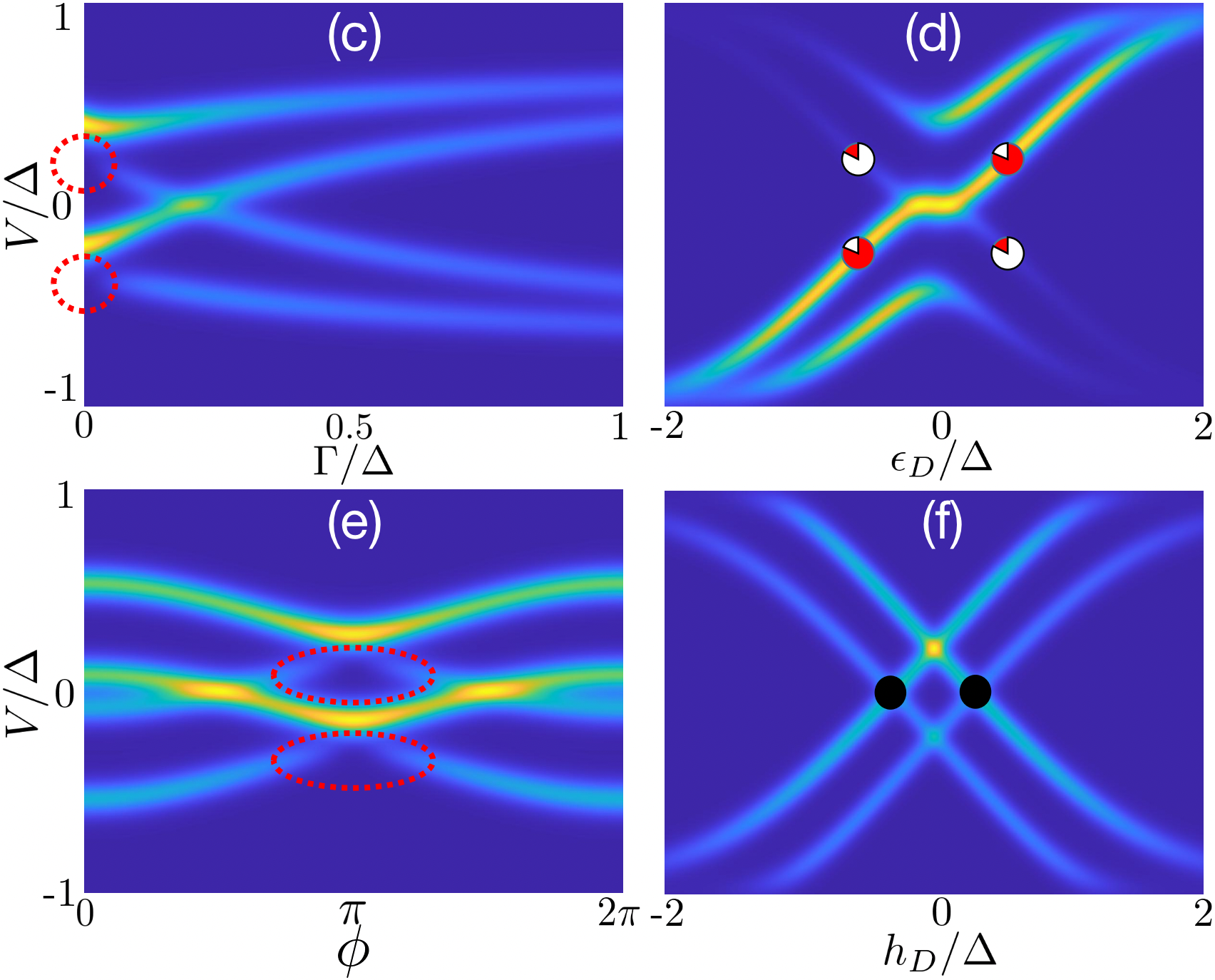}
\end{center}
\caption{(a) Sketch of a quantum dot (QD) formed in a nanowire and tunnel coupled to two superconductors (SCs). (b) The zero magnetic field differential conductance, $dI/dV$, as a function of the voltage difference, $V$. In the absence of dot-lead tunnel coupling, the tip electron can not tunnel couple to $\gamma_{s-}$ subgap quasiparticle (blue box), and hence there is not resonance at $V=-\epsilon_D$ (blue curve). While we find a pair of asymmetric resonances corresponding to two Andreev levels for $\Gamma\neq 0$ (red box and curve). The positive peak is higher than the negative one due to $\vert D_{1s+}\vert^2>\vert D_{1s-}\vert^2$, responsible for the particle-hole asymmetry in the subgap quasiparticles. (b-e) The finite magnetic field differential conductance, $dI/dV$, as a function of the voltage difference, $V$ and (c) $\Gamma$, (d) $\epsilon_D$,  (e) $\phi$, and (f) $h_D$, where other parameters are given by $\Gamma=0.2$, $\epsilon_D=0.1$, $\phi=\pi/2$, $h_D=0.3$, and $h_L=0$. }
\label{FIG6}
\end{figure}

To demonstrate the occupancy of negative quasiparticles based on our conjugate, we investigate the tunneling spectroscopy in a quantum-dot Josephson junction, plotted in Fig.~\ref{FIG6}(a). Next, we consider a hybrid quantum dot-superconducting lead structure. The pair potentials of the two superconductors, $\Delta e^{i\phi_j}$, has a phase difference $\phi = \phi_1 - \phi_2$, and the  tunnel coupling strength between quantum dot and superconducting leads is parameterized by $\Gamma$, which is assumed to be the same for left and right leads. $\epsilon_D$ ($\epsilon_{k}$) and $h_L$ ($h_D$) are the energy spectrum and Zeeman energy in quantum dot (superconducting leads), respectively. In our hybrid structure consisting of a quantum dot, nanowire, and superconductor \cite{deng2016majorana,lin2006intrinsic,persson2015scanning}, the quantum dot is spatially isolated from the superconductors, as depicted in Fig.~\ref{FIG6}(a). This spatial separation allows us to detect the distribution of the subgap quasiparticles between the quantum dot and the leads by directly measuring the differential conductance at a significant tip-dot distance \cite{ruby2015tunneling}.  The electron tunneling Hamiltonian between the quantum dot (red circle) and the tip (red triangle) can be expressed as follows
\begin{align} \label{glfdwd}
  V_{ep}&=\sum_{\vec{p}s} (t_dc^{\dagger}_{\vec{p}s}d^{}_s+h.c.)\\
  &=\sum_{l\vec{p}s} \left[t_d (u^{*}_{+})^{s s}_{l1}c^{\dagger}_{\vec{p}s}\gamma_{ls+}+t_ds(v_{+})^{-s -s}_{l1}c^{\dagger}_{\vec{p}s}\gamma_{ls-}+h.c.\right], \notag 
\end{align}
where $t_d$ is the tunnel coupling strength between dot and tip, and $c^{}_{\vec{p}s}$ is the field operator of electron in the probe with momentum $\vec{p}$, spin $s$, and energy $E_{\vec{p}}$.  We have expressed the dot field operator in terms of quasiparticle operators $d^{}_{s}=\sum_{l} (u^{*}_{+})^{s s}_{l1}\gamma_{ls+}+s(v_{+})^{-s -s}_{l1}\gamma_{ls-}$ which excludes the spin-flip effects from spin-orbit coupling and spin-dependent tunneling. Thus, the transition rate between  $\vert E_{ls\eta}  \rangle $  and $\vert E_{\vec{p}s} \rangle $ is given by  $T^{\vec{p}s}_{ls\eta}=t_dD_{ls\eta}$, where $D_{ls+}=(u^{*}_{+})^{ss}_{l1}$ and $D_{ls-}=s(v_{+})^{-s -s}_{l1}$. 

Here, we have kept the negative Bogoliubov quasiparticle operators ($\gamma_{ls-}$) in Eq. \eqref{glfdwd} so that the electron tunneling in quasiparticle representation has  $c^{\dagger}_{\vec{p}s}\gamma_{ls-}$ form  which can be effectively described by $f(E_{ls-})[1-f(E_{\vec{p}})]$ in scattering rate with $f(E)=1/(1+e^{E/k_BT})$ being the Fermi-Dirac distribution for temperature $\beta=1/k_BT$. 
Therefore, the tunneling current flowing from dot and tip reads as simply as the standard spectroscopy theory (see microscopic derivation in Appendix \ref{tunnelingcurrent})
\begin{align} \label{fvnjfvn}
    I_T&=\frac{2\pi e}{h}  \sum_{\vec{p} ls\eta} f(E_{ls\eta})[1-f(E_{\vec{p}}-V)]\vert T^{\vec{p}s}_{ls\eta}\vert^2 \delta(E_{\vec{p}}-E_{ls\eta})\notag\\
    &-\frac{2\pi e}{h} \sum_{\vec{p} ls\eta} f(E_{\vec{p}}-V)[1-f(E_{ls\eta})]\vert T^{\vec{p}s}_{ls\eta} \vert^2\delta(E_{\vec{p}}-E_{ls\eta}).
\end{align}
The first (second) term of the right-hand side of Eq. \eqref{fvnjfvn} corresponds to the contribution of tunnel currents from dot (tip) to tip (dot).  Then, we reach
\begin{align}
    I_T=-I^0_T \Gamma_t \sum_{ls\eta} \left[f\left(E_{ls\eta}\right)-f\left(E_{ls\eta}-V\right)\right] \vert D_{ls\eta}\vert^2,
\end{align}
with $I^0_T=2e/h$ and $\Gamma_t=\pi\nu_t\vert t_d\vert^2$, where $\nu_t$ is the density of state in the probe which is assumed to be energy-independent. Interestingly, we  incorporates both positive and negative quasiparticles, preserving the symmetry between spin-up and spin-down quasiparticles in the tunneling theory. Thus, the tunneling current, proportional to $\vert D_{ls\eta}\vert^2$, can intuitively contain the particle-hole information of the subgap quasiparticles \eqref{fvvldl}. 

We focus on the tunneling current assisted by the subgap energy levels by fixing our voltage biases within the superconducting gap, i.e., $\Delta>V>-\Delta$.  The corresponding differential conductance is given by $\frac{dI}{dV}\simeq-I_0 \Gamma_t \sum_{s\eta} \vert D_{1s\eta}\vert^2\frac{d}{dV}f\left(E_{1s\eta}-V\right)$. At zero temperature the differential conductance reduces to $\frac{dI}{dV}= -I_0 \Gamma_t \sum_{s} \vert D_{1s+}\vert^2\delta\left(E_{1s+}-V\right)-I_0 \Gamma_t \sum_{s} \vert D_{1s-}\vert^2\delta\left(E_{1s-}-V\right)$. The  first  terms describes the process of creating a positive subgap quasiparticle state by adding an  electron tunneling from the tip ($\gamma^{\dagger}_{1s+}c^{}_{\vec{p}s}$) and its conjugate process ($c^{\dagger}_{\vec{p}s}\gamma^{}_{1s+}$).   The second term corresponds to the creation of a negative subgap quasiparticle state by adding an electron tunneling from the tip ($\gamma^{\dagger}_{1s-}c^{}_{\vec{p}s}$) and its conjugate process ($c^{\dagger}_{\vec{p}s}\gamma^{}_{1s-}$), where the former ($\gamma^{\dagger}_{1s-}c^{}_{\vec{p}s}=\gamma^{}_{1-s+}c^{}_{\vec{p}s}$) can be reexplained as the formation of a Cooper pair with annihilation of a subgap quasiparticle and the transfer of an electron into the superconductor  and the later ($c^{\dagger}_{\vec{p}s}\gamma^{}_{1s-}=c^{\dagger}_{\vec{p}s}\gamma^{\dagger}_{1-s+}$)   represents the breakup of a Cooper pair with creation of a subgap quasiparticle and the transfer of an electron out of the superconductor~\cite{tinkham2004introduction}.

Let us begin with zero magnetic field case [Fig.~\ref{FIG6}(b)]. For $\Gamma=0$ (blue box and curve), we have $\gamma_{s+} = d_s$ and $\gamma_{s-} = d_{-s}^\dagger$ (see Appendix~\ref{quantumdot}), i.e., $\vert D_{1s+}\vert^2=1$ and $\vert D_{1s-}\vert^2=0$. The later unveils that the tip electron can not tunnel couple to $\gamma_{s-}$ subgap quasiparticle (blue box), and hence there is only one resonance  at $V=\epsilon_D$ (blue curve). While we find a pair of asymmetric resonances corresponding to two Andreev levels for $\Gamma\neq 0$ (red box and curve). The positive peak is higher than the negative one due to $\vert D_{1s+}\vert^2>\vert D_{1s-}\vert^2$, responsible for the particle-hole asymmetry in the Andreev bound states. In the presence of magnetic field, Figure~\ref{FIG6} plots $\text{dI/dV}$ as a function of $V$ and (c) $\Gamma$, (d) $\epsilon_D$,  (e) $\phi$, and (f) $h_D$. Two resonant peaks become weaker for larger $\Gamma$, where subgap quasiparticles have  less dot-electron population, i.e., $\vert D_{1s\eta}\vert^2$ [Fig.~\ref{FIG6}(c)]. For $\epsilon_D>0$ ($\epsilon_D<0$), the positive resonance is larger (smaller) than the negative, which reveals that positive Andreev level poses more (less) dot-electron population, i.e., $\vert D_{1s+}\vert^2>\vert D_{1s-}\vert^2$ ($\vert D_{1s+}\vert^2<\vert D_{1s-}\vert^2$), as shown by the partially-filled circles in Fig.~\ref{FIG6}(d). Besides, we find the spin-resolved Andreev levels cross zero-energy point [black circles Fig.~\ref{FIG6}(f)]. More interestingly, we find there are only two instead of four resonant peaks at $\phi=\pi$ [red circles in Fig. \ref{FIG6}(e)], because there exists zero dot-electron population, i.e., $\vert D_{1\uparrow-}\vert^2=\vert D_{1\downarrow-}\vert^2=0$, where the superconducting proximity effects from two superconductors interfere destructively.

In conclusion, we achieve tunnel spectroscopy asymmetry caused by the imbalanced particle-hole distribution of the subgap quasiparticles in the quantum-dot Josephson junction.  This kind of the tunnel spectroscopy asymmetry has been observed in hybrid superconductor–semiconductor nanostructure~\cite{deacon2010tunneling,lee2014spin} and superconductors with magnetic impurities \cite{yazdani1997probing,pan2000imaging,hanaguri2004checkerboard}.  Therefore, we believe that the occupancy of negative quasiparticle based on our conjugate loop  can be demonstrated by comparing the precise calculation and the quantitative measurement of gate-, phase-, and field-tunable tunnel spectroscopy. Moreover, our tunneling spectroscopy theory, as simple as the standard nonsuperconducting spectroscopy theory,   does not need the breakup and formation of the Cooper pairs.

\section{Conclusion and outlook} 

In summary, we have developed a general framework for dealing with the overcomplete basis set in the BdG formalism, that incorporates both positive and negative energy quasiparticle states and maintains the symmetry between spin-up and -down quasiparticles in scattering and tunneling theory. Here, we highlight the significant implications of our logical framework from both the theoretical concepts and the experimental predictions. Firstly,  we rigorously derive alla many-body eigenfunction of arbitrary nonuniform superconductors and uncover that superconducting eigenstates are full of superconducting spin clouds—the electron configuration within the Cooper-like pair of a nonuniform superconductor. We show how Bogoliubov excitations reconfigure these superconducting spin clouds. Secondly, we demonstrate a conjugate loop formed by the effective vacuum states of two orthonormal basis sets conjugate with each other, such as the Fermi sea and sky -- the effective vacuum states of positive and negative orthonormal basis sets, respectively. We emphasize the conceptual importance of the Fermi sky, which serves as the starting point for the filling of all negative quasiparticles within the Fermi sea.  Importantly, our conjugate loop provides insights into how to populate negative quasiparticles and achieves the one-body quasiparticle representation of the many-body superconducting state. This discovery provides a solid foundation for treating a many-body superconducting state as a one-body problem and thus proves valuable for addressing superconducting tunneling spectroscopy as simply as the standard nonsuperconducting spectroscopy theory. Thirdly, we demonstrate a gate-, field-, and phase-tunable tunnel spectroscopy asymmetry due to the particle-hole distribution imbalance of the subgap quasiparticles in the quantum-dot Josephson junction. 

Notably, we also derive the various transport properties of arbitrary superposition states from the wave-function perspective and demonstrate a novel supercurrent arising from the interference of distinct eigenstates that distinguishes with the supercurrent in the superconducting quantum interference device and exhibits oscillatory behaviour akin to the alternating current Josephson effect~\cite{zhang2024quantum}. Notably, the novel probability amplitude and oscillation frequency dependence of our interference supercurrrent suggests a groundbreaking quantum nondestructive measurement technique---the quantum noncollapsing measurement of the arbitrary superposition states of Andreev qubits~\cite{zhang2023Bogoliubovsea}. By following our logical framework, we have achieved a singlet-doublet quantum phase transition arising from the intricate competition between superconducting and spin-split proximity effects of superconducting leads rather than the Coulomb interaction of quantum dot \cite{zhang2023singlet}. Lastly, we have demonstrated how the singlet-doublet quantum phase transition induces additional asymmetry between the forward and reverse critical supercurrents, controlling the superconducting diode effect \cite{zhang2022enhancement}.
These significant findings underscore the exceptional power and implications of our logical framework, providing invaluable insights into the understanding and utilization of solid-state devices based on superconductivity. Moreover, our conjugate loop might lighten infinity issues of the Dirac sea in high-energy physics and build a connection between the worlds constructed by the particles and antiparticles~\cite{zhang2023Dirac}.

\section{Acknowledgement}
We sincerely thank J. Carlos Egues, Junwei Liu,  Gao Min Tang, and Chuan Chang Zeng for helpful and sightful discussions. This work is supported by the NSF of China (Grants Nos. 12321004, 12234003), the National Key R\&D Program of China (Grant No. 2020YFA0308800).

\appendix

\section{SYSTEM HAMILTONIAN} 
\label{ModelHamiltonian}

In this section,  we focus on investigating the properties of a superconducting system using a plan-wave representation, that encompasses a hybrid quantum dot tunnel coupled with two superconducting leads. 

Here we study a quantum dot and superconducting leads heterostructure, whoso Hamiltonian is given by
\begin{align} \label{fdvkfvk}
    H=H_L+H_D+H_T.
\end{align}
The quantum dot is described by a simple Hamiltonian as follows
\begin{align} \label{fdkvmdk}
    H_D=\sum_{s}(\epsilon_D+sh_D)d_{ s}^{\dagger}d_{ s}^{},
\end{align}
where the field operator $d_{ s}^{\dagger}$  create an electron of spin $s$ and energy $\epsilon_D+sh_D$ with $\epsilon_D$ and $h_D$ being the energy level and Zeeman energy, respectively.
The superconducting leads are described by the  Bardeen-Cooper-Schrieffer Hamiltonian 
\begin{align} \label{fdvlkkld}
    H_{L}&=\sum_{j\vec{k}s}(\epsilon_{j\vec{k}}+sh_L)c^{\dagger}_{j\vec{k}s}c^{}_{j\vec{k}s}\\
    &+\sum_{j\vec{k}}\left(\Delta_j c^{\dagger}_{j\vec{k}\uparrow}c^{\dagger}_{j-\vec{k}\downarrow}+h.c\right). \notag 
\end{align}
The field operator $c_{j\vec{k}s}^{\dagger}$ is the creation operator of the electron in superconductor $j=1,2$ with  spin $s$, wave vector $\vec{k}$, and energy $\epsilon_{j\vec{k}s}=\epsilon_{j\vec{k}}+sh_L$, where $\epsilon_{j\vec{k}}$ and $h_L$ are the corresponding energy spectrum and Zeeman energy, respectively. The attractive interaction is treated via a mean-field way, and $\Delta_{j}=\vert\Delta_j\vert e^{-i\phi_{j}}$ is the corresponding pairing potential with superconducting phase $\phi_j$ and gap $\vert\Delta_j\vert$. The quantum tunneling between superconducting leads and and the quantum dot is described as follows
\begin{align} \label{dvldl}
    H_T=\sum_{jn\vec{k}s}\left(t_jc^{\dagger}_{jn\vec{k}s}d^{}_{s}+h.c.\right).
\end{align}
The tunnel coupling amplitude $t_j$ is can be controlled by the corresponding electrostatic gate between quantum dot and superconducting lead.

Noting that the Hamiltonian of the hybrid system is quadratic, to numerically diagonalize the total Hamiltonian, we rewrite the total Hamiltonian \eqref{fdvkfvk} in an overcomplete basis set that contains both particle and hole operators of each electron 
\begin{align} \label{nambu}
    \Psi=
    \begin{bmatrix}
     d^{}_{\uparrow}\\ 
     -d^{\dagger}_{\downarrow}\\
     d^{}_{\downarrow}\\ 
     +d^{\dagger}_{\uparrow}
    \end{bmatrix}\bigoplus_{\vec{k}j}
    \begin{bmatrix}
      c^{}_{j\vec{k}\uparrow}\\
      -c^{\dagger}_{j-\vec{k}\downarrow}\\
      c^{}_{j\vec{k}\downarrow}\\
      +c^{\dagger}_{j-\vec{k}\uparrow}
    \end{bmatrix},
\end{align}
where $\bigoplus_{\vec{k}j} X_{\vec{k}j}$ concatenates $X_{\vec{k}j}$ vertically. Then, we can describe the Hamiltonian \eqref{fdvkfvk}   the Bogoliubov–de Gennes (BdG) Hamiltonian as follows
\begin{align} \label{fdvdjvkds}
    H= \frac{1}{2} \Psi^{\dagger}  \mathcal{H}_{\text{BdG}} \Psi+\mathcal{E},
\end{align}
where the half prefactor arises when we mathematically rewrite the total Hamiltonian \eqref{fdvkfvk} in terms of the overcomplete basis set  \eqref{nambu} \cite{bernevig2013topological}. The constant energy $\mathcal{E}=\mathcal{E}_{\epsilon}+\mathcal{E}_{\Delta}$ includes the contributions from energy spectrum $\epsilon_{j\vec{k}}$, given by $\mathcal{E}_{\epsilon}=\sum_{j\vec{k}}\epsilon_{j\vec{k}}$,  and pair potential $\Delta_j$, given by $\mathcal{E}_{\Delta}=\sum_{j\vec{k}}\vert \Delta_{j}\vert^2/g$, where $g$ is attractive interaction, e.g., the Gorkov contact pairing interaction and $\mathcal{E}_{\Delta}$ is the energy constant originating from the mean-field treatment of the attractive interaction~\cite{beenakker2005superconducting}.
Noting that spin is a good quantum number, the BdG matrix divided into two spin blocks as follows
\begin{align} \label{fdhvdfvjdo}
    \mathcal{H}_{\text{BdG}}=\left[\begin{array}{cccc} 
   \mathcal{H}_{\uparrow}  & 0 \\
    0 & \mathcal{H}_{\downarrow}
    \end{array}\right].
\end{align}
$\mathcal{H}^s$ is the corresponding Hamiltonian matrix for spin-$s$ space
\begin{align} \label{mvndefnvak}
    \mathcal{H}_s =
 \left[\begin{array}{cccc} 
   \epsilon_D\tau_z+sh_D  & \mathcal{H}_T^{\dagger} \\
    \mathcal{H}_T^{} & \mathcal{H}^{s}_L
    \end{array}\right].
\end{align}
The dot energy and Zeeman energy terms are described by the Pauli matrix $\tau_z$ and the identity matrix in the overcomplete basis set \eqref{nambu}, respectively. In our analysis, we focus on a single energy level on the dot, assuming that the spacing between levels is significantly larger than all other relevant energy scales. The matrix form of the BCS Hamiltonian for the superconducting lead, given by Eq.~\eqref{fdvlkkld}, is as follows 
\begin{align} \label{vfnvmk} 
\mathcal{H}^{s}_L=\bigotimes_{\vec{k}j}\begin{bmatrix}
      +\epsilon_{j\vec{k}}+sh_L & -\Delta_j \\
      -\Delta^*_j &  -\epsilon_{j-\vec{k}}+sh_L 
 \end{bmatrix},
\end{align}
where $\bigotimes_{\vec{k}jn} X_{\vec{k}jn}$ concatenates $X_{\vec{k}jn}$ diagonally. The matrix form of the quantum tunneling Hamitonian  \eqref{dvldl} is given by  
\begin{align}
    \mathcal{H}_T^{}=t_j  \bigoplus_{\vec{k}jn}\tau_{z},
\end{align}
where $\bigoplus_{\vec{k}jn} X_{\vec{k}jn}$ concatenates $X_{\vec{k}jn}$ vertically.

Numerically, we can diagonalize the matrix of the total BdG Hamiltonian \eqref{fdhvdfvjdo} as follows
\begin{align}
    U^{}_{}\mathcal{H}^{}_{\text{BdG}}U^{\dagger} _{}=\bigotimes_{ls\eta}[E_{ls\eta}],
\end{align}
where the transformation matrix $U^{}{}$ is diagonal in spin space akin to the matrix of the BdG Hamiltonian \eqref{fdhvdfvjdo} 
\begin{align}
    U _{}=\left[\begin{array}{cccc} 
   U_{\uparrow}  & 0 \\
    0 & U_{\downarrow}
    \end{array}\right].
\end{align}
Here, $U_{s}$ is a $2N\times 2N$ unitary matrix, with $N$ representing half the dimension of the Hamiltonian \eqref{mvndefnvak}, representing the size of the hybrid system. As a result, the BdG Hamiltonian \eqref{mvndefnvak} can be mathematically rewritten as:
\begin{align} \label{gfgblrgk}
    H= \frac{1}{2}\sum^{N}_{l=1}\sum_{s=\uparrow/\downarrow}\sum_{\eta=\pm} E_{ls\eta}\gamma_{ls\eta}^{\dagger} \gamma_{ls\eta }^{}+\mathcal{E},
\end{align}
Working in the overcomplete basis set including Nambu and spin space, we are required to add an additional index $\eta=+/-$ to label the high/low energy levels of each spin species ($E^{}_{ls+}>E^{}_{ls-}$), satisfying $E_{ls\eta}=-E_{l-s-\eta}$. Here, the larger $l$ corresponds to a higher energy level. The term with $l=1$ corresponds to the Andreev levels, which are subgap energy levels of the hybrid quantum dot and superconductor system. All other energy levels ($l>1$) are referred to as Bogoliubov levels, representing the out-of-gap energy levels of the hybrid system. The quasiparticle operator $\gamma_{ls\eta }^{}$ with energy $E_{ls\eta}$ can be expressed as a unit vector in the overcomplete basis set of the hybrid system:
\begin{align} \label{tfvvldl} 
     \gamma_{ls\eta}= \sum_{k}\left[(u_{s\eta})^{ }_{lk}c^{}_{ks}+(v_{s\eta})^{}_{lk}(-sc^{\dagger}_{k-s})\right].
\end{align}
In Eq. \eqref{tfvvldl}, we define $c_{ks}$ such that $c_{1s}=d_s$ and $c_{ks}=c_{j\vec{k}s}$ for all $k>1$. The coefficients $(u_{s\eta})^{ }{lk}$ and $(v{s\eta})^{}{lk}$, obtained from the unitary matrix $U^s$, describe the electron and hole distributions of the quasiparticles $(\gamma{ls\eta})$, respectively.

The advantages to using the overcomplete basis set \eqref{nambu} is the convenient description of superconductivity in quadratic $\Psi^{\dagger} \Psi$ form easy for diagonalization  [Eq. \eqref{fdvdjvkds}]. Especially, it is necessary to use an overcomplete basis set to obtain quadratic and diagnalizable $\Psi^{\dagger} \Psi$ form in the presence of both superconductivity and the spin-flip from the spin-orbit coupling and quantum tunneling.  The penalties we pay for working with an overcomplete basis set 
are the additional need to keep track of the overlap matrix of the BdG Hamiltonian and the basis states of the overcompleter basis set, because the overcomplete basis set are not independent of each other anymore. For example, the electron and hole distributions of the quasiparticles always satisfy the orthogonal and normalization conditions as a result of eigen calculation 
\begin{align}
 \sum_{k}\left[(u^{*}_{s\eta})_{lk}(u^{}_{s'\eta'})_{l'k}+(v^{*}_{s\eta})_{lk}(v^{}_{s'\eta'})_{l'k}\right]=\delta_{ll'}\delta_{ss'}\delta_{\eta\eta'},
\end{align}
but the quasiparticle operators \eqref{tfvvldl} do not always satisfy the anticommutation relations. This is because the matrix of the BdG Hamiltonian is based on an overcomplete basis set \eqref{nambu}, whose elements are not independent of each other.  By substitution of Eq. \eqref{tfvvldl},  we can calculate the following anticommutation 
\begin{align} \label{fvnfjvn}
    \{\gamma^{}_{ls\eta},&\gamma^{}_{l-s-\eta}\}=\sum_{k}(u_{s\eta})^{ }_{lk}(sv_{-s-\eta})^{}_{lk}\left\{c^{}_{ks},c^{\dagger}_{ks}\right\}\\
    &+(v_{s\eta})^{}_{lk}(-su_{-s-\eta})^{ }_{lk}\left\{c^{\dagger}_{k-s},c^{}_{k-s}\right\} \notag \\
    &=\left[(u^{}_{s\eta})_{lk}(u^{*}_{s\eta})_{lk}+(v_{s\eta})_{lk}(v^{*}_{s\eta})_{lk}\right]=1, \notag 
\end{align}
where we use the fact that the electron and hole distributions satisfy 
\begin{align} \label{tpfvvldlr}
    (u^{*}_{s\eta})^{}_{lk}=s(v_{-s-\eta})^{ }_{lk}.
\end{align}
The above identity \eqref{tpfvvldlr} is guaranteed by the particle-hole or charge-conjugation "symmetry", which is artificially introduced by rewriting the orignal Hamiltonian into the quadratic form of the BdG Hamiltonian in terms of the overcomplete basis set \cite{bernevig2013topological}. Therefore, the quasiparticles \eqref{tfvvldl} are overcomplete bases and satisfy  conjugate relations 
\begin{equation} \label{atpfvvldl}
    \gamma^{\dagger}_{ls\eta}= \gamma^{}_{l-s-\eta}.
\end{equation}
Inverting Eq. \eqref{fvvldl}, we can express $c^{}_{ks}$ with $\gamma_{ls\eta}$ as follows
\begin{align} \label{fdnvfkgdr}
    c^{}_{ks}=\sum_{k} (u^{*}_{s+})^{}_{lk}\gamma_{ls+}+s(v_{-s+})^{}_{lk}\gamma_{ls-},
\end{align}
where we have used the identity \eqref{tfvvldl} to convert $(u^{*}_{s-})^{}_{lk}$ into $s(v_{-s+})^{ }_{lk}$.

Each spinful Andreev quasiparticle always has its conjugate alternative with opposite spin and energy, as shown by the conjugate relation \eqref{atpfvvldl}. Thus,  we can have $4^{N}$ orthonormal basis sets (OBSs). In principle, we can choose any OBS. The Datta-like OBS, $(\gamma^{}_{ls+},\gamma^{}_{ls-})$ for all $l$ \cite{datta1996scattering,datta1999can}, corresponds to the Hamiltonian
\begin{align} \label{ndrvdknkdf}
    H=\sum^{N}_{l=1}\sum_{\eta=+/-} E^{}_{ls\eta}\gamma_{ls\eta}^{\dagger} \gamma_{ls\eta }^{}+\mathcal{E}_{s}.
\end{align} 
$\mathcal{E}_{\uparrow}=\mathcal{E}+\frac{1}{2}\sum_{l\eta}E_{l\downarrow\eta}$ and $\mathcal{E}_{\downarrow}=\mathcal{E}+\frac{1}{2}\sum_{l\eta}E_{l\uparrow\eta}$ are the energy constants generated when we rewrite the Hamiltonian \eqref{gfgblrgk} into the Hamiltonian \eqref{ndrvdknkdf} with an OBS $(\gamma_{l\uparrow+},\gamma_{l\uparrow-})$ and $(\gamma_{l\downarrow+},\gamma_{l\downarrow-})$, respectively. During the rewriting, we will use the conjugate relation \eqref{atpfvvldl} and $E_{ls\eta}=-E_{l-s-\eta}$ to remove the prefactor $1/2$ in Eq. \eqref{gfgblrgk}. Therefore, quasiparticle $\gamma^{\dagger}_{ls\eta}$ always have well defined energy $E^{\dagger}_{ls\eta}$ rather than $E^{\dagger}_{ls\eta}/2$.  Alternatively, we can also choose textbook basis states  $(\gamma^{}_{l\uparrow\eta},\gamma^{}_{l\downarrow\eta})$ for all $l$. Thus, the total Hamiltonian  \eqref{gfgblrgk} can be rewritten into 
\begin{align}  \label{fvdkvmdkfmv}
H&=\sum^{N}_{l=1}\sum_{s=\uparrow/\downarrow} E^{}_{ls\eta}\gamma_{ls\eta}^{\dagger} \gamma_{ls\eta }^{}+\mathcal{E}_{\eta},
\end{align} 
where $\mathcal{E}_{\eta}=\mathcal{E}+\frac{1}{2}\sum_{ls}E_{ls-\eta}$ are the energy constants generated when we rewrite the Hamiltonian \eqref{gfgblrgk} into the Hamiltonian \eqref{fvdkvmdkfmv}. Of course, we can also choose the excitation OBS -- $\{\gamma^{}_{ls\eta}\}$ for all $E_{ls\eta}>0$, 
\begin{align}  \label{yfvfldlfl}
    H&=\sum_{E_{ls\eta}>0} E^{}_{ls\eta}\gamma_{ls\eta}^{\dagger} \gamma_{ls\eta }^{}+\mathcal{E}_{G},
\end{align} 
where $\mathcal{E}_{G}=\mathcal{E}+\frac{1}{2}\sum_{E_{ls\eta}<0}E_{ls\eta}$ is the energy constant generated when we rewrite the Hamiltonian \eqref{gfgblrgk} into the Hamiltonian \eqref{yfvfldlfl} and is exact ground-state energy.

\section{TIGHT-BINDING MODEL}
\label{tightbindingmodel}

In the section at hand, we focus on investigating the properties of a superconducting system using a tight-binding model. The model encompasses a hybrid quantum dot coupled with superconducting leads. To simplify the analysis, we consider a one-dimensional tight-binding model that does not account for orbital effects.

We use a one-dimensional tight-binding 
model \cite{davydova2022universal,zhang2022enhancement} to describe our system
\begin{align} \label{maindvmdl}
    H&=\sum^{N}_{n=1}\sum_{j=L,R}\sum_{s=\uparrow,\downarrow}\epsilon_{jns}c^{\dagger}_{jns}c^{}_{jns}+\sum_{s=\uparrow,\downarrow}(\epsilon_D+sh_D)d^{\dagger}_{s}d^{}_{s}\notag  \\
   &+\sum^{N-1}_{n=1}\sum_{j=L,R}\sum_{s=\uparrow,\downarrow}\left(t_{0}c^{\dagger}_{jns}c^{}_{jn+1s}+h.c.\right)\notag \\
&+\sum^{N}_{n=1}\sum_{j=L,R}\left(\Delta^{j}_{n}c^{\dagger}_{jn\uparrow}c^{\dagger}_{jn\downarrow}+h.c. \right)\\
 &+t\sum_{ss'}\left[c^{\dagger}_{LNs} \mathcal{U}_{ss'}(\theta_{so})d_{s'}+d^{\dagger}_{s} \mathcal{U}_{ss'}(\theta_{so})c_{R1s'} +h.c. \right]\notag
    .
\end{align} 
In the system described, the field operator $c_{jns}^{}$ annihilates an electron in the $j$-th lead with spin $s$ and energy $\epsilon_{jns}=\epsilon_{jn}+sh_{L}$. Here, $\epsilon_{jn}$ represents the site energy in the lead, and $h_{L}$ is the Zeeman energy. The leads are labeled as $L$ for the left lead and $R$ for the right lead, and each lead consists of $N$ sites. The field operator $d_s^{}$ annihilates an electron in the quantum dot with spin $s$ and energy $\epsilon_D+sh_D$. $\epsilon_D$ represents the dot energy, and $h_D$ is the Zeeman energy for the dot. The spin rotation matrix $\mathcal{U}(\theta_{\text{so}})=e^{i\theta_{\text{so}} s^{y}/2}$ describes the spin-orbit-induced rotation in the tunnel coupling between the dot and the last (first) site $c_{LNs}$ ($c_{R1s}$) of the left (right) lead. The strength of this coupling is denoted by $t$. The parameter $t_0$ represents the nearest-neighbor hopping amplitude in the leads. In the left (right) lead, the Fulde-Ferrell-type order parameter is considered: $\Delta^{L}_n=\Delta e^{-i\frac{\phi}{2}+2iqna}$ ($\Delta^{R}_n=\Delta e^{+i\frac{\phi}{2}+2iqna}$). Here, $\Delta$ is the amplitude of the order parameter, $\phi$ is the global phase difference between the two leads, $q$ is the momentum of the Cooper pair, and $a$ is the lattice constant. The Fulde-Ferrell-type order parameter allows for finite-$q$ Cooper pairs, which can be realized through direct current injection or by screening currents through the Meissner effect.

We rewrite the total Hamiltonian \eqref{maindvmdl} in the Nambu space of the hybrid quantum dot and superconducting lead system, i.e., an overcomplete basis set that contains both particle and hole operators of each electron as follows  
\begin{align} \label{tnambu}
    \Psi=
    \begin{bmatrix}
      d^{}_{\uparrow}\\
      d^{}_{\downarrow}\\
      -d^{\dagger}_{\downarrow}\\
      d^{\dagger}_{\uparrow}
    \end{bmatrix} 
   \oplus
    \left(\bigoplus_{n}
    \begin{bmatrix}
      c^{}_{Ln\uparrow}\\
      c^{}_{Ln\downarrow}\\
      -c^{\dagger}_{Ln\downarrow}\\
      c^{\dagger}_{Ln\uparrow}
    \end{bmatrix} \right)
    \oplus
   \left(\bigoplus_{n}
    \begin{bmatrix}
      c^{}_{Rn\uparrow}\\
      c^{}_{Rn\downarrow}\\
      -c^{\dagger}_{Rn\downarrow}\\
      c^{\dagger}_{Rn\uparrow}
    \end{bmatrix} \right).
\end{align}
This overcomplete basis set enables the diagonalizable description of the coexistence of  superconductivity and spin-orbit coupling. 
Using the Nambu spinor $\Psi$, we recast the total Hamiltonian \eqref{maindvmdl} as follows
\begin{align} \label{fvdvldv}
    H= \frac{1}{2} \Psi^{\dagger}  \mathcal{H}_{\text{BdG}} \Psi+\mathcal{E},
\end{align}
where  the BdG Hamiltonian matrix reads
\begin{align} \label{fdnkvamk}
    \mathcal{H}_{\text{BdG}}=\left[\begin{array}{cccc} 
    \mathcal{H}_{D}  & \mathcal{T}_{L} & \mathcal{T}_{R} \\
   \mathcal{T}^{\dagger}_{L}& \mathcal{H}_{L}  & 0 \\
   \mathcal{T}^{\dagger}_{R}& 0 & \mathcal{H}^{\dagger}_{R}
    \end{array}\right],
\end{align}
and  $\phi$-independent constant is given by 
\begin{equation}
 \mathcal{E}=\sum_{jn}\epsilon_{jn},
 \label{constant-E},
 \end{equation}
The factor of 1/2 in Eq. \eqref{fvdvldv} arises from the artificial doubling in the BdG formalism \cite{bernevig2013topological}.
 The  quantum dot is described by the non-interacting Hamiltonian 
\begin{align}
   \mathcal{H}_{D}= 
  \begin{bmatrix}
    \epsilon_D +h_D & 0 &  0 & 0\\
    0 & \epsilon_D -h_D&  0 & 0\\
    0 & 0 & -\epsilon_D +h_D  & 0\\
    0 & 0 &0 &  -\epsilon_D - h_D
    \end{bmatrix}_{4\times 4},
    \label{dot-H}
\end{align}
where $\tau_z$ and $s^{z}$ are Pauli matrices acting within the Nambu and spin spaces, respectively. 
The Hamiltonian $\mathcal{H}_j$ describes the superconducting leads $j=L,R$,
\begin{align} \label{tvfnvmk} 
    \mathcal{H}_j=
   \begin{bmatrix}
       \mathcal{H}_{j1} & \mathcal{T}_{0} & & & \\
      \mathcal{T}^{\dagger}_{0} & \mathcal{H}_{j2} & \mathcal{T}_{0} & & & \\
        & \mathcal{T}^{\dagger}_{0} & \mathcal{H}_{j3} & \ddots & \\
        &  & \ddots & \ddots & \mathcal{T}_{0} \\
        &  & & \mathcal{T}^{\dagger}_{0} & \mathcal{H}_{jN}
 \end{bmatrix}_{4N\times 4N},
\end{align}
where $\mathcal{T}_{0}$ is a tunnel-coupling matrix 
\begin{align} \label{tybgbfmk}
    \mathcal{T}_{0}= 
    \begin{bmatrix}
    t_0 & 0 &  0 & 0\\
    0 & t_0 &  0 & 0\\
    0 & 0 & -t_0  & 0\\
    0 & 0 &0 &  -t_0
    \end{bmatrix}_{4\times 4}
\end{align}
and $\mathcal{H}_{jn}$ is the Hamiltonian for lead $j$ at site $n$
\begin{align} \label{vnskfm}
    \mathcal{H}_{jn}=
    \begin{bmatrix}
    \epsilon_{jn\uparrow} & 0 &  -\Delta^j_n & 0\\
    0 & \epsilon_{jn\downarrow} &  0 & -\Delta^j_n\\
    -(\Delta^j_n)^* & 0 & -\epsilon_{jn\downarrow}  & 0\\
    0 & -(\Delta^j_n)^* &0 &  -\epsilon_{jn\uparrow}
    \end{bmatrix}_{4\times 4}.
\end{align}
For simplicity, in the following analysis, we assume that the only distinction between the left and right superconducting leads is in the respective phases $\phi_L$ and $\phi_R$. Other than that, the leads are considered identical. The tunnel-coupling matrix between the quantum dot and the left and right leads are denoted as follows
\begin{align} \label{fvmfkll}
    \mathcal{T}_L&= t\begin{bmatrix}
      0 &  \cdots & 0 &  U(\theta_{so})
 \end{bmatrix}_{4\times 4N},
\end{align}
\begin{align} \label{fvmfklr}
    \mathcal{T}_R&= t\begin{bmatrix}
      U^{\dagger}(\theta_{so}) &  0 & \cdots & 0
 \end{bmatrix}_{4\times 4N}.
\end{align}
The spin-orbit effect in the tunneling processes, both for the left and right leads, is accounted for by the unitary matrix 
\begin{equation} \label{fmmvfkd}
U(\theta_{so})=\begin{bmatrix}
\mathcal{U}_{}(\theta_{so})& \\ &-\mathcal{U}_{}(\theta_{so})\end{bmatrix}_{4\times 4},
\end{equation}
where $\mathcal{U}(\theta_{so})$ represents the spin rotation caused by the spin-orbit coupling during the tunneling between the quantum dot and the superconducting leads. Specifically, $\mathcal{U}(\theta_{so})$ is defined as $e^{i\theta s^{y}/2}$, where $\theta$ represents the spin-orbit coupling strength and $s^{y}$ denotes the y-component of the spin operator. 

By numerically diagonalizing the BdG matrix \eqref{fdnkvamk}, we can construct the Bogoliubov quasiparticle operators $\gamma^{}_{l\sigma\eta}(q,\phi_L,\phi_R)\rightarrow \gamma^{}_{l\sigma\eta}$, which depend on the Cooper pair momentum $q$ as well as the phase differences $\phi_L$ and $\phi_R$. These operators can be represented as unit vectors in the overcomplete basis set \eqref{tnambu} and are given by the expression
\begin{equation}
\gamma^{}_{l\sigma\eta}=\sum^{8N+4}_{m=1}u_{l,m}\Psi_{m}, 
\end{equation}
i.e., 
\begin{equation} \label{fvvldle}
    \gamma_{l\sigma\eta}= \sum^{N}_{n=1}\sum_{s=\uparrow,\downarrow}\left[(u_{\eta})^{\sigma s}_{ln}c^{}_{ns}+(v_{\eta})^{\sigma s}_{ln}(-sc^{\dagger}_{n-s})\right].
\end{equation}
Here, $\Psi_{m}$ denotes the $m$-th component of $\Psi$ in Eq. \eqref{tnambu}, and $[u_{l,1},u_{l,2},\cdots, u_{l,8N+4}]$ represents the $l$-th eigenvector that diagonalizes the BdG Hamiltonian \eqref{fdnkvamk}. The coefficients $u_{l,m}$ satisfy the normalization condition
\begin{align}
   \sum^{8N+4}_{m=1}\vert u_{l,m}\vert^2=1.
\end{align}
By recasting the total Hamiltonian \eqref{fvdvldv} in the form of Eq. \eqref{fdbldfl} in the main text, we have:
\begin{align} \label{gdrtfgblrgk}
    H= \frac{1}{2}\sum^{2N+1}_{l=1}\sum_{\sigma=\Uparrow/\Downarrow}\sum_{\eta=\pm} E_{l\sigma\eta}\gamma_{l\sigma\eta}^{\dagger} \gamma_{l\sigma\eta }^{}+\mathcal{E}.
\end{align} 
In the above equation, the Bogoliubov quasiparticle energies $E_{l\sigma\eta}$ depend solely on the phase difference $\phi=\phi_R-\phi_L$. Additionally, these eigenenergies satisfy the relationship $E_{l\sigma\eta}=-E_{l-\sigma-\eta}$ due to particle-hole symmetry.

The quasiparticle operators $(\gamma_{l\sigma\eta})$ for all $l$, $\sigma$, and $\eta$ form an overcomplete basis set that consists of two mutually conjugate OBSs. The quasiparticle operators described by Eq. \eqref{fvvldle} satisfy the conjugate relation
\begin{equation} \label{tpfvvldle}
\gamma_{l\sigma\eta}^\dagger = \gamma_{l-\sigma-\eta}.
\end{equation}
As a consequence of this conjugate relation, the electron ($u_\eta$) and hole ($v_\eta$) distributions satisfy the following relation
\begin{equation}
(u_\eta^*)^{\sigma s}_{ln} = s(v_{-\eta})^{-\sigma -s}_{ln}.
\end{equation}
The conjugate relation~\eqref{tpfvvldle} offers an intuitive and comprehensive illustration of the artificial redundancy in the BdG formalism from a quasiparticle operator point of view. It is important to note that the ubiquitous conjugation relation is different from the demanding requirement of the Majorana fermion, i.e., $\gamma_{ls\eta}^\dagger = \gamma_{ls\eta}$. 
By inverting Eq. \eqref{fvvldle}, we can express the electron operator $c_{ns}$ in terms of the $\gamma_{ls\eta}$ operators as follows
\begin{align} \label{fdnvfkgd}
    c^{}_{ns}=\sum_{l\sigma} (u^{*}_{+})^{\sigma s}_{ln}\gamma_{l\sigma+}+s(v_{+})^{-\sigma -s}_{ln}\gamma_{l\sigma-}.
\end{align}

\section{Logic} \label{logic}

We present our logic for the usage of the overcomplete basis set including two OBSs conjugate with each other 
\begin{itemize}
    \item STEP I): divide the overcomplete basis set [$(\gamma^{}_{l\sigma\eta})$ for all $l$, $\sigma$, and $\eta$] into two OBSs conjugate with each other; 
    \item STEP II): derive the  EVSs of these two OBSs; 
    \item STEP III): ground state is a filled sea of all negative basis states in the chosen OBS from the corresponding EVS.
\end{itemize}
Then we attain the following three properties 
\begin{itemize}
    \item PROPERTY I): the  EVSs of any two OBSs conjugate with each other form a conjugate loop;
    \item PROPERTY II): there are $2^N$ possible divisions of the two OBSs conjugate with each other. When a pair of basis states is exchanged between the two OBSs, the fermionic parity  of the EVSs switches.
    \item PROPERTY III): The even-parity eigenstates are full of Bogoliubov-like singlets -- a superposition of vacuum state and Cooper-like pair, while the odd-parity eigenstates contain not only plenty of Bogoliubov-like singlets but also a pseudospin cloud.
\end{itemize} 
To gain an intuitive understanding of this logic, we provide examples starting with a simple quantum dot case (Sec.~\ref{quantumdot}),  followed by normal metal (Sec.~\ref{newlogic}) and superconductor (Sec.~\ref{superconductor}) cases. 

\subsection{Quantum dot case} \label{quantumdot}

It is illustrative to start with the  quantum dot without the superconducting leads. The BdG Hamiltonian can be expressed in terms of the overcomplete basis set $(\gamma_{\uparrow+},\gamma_{\downarrow+},\gamma_{\uparrow-},\gamma_{\downarrow-})$. The Hamiltonian takes the form
\begin{align} \label{mfdvkvkdk}
    H_D=\frac{1}{2}\sum_{s\eta}E_{s\eta}\gamma^{\dagger}_{s\eta}\gamma^{}_{s\eta}+\mathcal{E}^{D},
\end{align}
with $\mathcal{E}^{D}=\epsilon_D$. Without loss of generality, let's consider the case where $\epsilon_D$ and $h_D$ are both positive. In this case, we can identify $\gamma_{s+}$ as the electron operator $d_s$ ($\gamma_{s+} = d_s$) and $\gamma_{s-}$ as the hole operator $d_s^\dagger$ ($\gamma_{s-} = d_{-s}^\dagger$). The single-particle energy spectrum $E_{s\eta}$ is given by $E_{s\eta} = \eta|\epsilon_D| + sh_D$, and it satisfies the relation $E_{-s-\eta} = -E_{s\eta}$. The field operators $\gamma_{s\eta}^\dagger$ and $\gamma_{-s-\eta}$ are conjugates of each other according to
\begin{align} \label{mfdflal}
    \gamma^{\dagger}_{s\eta}=\gamma_{-s-\eta}.
\end{align}
Each spin-$s$ electron has both particle ($d_s = \gamma_{s+}$) and hole ($d_s^\dagger = \gamma_{-s-}$) representations. Therefore, there are four OBSs that can be constructed: two spin OBSs $(\gamma_{\uparrow+},\gamma_{\downarrow+})$ and $(\gamma_{\uparrow-},\gamma_{\downarrow-})$, as well as two Nambu OBSs $(\gamma_{\uparrow+},\gamma_{\uparrow-})$ and $(\gamma_{\downarrow+},\gamma_{\downarrow-})$. However, it is important to note that  $(\gamma_{\uparrow+},\gamma_{\downarrow-})$ and $(\gamma_{\uparrow-},\gamma_{\downarrow+})$ are not valid OBSs because the operators within each pair do not satisfy the anticommutation relation ($\{\gamma_{\uparrow+},\gamma_{\downarrow-}\} \neq 0$ and $\{\gamma_{\uparrow-},\gamma_{\downarrow+}\} \neq 0$). It is important to emphasize that the choice of which OBS to use does not affect the physical properties of the quantum dot. The quantum dot itself is unaware of the specific OBS chosen, and all four OBSs will yield the same ground and excitation states. This highlights the arbitrariness in the choice of OBS and the consistency of physical results regardless of the chosen representation.

\subsubsection{Spin OBSs}

STEP I): Let's begin by considering two spin OBSs $(\gamma_{\uparrow+},\gamma_{\downarrow+})$ and  $(\gamma_{\uparrow-},\gamma_{\downarrow-})$. We want to express the BdG Hamiltonian \eqref{mfdvkvkdk} in terms of these chosen OBSs
\begin{align} \label{mfvdkfvdk}
    H_D=\sum_{s}E_{s+}\gamma^{\dagger}_{s+}\gamma^{}_{s+}+\mathcal{E}^{D}_{+}=\sum_{s}E_{s-}\gamma^{\dagger}_{s-}\gamma^{}_{s-}+\mathcal{E}^{D}_{-}.
\end{align}
This transformation is accomplished by utilizing the conjugate relation  \eqref{mfdflal}.
In this case, we have $\mathcal{E}^{D}_{+}=0$ and $\mathcal{E}^{D}_{-}=2\epsilon_D$ as the energy constants obtained when rewriting the BdG Hamiltonian \eqref{mfdvkvkdk} in terms of the OBSs $(\gamma_{\uparrow+},\gamma_{\downarrow+})$ and $(\gamma_{\uparrow-},\gamma_{\downarrow-})$, respectively.

STEP II):  Next, we determine the  EVSs of the two spin OBSs $(\gamma_{\uparrow+},\gamma_{\downarrow+})$ and  $(\gamma_{\uparrow-},\gamma_{\downarrow-})$, defined by the equations
\begin{align}
    \gamma_{s+}\vert V\rangle_{+}=0 \text{ for all } s=\uparrow/\downarrow,
\end{align}
\begin{align}
    \gamma_{s-}\vert V\rangle_{-}=0 \text{ for all } s=\uparrow/\downarrow.
\end{align}
By solving these equations, we can find the corresponding  EVSs
\begin{align} \label{fdvnknfvlkp}
    \vert V\rangle_{+}=\vert 0\rangle,
\end{align}
\begin{align} \label{fdvnknfvlkm}
    \vert V\rangle_{-}=d^{\dagger}_{\uparrow}d^{\dagger}_{\downarrow}\vert 0\rangle.
\end{align}
Here, $\vert 0\rangle$ represents the vacuum state with no electrons. Thus, $\mathcal{E}^{D}_{+}$ and $\mathcal{E}^{D}_{-}$ correspond to the energy of the EVSs $\vert V\rangle_{+}$ and $\vert V\rangle_{-}$, respectively.
It is important to note that the EVSs of different OBSs may differ, and therefore, the EVS does not always represent the ground state of the system.

\begin{figure*}
\begin{center}
\includegraphics[width=0.75\linewidth]{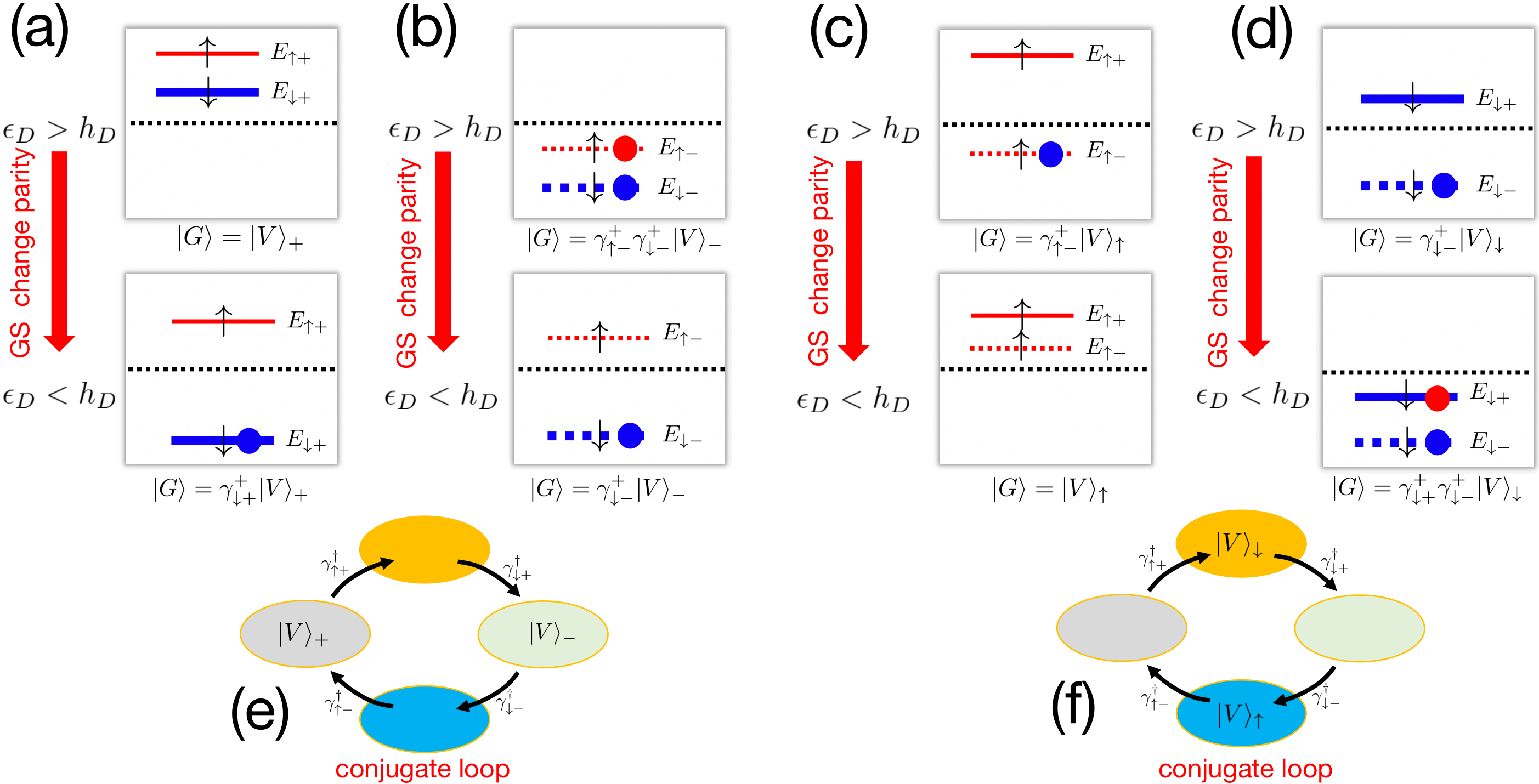}
\end{center}
\caption{The QPT of the spin-split quantum dot in different OBSs (a) $(\gamma_{\uparrow+},\gamma_{\downarrow+})$, (b) $(\gamma_{\uparrow-},\gamma_{\downarrow-})$, (c) $(\gamma_{\uparrow+},\gamma_{\uparrow-})$, and (d) $(\gamma_{\downarrow+},\gamma_{\downarrow-})$. The ground state (GS) is equal to adding all negative basis states to the corresponding EVS and undergoes a parity changes at $\epsilon_D=h_D$, where both $\epsilon_D$ and $h_D$ is asssumed to be positive. (e,f) The  EVSs of two OBSs conjugate with each other, for example (e) $\vert V\rangle_{\pm}$ [Eq. \eqref{fdkfdk}] and (f) $\vert V\rangle_{\uparrow/\downarrow}$ [Eq. \eqref{fdtykfdk}], form a conjugate loop.
}
\label{FIGQD} 
\end{figure*}

STEP III):  Now, let us consider the spin OBS $(\gamma_{\uparrow+},\gamma_{\downarrow+})=(d_{\uparrow},d_{\downarrow})$  with the EVS $\vert V\rangle_{+}=\vert 0\rangle$ (electron picture). The quantum dot has four eigenstates: $\vert V \rangle_+=\vert 0\rangle$, $\gamma^{\dagger}_{\uparrow+}\vert V \rangle_+=d^{\dagger}_{\uparrow}\vert 0\rangle$, $\gamma^{\dagger}_{\downarrow+}\vert V \rangle_+=d^{\dagger}_{\downarrow}\vert 0\rangle$, and $\gamma^{\dagger}_{\uparrow+}\gamma^{\dagger}_{\downarrow+}\vert V \rangle_+=d^{\dagger}_{\uparrow}d^{\dagger}_{\downarrow}\vert 0\rangle$. The corresponding eigenenergies are $\mathcal{E}^{D}_{+}=0$, $\mathcal{E}^{D}_{+}+E_{\uparrow+}=\epsilon_D+h_D$, $\mathcal{E}^{D}_{+}+E_{\downarrow+}=\epsilon_D-h_D$, and $\mathcal{E}^{D}_{+}+E_{\uparrow+}+E_{\downarrow+}=2\epsilon_D$,  respectively.  The ground state is the lowest energy state among these four eigenstates, which requires adding all negative single particles of the spin OBS $(\gamma_{\uparrow+},\gamma_{\downarrow+})$ into the EVS $\vert V\rangle_{+}$, see Fig.~\ref{FIGQD}(a). If $\epsilon_D>h_D>0$, both $E_{\uparrow+}$ and $E_{\downarrow+}$ are positive, and hence the ground state is equal to  EVS $\vert V\rangle_{+}$, i.e., $\vert G\rangle=\vert V\rangle_{+}=\vert 0\rangle$.  However, if $\epsilon_D<h_D$, $E_{\uparrow+}$ remains positive while $E_{\downarrow+}$ becomes negative. In this case, we need to add the single particle with energy $E_{\downarrow+}$ to the EVS $\vert V\rangle_{+}$. As a result, we obtain an odd-parity ground state $\vert G\rangle=\gamma^{\dagger}_{\downarrow+}\vert V\rangle_{+}=d^{\dagger}_{\downarrow}\vert 0\rangle$. Therefore, the fermionic parity of the ground state changes when the single-particle energy level $E_{\downarrow+}$ crosses zero.

Next, let us consider  another spin OBS  $(\gamma_{\uparrow-},\gamma_{\downarrow-})=(d^{\dagger}_{\downarrow},d^{\dagger}_{\uparrow})$ with  EVS $\vert V\rangle_{-}=d^{\dagger}_{\uparrow}d^{\dagger}_{\downarrow}\vert 0\rangle$ (hole picture). The quantum dot has four eigenstates: $\vert V \rangle_-=d^{\dagger}_{\uparrow}d^{\dagger}_{\downarrow}\vert 0\rangle$, $\gamma^{\dagger}_{\uparrow-}\vert V \rangle_-=d^{\dagger}_{\uparrow}\vert 0\rangle$, $\gamma^{\dagger}_{\downarrow-}\vert V \rangle_-=d^{\dagger}_{\downarrow}\vert 0\rangle$, and $\gamma^{\dagger}_{\uparrow-}\gamma^{\dagger}_{\downarrow-}\vert V \rangle_-=\vert 0\rangle$. The corresponding eigenenergies are $\mathcal{E}^{D}_{-}=2\epsilon_D$, $\mathcal{E}^{D}_{-}+E_{\uparrow-}=\epsilon_D+h_D$, $\mathcal{E}^{D}_{-}+E_{\downarrow-}=\epsilon_D-h_D$, and $\mathcal{E}^{D}_{-}+E_{\uparrow-}+E_{\downarrow-}=0$,  respectively. In the hole picture, similar to the electron picture, the ground state is obtained by adding all negative single particles of the chosen spin OBS $(\gamma_{\uparrow-},\gamma_{\downarrow-})$ into the corresponding  EVS $\vert V\rangle_{-}$, see Fig. \ref{FIGQD} (b). When $\epsilon_D>h_D>0$, both $E_{\uparrow-}$ and $E_{\downarrow-}$ are negative, and the ground state requires adding both $E_{\uparrow-}$ and $E_{\downarrow-}$ single particles to the EVS $\vert V\rangle_{-}$. Thus, the ground state is given by $\vert G\rangle=\gamma^{\dagger}_{\uparrow-}\gamma^{\dagger}_{\downarrow-}\vert V\rangle_{-}=\vert 0\rangle$, where $\gamma^{\dagger}_{\uparrow-}=d_{\downarrow}$ and $\gamma^{\dagger}_{\downarrow-}=d_{\uparrow}$ annihilate the electrons in the EVS $\vert V\rangle_{-}$. When $\epsilon_D<h_D$, $E_{\uparrow-}$ becomes positive while $E_{\downarrow-}$ remains negative. In this case, adding the negative quasiparticle $E_{\downarrow-}$ to the EVS $\vert V\rangle_{-}$ is equivalent to removing a spin-up electron from $\vert V\rangle_{-}$. Thus, the ground state becomes an odd-parity state given by $\vert G\rangle=\gamma^{\dagger}_{\downarrow-}\vert V\rangle_{-}=d^{\dagger}_{\downarrow}\vert 0\rangle$. It is worth noting that the ground state undergoes a fermionic parity change when the single-particle energy level $E_{\uparrow-}$ crosses zero.

Although the EVSs of the two spin OBSs $(\gamma_{\uparrow+},\gamma_{\downarrow+})$ and $(\gamma_{\uparrow-},\gamma_{\downarrow-})$ differ in energy ($\mathcal{E}_{+}\neq \mathcal{E}_{-}$) and wave function ($\vert V\rangle_{+}\neq \vert V\rangle_{-}$), the quantum dot possesses the same four eigenstates, including the ground state $\vert G\rangle$ and three excitation states. To obtain the ground state in each picture, one needs to add all negative single particles of the chosen spin OBS to the corresponding EVS. Furthermore, the same eigenstate of the quantum dot can be represented differently using different spin SBSs and EVSs. For example, $\vert 0\rangle=\vert V \rangle_+=\gamma^{\dagger}_{\uparrow-}\gamma^{\dagger}_{\downarrow-}\vert V \rangle_-$. Additionally, the fermionic parity of the EVSs $\vert V\rangle_{\eta}$ is even, while the ground state's fermionic parity becomes odd when $\vert h_D\vert >\vert\epsilon_D\vert$ [as shown in Figs. \ref{FIGQD} (a) and (b)]. This change in fermionic parity occurs when the single-particle energy levels cross the zero-energy point.

PROPERTY I): Let's analyze the relation between the two EVSs. When $h_D=0$, the ground state can be expressed in terms of the two spin OBSs as $\vert G\rangle=\vert V\rangle_+=\gamma^{\dagger}_{\uparrow-}\gamma^{\dagger}_{\downarrow-}\vert V\rangle_{-}$. This implies that the EVSs $\vert V\rangle_+$ and $\vert V\rangle_-$ form a conjugate loop, as illustrated in Fig. \ref{FIGQD}(e). The relation between the EVSs can be written as
\begin{align} \label{fdkfdk}
    \vert V\rangle_{\eta}= \gamma^{\dagger}_{\uparrow-\eta}\gamma^{\dagger}_{\downarrow-\eta}\vert V\rangle_{-\eta}.
\end{align}
This relation arises due to the conjugate relation \eqref{mfdflal}. It is important to note that this relation implies that  EVS $\vert V\rangle_{\eta}$ is filled from another  EVS $\vert V\rangle_{-\eta}$ instead of the vacuum of electrons $\vert 0\rangle$.

\subsubsection{Nambu OBSs}

STEP I): By exchanging a pair of basis states in the spin OBSs, we arrive at the Nambu OBSs $(\gamma_{\uparrow+},\gamma_{\uparrow-})$ and $(\gamma_{\downarrow+},\gamma_{\downarrow-})$. These basis sets select a specific spin species and combine the electron and hole pictures. By applying the conjugate relation \eqref{mfdflal}, we can obtain the corresponding Hamiltonian from the BdG Hamiltonian \eqref{mfdvkvkdk} as follows
\begin{align} \label{gfjbofg}
H_D=\sum_{\eta}E_{\uparrow\eta}\gamma^{\dagger}_{\uparrow\eta}\gamma^{}_{\uparrow\eta}+\mathcal{E}^{D}_{\uparrow}=\sum_{\eta}E_{\downarrow\eta}\gamma^{\dagger}_{\downarrow\eta}\gamma^{}_{\downarrow\eta}+\mathcal{E}^{D}_{\downarrow}.
\end{align}
Here, $\mathcal{E}^{D}_{\uparrow}=\epsilon_D-h_D$ and $\mathcal{E}^{D}_{\downarrow}=\epsilon_D+h_D$ are energy constants that arise when we mathematically rewrite the BdG Hamiltonian \eqref{mfdvkvkdk} into the Hamiltonian \eqref{gfjbofg} using the Nambu OBSs $(\gamma_{\uparrow+},\gamma_{\uparrow-})$ and $(\gamma_{\downarrow+},\gamma_{\downarrow-})$, respectively.

STEP II): The  EVSs for the Nambu OBSs  $(\gamma_{\uparrow+},\gamma_{\uparrow-})$ and $(\gamma_{\downarrow+},\gamma_{\downarrow-})$ are, respectively, defined as follows
\begin{align}
    \gamma_{\uparrow\eta}\vert V\rangle_{\uparrow}=0 \text{ for all } \eta=+/-,
\end{align}
\begin{align}
    \gamma_{\downarrow\eta}\vert V\rangle_{\downarrow}=0 \text{ for all } \eta=+/-.
\end{align}
It is straightforward to obtain the EVSs
\begin{align} \label{gfkmdvkp}
    \vert V\rangle_{\uparrow}=d^{\dagger}_{\downarrow}\vert 0\rangle,
\end{align}
\begin{align} \label{gfkmdvkm}
    \vert V\rangle_{\downarrow}=d^{\dagger}_{\uparrow}\vert 0\rangle.
\end{align}
Thus, $\mathcal{E}^{D}_{\uparrow}$ and $\mathcal{E}^{D}_{\downarrow}$ represent the energies of the EVSs $\vert V\rangle_{\uparrow}$ and $\vert V\rangle_{\downarrow}$, respectively.

STEP III): Let us consider the Nambu OBSs $(\gamma_{\uparrow+},\gamma_{\uparrow-})=(d_{\uparrow},d^{\dagger}_{\downarrow})$   with the corresponding EVS $\vert V\rangle_{\uparrow}=d^{\dagger}_{\downarrow}\vert 0\rangle$. The quantum dot has four eigenstates:  $\vert V \rangle_{\uparrow}=d^{\dagger}_{\downarrow}\vert 0\rangle$, $\gamma^{\dagger}_{\uparrow+}\vert V \rangle_{\uparrow}=d^{\dagger}_{\uparrow}d^{\dagger}_{\downarrow}\vert 0\rangle$, $\gamma^{\dagger}_{\uparrow-}\vert V \rangle_{\uparrow}=\vert 0\rangle$, and $\gamma^{\dagger}_{\uparrow+}\gamma^{\dagger}_{\uparrow-}\vert V \rangle_{\uparrow}=d^{\dagger}_{\uparrow}\vert 0\rangle$. The corresponding energies are given by $\mathcal{E}^{D}_{\uparrow}=\epsilon_D-h_D$, $\mathcal{E}^{D}_{\uparrow}+E_{\uparrow+}=2\epsilon_D$, $\mathcal{E}^{D}_{\uparrow}+E_{\uparrow-}=0$, and $\mathcal{E}^{D}_{\uparrow}+E_{\uparrow+}+E_{\uparrow-}=\epsilon_D+h_D$,  respectively.  To determine the ground state, we need to find the lowest energy state among these four states. This involves adding all the negative single particles of the Nambu OBS  $(\gamma_{\uparrow+},\gamma_{\uparrow-})$ into the corresponding EVS $\vert V\rangle_{\uparrow}$, see Fig. \ref{FIGQD} (c). For the case where $\epsilon_D>h_D>0$, we have $E_{\uparrow+}>0$ and $E_{\uparrow-}<0$. Therefore, we need to add $E_{\uparrow-}$ quasiparticle to the EVS $\vert V\rangle_{\uparrow}$, which corresponds to removing a spin-down electron from $\vert V\rangle_{\uparrow}$. As a result, the ground state in the even-parity sector is given by $\vert G\rangle=\gamma^{\dagger}_{\uparrow-}\vert V\rangle_{\uparrow}=\vert 0\rangle$. On the other hand, for the case where $\epsilon_D<h_D$, we have $E_{\uparrow+}>0$ and $E_{\uparrow-}>0$. In this scenario, the ground state is simply equal to the EVS $\vert V\rangle_{\uparrow}$, i.e., $\vert G\rangle=\vert V\rangle_{\uparrow}=d^{\dagger}_{\downarrow}\vert 0\rangle$.

Next, we consider another Nambu OBS $(\gamma_{\downarrow+},\gamma_{\downarrow-})=(d^{}_{\downarrow},d^{\dagger}_{\uparrow})$ with the corresponding EVS $\vert V\rangle_{\downarrow}=d^{\dagger}_{\uparrow}\vert 0\rangle$. The quantum dot has four eigenstates: $\vert V \rangle_{\downarrow}=d^{\dagger}_{\uparrow}\vert 0\rangle$, $\gamma^{\dagger}_{\downarrow+}\vert V \rangle_{\downarrow}=d^{\dagger}_{\downarrow}d^{\dagger}_{\uparrow}\vert 0\rangle$, $\gamma^{\dagger}_{\downarrow-}\vert V \rangle_{\downarrow}=\vert 0\rangle$, and $\gamma^{\dagger}_{\downarrow+}\gamma^{\dagger}_{\downarrow-}\vert V \rangle_{\downarrow}=d^{\dagger}_{\uparrow}\vert 0\rangle$. The corresponding energies are $\mathcal{E}^{D}_{\downarrow}=\epsilon_D+h_D$, $\mathcal{E}^{D}_{\downarrow}+E_{\downarrow+}=2\epsilon_D$, $\mathcal{E}^{D}_{\downarrow}+E_{\downarrow-}=0$, and $\mathcal{E}^{D}_{\downarrow}+E_{\downarrow+}+E_{\downarrow-}=\epsilon_D-h_D$,  respectively. Similar to before, we find the lowest energy state among these four states to determine the ground state. This time, we add all the negative single-particle energies of the Nambu OBS $(\gamma_{\downarrow+},\gamma_{\downarrow-})$ into the EVS $\vert V\rangle_{\downarrow}$, as plotted in Fig. \ref{FIGQD} (d). For the case where $\epsilon_D>h_D>0$, we have $E_{\downarrow+}>0$ and $E_{\downarrow-}<0$. Thus, we need to add $E_{\downarrow-}$ single particle to the EVS $\vert V\rangle_{\downarrow}$, which is equivalent to removing the spin-up electron from the EVS $\vert V\rangle_{\downarrow}$. As a result, the even-parity ground state is given by $\vert G\rangle=\gamma^{\dagger}_{\downarrow-}\vert V\rangle_{\downarrow}=\vert 0\rangle$. On the other hand, for the case where $\epsilon_D<h_D$, we have $E_{\downarrow+}<0$ and $E_{\downarrow-}<0$. In this scenario, we need to add both $E_{\downarrow+}$ and $E_{\downarrow-}$ quasiparticles to the eigenstate $\vert V\rangle_{\downarrow}$, which correspond to removing the spin-up electron and adding the spin-down electron to $\vert V\rangle_{\downarrow}$, respectively. Thus, the odd-parity ground state is given by $\vert G\rangle=\gamma^{\dagger}_{\downarrow+}\gamma^{\dagger}_{\downarrow-}\vert V\rangle_{\downarrow}=d^{\dagger}_{\downarrow}\vert 0\rangle$.

Though the  EVSs of the two Nambu OBSs $(\gamma_{\uparrow+},\gamma_{\uparrow-})$ and $(\gamma_{\downarrow+},\gamma_{\downarrow-})$ differ in terms of energy ($\mathcal{E}_{\uparrow}\neq \mathcal{E}_{\downarrow}$) and wave function ($\vert V\rangle_{\uparrow}\neq \vert V\rangle_{\downarrow}$), it is possible to have the same ground state $\vert G \rangle$ and three excitation states. The ground state $\vert G \rangle$ is obtained by populating all negative single-particle states of the chosen OBS within the corresponding EVS. It's important to note that the same quantum dot eigenstate can be represented differently using different OBS and EVS. For instance, $\vert 0\rangle=\gamma^{\dagger}_{\uparrow-}\vert V \rangle_{\uparrow}=\gamma^{\dagger}_{\downarrow-}\vert V \rangle_{\downarrow}$. This highlights that a single physical state can be described using distinct representations based on the chosen OBS and EVS. Furthermore, the fermionic parity of the EVSs $\vert V\rangle_{s}$ is odd, while the fermionic parity of the ground state becomes even when $h_D<\epsilon_D$ [referring to Figs. \ref{FIGQD} (c) and (d)]. As a result, the fermionic parity of the ground state undergoes a change when the quantum dot levels cross zero-energy.

PROPERTY I): Then, we try to find the relation between these two EVSs. When $h_D=0$, the ground state can be expressed as $\vert G \rangle = \gamma^{\dagger}_{\uparrow-}\vert V\rangle_{\uparrow} = \gamma^{\dagger}_{\downarrow-}\vert V\rangle_{\downarrow}$. This implies that the EVSs $\vert V\rangle_{\uparrow}$ and $\vert V\rangle_{\downarrow}$ form a conjugate loop, as shown in Fig. \ref{FIGQD} (f)
\begin{align}\label{fdtykfdk}
    \vert V\rangle_{s}= \gamma^{\dagger}_{-s+}\gamma^{\dagger}_{-s-}\vert V\rangle_{-s}.
\end{align}
In other words, the EVS $\vert V\rangle_{s}$ is filled from another EVS $\vert V\rangle_{-s}$ due to the conjugate relation \eqref{mfdflal}.

PROPERTY II): By comparing EVSs \eqref{fdvnknfvlkp} and \eqref{fdvnknfvlkm} with EVSs \eqref{gfkmdvkp} and \eqref{gfkmdvkm}, we can easily prove that $\vert V\rangle_{\uparrow}=\gamma^{\dagger}_{\downarrow+}\vert V\rangle_{+}$ and $\vert V\rangle_{\downarrow}=\gamma^{\dagger}_{\uparrow-}\vert V\rangle_{-}$. This implies that when a basis state is exchanged, the fermionic parity of the EVSs changes.

\subsection{Normal metal case} \label{newlogic}
In this subsection, we use our logic in normal metal. The normal metal  behaves as many quantum dots, which is described by the Hamiltonian
\begin{align} \label{fsgnhh}
    H_{L}=\sum_{\vec{k}s}(\epsilon_{\vec{k}}+sh_L)c^{\dagger}_{\vec{k}s}c^{}_{\vec{k}s}.
\end{align}
Next, we work in Nambu space, which doubles the Hilbert space, and we obtain the BdG Hamiltonian in overcomplete basis set 
\begin{align} \label{gfeyeegblrgk}
    H_L= \frac{1}{2}\sum_{\vec{k}}\sum_{s=\uparrow/\downarrow}\sum_{\eta=\pm} E_{\vec{k}s\eta}\gamma_{\vec{k}s\eta}^{\dagger} \gamma_{\vec{k}s\eta }^{}+\mathcal{E},
\end{align}
with $\mathcal{E}=\sum_{\vec{k}}\epsilon_{\vec{k}}$. Additional index $\eta=+/-$ labels the high/low energy levels of each spin species. The quasiparticle  energy spectrum is given by  $E^{}_{\vec{k}s\eta}=sh_L+\eta\vert\epsilon_{\vec{k}}\vert$ and the corresponding  operators read 
\begin{align} \label{ufdvmkdkplus} 
    \gamma^{\dagger}_{\vec{k}s+}=u_{\vec{k}}c^{\dagger}_{\vec{k}s}+sv^{}_{\vec{k}}c^{}_{-\vec{k}-s},
\end{align}
\begin{align} \label{ufdvmkdkminus}
    \gamma^{\dagger}_{\vec{k}s-}=-s v^{*}_{\vec{k}}c^{\dagger}_{\vec{k}s}+u_{\vec{k}}c^{}_{-\vec{k}-s},
\end{align}
with
\begin{align} \label{ufvdnvku}
    u_{\vec{k}}=\left\{\begin{matrix}
      1,& \epsilon_{\vec{k}}>0;\\
      0, & \epsilon_{\vec{k}}\leq 0.
    \end{matrix}\right.,
\end{align}
\begin{align} \label{ufvdnvkv}
    v_{\vec{k}}=\left\{\begin{matrix}
      0,& \epsilon_{\vec{k}}>0;\\
      1, & \epsilon_{\vec{k}}\leq 0.
    \end{matrix}\right..
\end{align}
The quasiparticle operators satisfy a conjugate relation 
\begin{align} \label{uutpfvvldl}
    \gamma^{\dagger}_{\vec{k}s\eta}&=  \gamma^{}_{-\vec{k}-s-\eta}.
\end{align}
Each spinful  quasiparticle always has its conjugate alternative with opposite spin and energy. Thus,  we can have $4^{N}$ combinations of complete and orthogonal quaiparticle bases where $N$ describes the size of the normal metal (or number of the kinetic momentum). In principle, we can choose any OBS, for instance spin-momentum OBSs  $(\gamma^{}_{\vec{k}\uparrow\eta},\gamma^{}_{\vec{k}\downarrow\eta})$ for all $\vec{k}$  as well as  Nambu-momentum OBSs  $(\gamma^{}_{\vec{k}s+},\gamma^{}_{\vec{k}s-})$ for all $\vec{k}$. 
We will see details in the following subsections.

\subsubsection{Spin-momentum OBSs}
STEP I): Let us first study the spin-momentum OBSs $(\gamma^{}_{\vec{k}\uparrow\eta},\gamma^{}_{\vec{k}\downarrow\eta})$ for all $\vec{k}$, which satisfy the anticommutation relation. The corresponding Hamiltonian is given by 
\begin{align} \label{ghkbgkb}
    H_L&=\sum_{\vec{k}s}E_{\vec{k}s+}\gamma^{\dagger}_{\vec{k}s+}\gamma^{}_{\vec{k}s+}+\mathcal{E}^{L}_{+}\\
    &=\sum_{\vec{k}s}E_{\vec{k}s-}\gamma^{\dagger}_{\vec{k}s-}\gamma^{}_{\vec{k}s-}+\mathcal{E}^{L}_{-}.\notag 
\end{align}
$\mathcal{E}^{L}_{+}=\mathcal{E}+\frac{1}{2}\sum_{\vec{k}s} E_{\vec{k}s-}$ and $\mathcal{E}^{L}_{-}=\mathcal{E}+\frac{1}{2}\sum_{\vec{k}s} E_{\vec{k}s+}$ are the energy constants generated when we rewrite the origin quantum dot Hamiltonian \eqref{fsgnhh} into the Hamiltonian \eqref{ghkbgkb} with the spin-momentum OBSs $(\gamma_{\vec{k}\uparrow+},\gamma_{\vec{k}\downarrow+})$ and $(\gamma_{\vec{k}\uparrow-},\gamma_{\vec{k}\downarrow-})$, respectively.

STEP II):  The EVSs for $\eta=\pm$ quasiparticles, defined by  $\gamma^{}_{\vec{k}s\eta}\vert V\rangle_{\eta}=0$ for all momentum $\vec{k}$ and spins $s=\uparrow/\downarrow$, respectively,
are given by 
\begin{align} \label{upmfvdkdmfk1}
     \vert V\rangle_+=\prod_{\epsilon_{\vec{k}}\leq 0} c_{\vec{k} \uparrow}^{\dagger} c_{\vec{k} \downarrow}^{\dagger}|0\rangle,
\end{align}
\begin{align} \label{upmfvdkdmfk2}
   \vert V\rangle_-=\prod_{\epsilon_{\vec{k}}>0} c_{\vec{k} \uparrow}^{\dagger} c_{\vec{k} \downarrow}^{\dagger}|0\rangle.
\end{align}
Thus, $\mathcal{E}^{L}_{+}$ and $\mathcal{E}^{L}_{-}$ is the energy of the EVSs $\vert V\rangle_{+}$ and $\vert V\rangle_{-}$, respectively. 

\begin{figure}
\begin{center}
\includegraphics[width=1\linewidth]{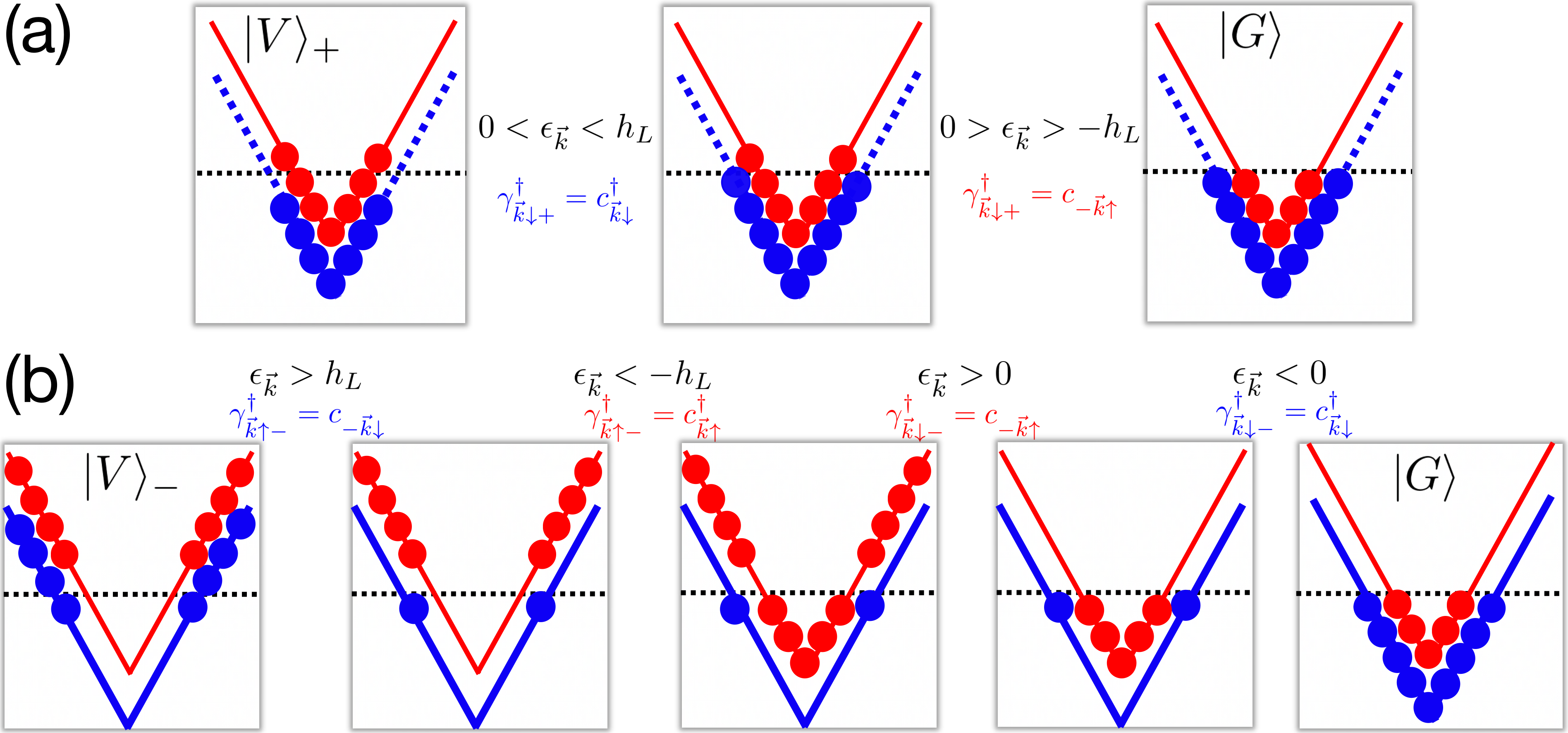}
\end{center}
\caption{Illustration of Eq. \eqref{ufvklal}. Panels (a) and (b) correspond to the construction of the ground state from Fermi sea and sky, respectively. 
}
\label{FIGNL} 
\end{figure}

STEP III): Let us calculate the ground state in term of the spin-momentum OBS  $(\gamma^{}_{\vec{k}\uparrow+},\gamma^{}_{\vec{k}\downarrow+})$. The normal metal has $4^{N}=\sum^{2N}_{n=0}C(2N,n)$ eigenstates, where $C(2N,n)=(2N)!/[(2N-n)!n!]$ is  combination formula. These eigenstates include $C(2N,0)$ EVS $\vert V \rangle_+$, $C(2N,1)$ one-quasiparticles states $\gamma^{\dagger}_{\vec{k}_1s_1+}\vert V \rangle_+$, $C(2N,2)$ two-quasiparticles states $\gamma^{\dagger}_{\vec{k}_1s_1+}\gamma^{\dagger}_{\vec{k}_2s_2+}\vert V \rangle_+$, and so on, up to  $C(2N,2N)$ $2N$-quasiparticles states $(\prod_{\vec{k}s}\gamma^{\dagger}_{\vec{k}s+})\vert V \rangle_+$. The corresponding energies of these eigenstates are $\mathcal{E}^{L}_{+}$, $\mathcal{E}^{L}_{+}+E_{\vec{k}_1s_1+}$, $\mathcal{E}^{L}_{+}+E_{\vec{k}_1s_1+}+E_{\vec{k}_2s_2+}$, and so on, up to $\mathcal{E}^{L}_{+}+\sum_{\vec{k}s}E_{\vec{k}s+}$,  respectively, where $\vec{k}_i,s_i$ are required to satisfy the Pauli exclusion principle  for many quasiparticles. The ground state is the lowest energy state  among these $4^{N}$ states. In this case, the ground state is the Fermi sea, which corresponds to filling all the negative-energy quasiparticles in the OBS $(\gamma_{\vec{k}\uparrow+},\gamma_{\vec{k}\downarrow+})$ starting from the EVS $\vert V \rangle_+$. Similarly, for the OBS $(\gamma_{\vec{k}\uparrow-},\gamma_{\vec{k}\downarrow-})$, the ground state is given by filling all the negative-energy  quasiparticles in the OBS $(\gamma_{\vec{k}\uparrow-},\gamma_{\vec{k}\downarrow-})$ starting from the EVS $\vert V \rangle_-$. Therefore,  we obtain the  ground state
\begin{align} \label{ufvklal}
    \vert G \rangle=\left(\prod_{E^{}_{\vec{k}s+}<0}\gamma^{\dagger}_{\vec{k}s+}\right)\vert V\rangle_+=\left(\prod_{E^{}_{\vec{k}s-}<0}\gamma^{\dagger}_{\vec{k}s-}\right)\vert V\rangle_-.
\end{align}
When we work in the spin-momentum OBS $(\gamma^{}_{\vec{k}\uparrow+},\gamma^{}_{\vec{k}\downarrow+})$, this product  adds spin-down electrons with $0<\epsilon_{\vec{k}}<h_L$ ($\gamma^{\dagger}_{\vec{k}\downarrow+}=c^{\dagger}_{\vec{k}\downarrow}$) and removes spin-up electrons with energy in the interval $-h_L<\epsilon_{-\vec{k}}<0$ ($\gamma^{\dagger}_{\vec{k}\downarrow+}=c^{}_{-\vec{k}\uparrow}$) [Fig. \ref{FIGNL} (a)]. When we work in another spin-momentum OBS  $(\gamma^{}_{\vec{k}\uparrow-},\gamma^{}_{\vec{k}\downarrow-})$, this product removes spin-down electrons with energy in the interval $\epsilon_{\vec{k}}>h_L$ ($\gamma^{\dagger}_{\vec{k}\uparrow-}=c^{}_{-\vec{k}\downarrow}$), adds spin-up electrons with $0<\epsilon_{\vec{k}}<-h_L$ ($\gamma^{\dagger}_{\vec{k}\uparrow-}=c^{\dagger}_{\vec{k}\uparrow}$), removes spin-up electrons with energy in the interval $\epsilon_{\vec{k}}>0$ ($\gamma^{\dagger}_{\vec{k}\downarrow-}=c^{}_{-\vec{k}\uparrow}$), and adds spin-down electrons with $\epsilon_{\vec{k}}<0$ ($\gamma^{\dagger}_{\vec{k}\downarrow-}=c^{\dagger}_{\vec{k}\downarrow}$) [Fig. \ref{FIGNL} (b)]. These two combinations of the complete and orthogonal basis states obtain the same ground state as shown in Fig. \ref{FIGNL}.

PROPERTY I): It is clear that we can add any $E_{\vec{k}s\eta}$ quasiparticle into the EVS $\vert V\rangle_{\eta}$, which, according to conjugate relation \eqref{uutpfvvldl}, is equal to removing a $E_{-\vec{k}-s-\eta}$ quasiparticle. Thus, $\vert V\rangle_+$ ($\vert V\rangle_-$) must contain all $\eta=-$ ($\eta=+$) quasiparticles. 
For $h_L=0$, the ground state can be expressed in terms of two combinations of the basis states $\vert G
\rangle=\vert V\rangle_+=\prod_{\vec{k}}\gamma^{}_{\vec{k}\uparrow-}\gamma^{}_{\vec{k}\downarrow-}\vert V\rangle_-$.  
Thus, we discover the EVSs \eqref{upmfvdkdmfk1} and \eqref{upmfvdkdmfk2} relate to each other through
\begin{align} \label{upmvfnjk1}
\vert V\rangle_{\eta}=\prod_{ \vec{k}} \gamma_{ \vec{k} \uparrow-\eta}^{\dagger}\gamma_{-\vec{k} \downarrow-\eta}^{\dagger} \vert V\rangle_{-\eta}.
\end{align}
The EVS  $\vert V\rangle_{+}$ is  a filled sea of negative quasiparticle states for the whole Brillouin zone with respect to their EVS $\vert V\rangle_{-}$, rather than the vacuum of electrons $\vert 0\rangle$. By substitution of Eqs. \eqref{ufdvmkdkplus} and  \eqref{ufdvmkdkminus}, we can quickly check
\begin{align} \label{fdvmkdfmk}
    \prod_{ \vec{k}} \gamma_{ \vec{k} \uparrow-}^{\dagger}\gamma_{-\vec{k} \downarrow-}^{\dagger} \vert 0\rangle=0.
\end{align}
Notable, the wrong usage of the EVS will generate meaningless system state.

\subsubsection{Nambu-momentum OBSs}

STEP I): Then, we study the Nambu-momentum OBSs  $(\gamma^{}_{\vec{k}s+},\gamma^{}_{\vec{k}s-})$ for the whole Brillouin zone, which satisfy the anticommutation relation. The corresponding Hamiltonian is given by 
\begin{align} \label{fvdmfkvmk}
    H_L&=\sum_{\vec{k}\eta}E_{\vec{k}\uparrow\eta}\gamma^{\dagger}_{\vec{k}\uparrow\eta}\gamma^{}_{\vec{k}\uparrow\eta}+\mathcal{E}^{L}_{\uparrow}\\
    &=\sum_{\vec{k}\eta}E_{\vec{k}\downarrow\eta}\gamma^{\dagger}_{\vec{k}\downarrow\eta}\gamma^{}_{\vec{k}\downarrow\eta}+\mathcal{E}^{L}_{\downarrow}.\notag 
\end{align}
$\mathcal{E}^{L}_{\uparrow}=\mathcal{E}+\frac{1}{2}\sum_{\vec{k}\eta} E_{\vec{k}\downarrow\eta}$ and $\mathcal{E}^{L}_{\downarrow}=\mathcal{E}+\frac{1}{2}\sum_{\vec{k}\eta} E_{\vec{k}\uparrow\eta}$ are the energy constants generated when we rewrite the origin quantum dot Hamiltonian \eqref{fsgnhh} into the Hamiltonian \eqref{fvdmfkvmk} with the chosen OBSs $(\gamma_{\vec{k}\uparrow+},\gamma_{\vec{k}\uparrow-})$ and $(\gamma_{\vec{k}\downarrow+},\gamma_{\vec{k}\downarrow-})$, respectively.

STEP II): The EVSs for $s=\uparrow/\downarrow$ quasiparticles, defined by  $\gamma^{}_{\vec{k}s\eta}\vert V\rangle_{s}=0$ for all momentum $\vec{k}$ and Nambu $\eta$, respectively,
are given by \cite{datta1996scattering,datta1999can}
\begin{align} \label{urpmfvdkdmfk1}
     \vert V\rangle_{\uparrow}=\prod_{\vec{k}}c_{\vec{k} \downarrow}^{\dagger} |0\rangle,
\end{align}
\begin{align} \label{urpmfvdkdmfk2}
   \vert V\rangle_{\downarrow}=\prod_{\vec{k}}c_{\vec{k} \uparrow}^{\dagger}|0\rangle.
\end{align}
Thus, $\mathcal{E}^{L}_{\uparrow}$ and $\mathcal{E}^{L}_{\downarrow}$ is the energy of the EVSs $\vert V\rangle_{\uparrow}$ and $\vert V\rangle_{\downarrow}$, respectively. 
The EVS $\vert V\rangle_{\uparrow}$ ($\vert V\rangle_{\downarrow}$) contains all spin down (up) electrons. 

\begin{figure}[t]
\begin{center}
\includegraphics[width=1\linewidth]{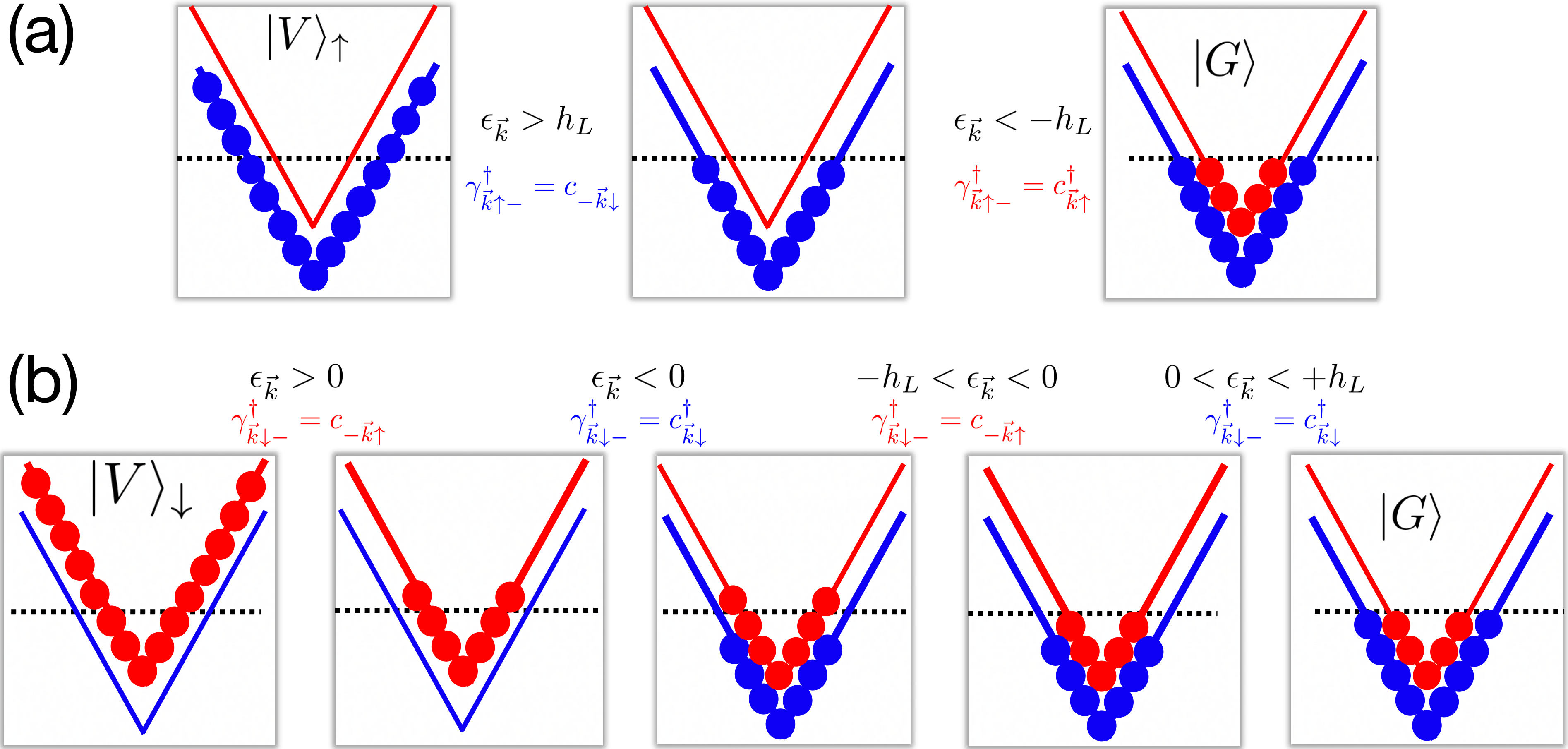}
\end{center}
\caption{Illustration of Eq. \eqref{ufksdlfd}. Panels (a) and (b) correspond to the construction of the ground state from EVSs $\vert V\rangle_{\uparrow}$ and $\vert V\rangle_{\downarrow}$, respectively. 
}
\label{FIGNLS} 
\end{figure}

STEP III): Let us express the  ground state using the Nambu-momentum OBSs -- $(\gamma^{}_{\vec{k}\uparrow+},\gamma^{}_{\vec{k}\uparrow-})$ for all $\vec{k}$. Note that there are $4^{N}$ eigenstates: $C(2N,0)$ EVS $\vert V \rangle_{\uparrow}$, $C(2N,1)$ one-quasiparticle states $\gamma^{\dagger}_{\vec{k}_1\uparrow\eta_1}\vert V \rangle_{\uparrow}$, $C(2N,2)$ two-quasiparticle states $\gamma^{\dagger}_{\vec{k}_1\uparrow\eta_1}\gamma^{\dagger}_{\vec{k}_2\uparrow\eta_2}\vert V \rangle_{\uparrow}$, and so on, up to $C(2N,2N)$ $2N$-quasiparticle state $(\prod_{\vec{k}\eta}\gamma^{\dagger}_{\vec{k}\uparrow\eta})\vert V \rangle_{\uparrow}$. The corresponding energies are $\mathcal{E}^{L}_{\uparrow}$, $\mathcal{E}^{L}_{\uparrow}+E_{\vec{k}_1\uparrow\eta_1}$, $\mathcal{E}^{L}_{\uparrow}+E_{\vec{k}_1\uparrow\eta_1}+E_{\vec{k}_2\uparrow\eta_2}$, and so on, up to  $\mathcal{E}^{L}_{\uparrow}+\sum_{\vec{k}\eta}E_{\vec{k}\uparrow\eta}$,  respectively. Here, $\vec{k}_i,\eta_i$ are required to satisfy the Pauli exclusion principle  for many  quasiparticles. The ground state is the state with the lowest energy among these $4^{N}$ eigenstates. It corresponds to filling all negative-energy  quasiparticles of the Nambu-momentum OBS $(\gamma^{}_{\vec{k}\uparrow+},\gamma^{}_{\vec{k}\uparrow-})$ into the EVS $\vert V\rangle_{\uparrow}$. The same applies to the Nambu-momentum OBS $(\gamma^{}_{\vec{k}\downarrow+},\gamma^{}_{\vec{k}\downarrow-})$ for all $\vec{k}$, using the EVS \eqref{urpmfvdkdmfk2}. Therefore, we reach
\begin{align} \label{ufksdlfd}
    \vert G\rangle =\left(\prod_{E^{}_{\vec{k}\uparrow\eta}<0}\gamma^{\dagger}_{\vec{k}\uparrow\eta}\right)\vert V\rangle_{\uparrow}=\left(\prod_{E^{}_{\vec{k}\downarrow\eta}<0}\gamma^{\dagger}_{\vec{k}\downarrow\eta}\right)\vert V\rangle_{\downarrow}.
\end{align}
When we work in the Nambu-momentum OBS $(\gamma^{}_{\vec{k}\uparrow+},\gamma^{}_{\vec{k}\uparrow-})$, this product removes spin-down electrons with $\epsilon_{\vec{k}}>h_L$ ($\gamma^{\dagger}_{\vec{k}\uparrow-}=c^{\dagger}_{-\vec{k}\downarrow}$) and adds spin-up electrons with energy in the interval $\epsilon_{\vec{k}}<-h_L$ ($\gamma^{\dagger}_{\vec{k}\uparrow-}=c^{\dagger}_{\vec{k}\uparrow}$) [Fig. \ref{FIGNLS} (a)]. When we work in another Nambu-momentum OBS $(\gamma^{}_{\vec{k}\downarrow+},\gamma^{}_{\vec{k}\downarrow-})$, this product removes spin-down electrons with energy in the interval $\epsilon_{\vec{k}}>0$ ($\gamma^{\dagger}_{\vec{k}\downarrow-}=c^{}_{-\vec{k}\uparrow}$), adds spin-up electrons with $\epsilon_{\vec{k}}<0$ ($\gamma^{\dagger}_{\vec{k}\downarrow-}=c^{\dagger}_{\vec{k}\downarrow}$), removes spin-up electrons with energy in the interval $-h_L<\epsilon_{\vec{k}}<0$ ($\gamma^{\dagger}_{\vec{k}\downarrow-}=c^{}_{-\vec{k}\uparrow}$), and adds spin-down electrons with $0<\epsilon_{\vec{k}}<+h_L$ ($\gamma^{\dagger}_{\vec{k}\downarrow-}=c^{\dagger}_{\vec{k}\downarrow}$) [Fig. \ref{FIGNLS} (b)]. We find all OBSs obtain the same ground state as shown in Figs. \ref{FIGNL} and \ref{FIGNLS}.

PROPERTY I): Then, we try to find the relation of two EVSs \eqref{urpmfvdkdmfk1} and \eqref{urpmfvdkdmfk2}. For $h_D=0$, the ground state can be expressed in terms of two combinations of the basis states $\vert G
\rangle=\prod_{\vec{k}}\gamma^{\dagger}_{\vec{k}\uparrow-}\vert V\rangle_{\uparrow}=\prod_{\vec{k}}\gamma^{\dagger}_{\vec{k}\downarrow-}\vert V\rangle_{\downarrow}$.  Thus, EVSs $\vert V\rangle_{\uparrow}$ and $\vert V\rangle_{\downarrow}$ form a conjugate loop
\begin{align} \label{ufdkfmvdk}
    \vert V\rangle_{s}=\prod_{ \vec{k}}\gamma^{\dagger}_{\vec{k}-s+}\gamma^{\dagger}_{\vec{k}-s-}\vert V\rangle_{-s}.
\end{align}
Thus, for each spin, the EVS $\vert V\rangle_{s}$ contains all remaining quasiparticles dual to the chosen OBSs $(\gamma^{}_{\vec{k}s+},\gamma^{}_{\vec{k}s-})$ for all $\vec{k}$.

\subsection{Superconductor case} 
\label{superconductor}
Let us generalize our new logic to the superconductor case. The BdG  Hamiltonian, expressed in terms of an overcomplete basis set $(\gamma^{}_{\vec{k}\uparrow+},\gamma^{}_{\vec{k}\downarrow-},\gamma^{}_{\vec{k}\downarrow+},\gamma^{}_{\vec{k}\downarrow-})$ for all $\vec{k}$, is given by
\begin{align} \label{mgfeegblrgk}
    H_L= \frac{1}{2}\sum_{\vec{k}}\sum_{s=\uparrow/\downarrow}\sum_{\eta=\pm} E_{\vec{k}s\eta}\gamma_{\vec{k}s\eta}^{\dagger} \gamma_{\vec{k}s\eta }^{}+\mathcal{E},
\end{align}
with $\mathcal{E}=\sum_{\vec{k}}\epsilon_{\vec{k}}$. Electrons ($c^{\dagger}_{\vec{k}s}$) interact with holes ($-sc^{}_{-\vec{k}-s}$) to generate Bogoliubov quasiparticles with energy levels $E^{}_{\vec{k}s\eta}=sh_L+\eta\sqrt{\Delta^2+\epsilon^2_{\vec{k}}}$, as well as field operators
\begin{align} \label{fdvmdfvdkmplusq}
    \gamma^{\dagger}_{\vec{k}s+}=u_{\vec{k}}c^{\dagger}_{\vec{k}s}+sv^{}_{\vec{k}}c^{}_{-\vec{k}-s},
\end{align}
\begin{align} \label{fdvmdfvdkmminuq}
    \gamma^{\dagger}_{\vec{k}s-}=-s v^{*}_{\vec{k}}c^{\dagger}_{\vec{k}s}+u_{\vec{k}}c^{}_{-\vec{k}-s},
\end{align}
where the Bogoliubov coefficients are given by $u_{\vec{k}}=\frac{1}{\sqrt{2}}(1+\epsilon_{\vec{k}}/\sqrt{\Delta^2+\epsilon^2_{\vec{k}}})^{1/2}$ and $v_{\vec{k}}=\frac{1}{\sqrt{2}}(1-\epsilon_{\vec{k}}/\sqrt{\Delta^2+\epsilon^2_{\vec{k}}})^{1/2}$. We here also show the Bogoliubov transformation
\begin{align}
    c_{\vec{k}s}=u_{\vec{k}} \gamma_{\vec{k}s+}-sv^{*}_{\vec{k}} \gamma^{}_{\vec{k}s-},
\end{align}
\begin{align}
    c^{}_{-\vec{k}-s}=u_{\vec{k}} \gamma^{\dagger}_{\vec{k}s-}+sv^{*}_{\vec{k}} \gamma^{\dagger}_{\vec{k}s+}.
\end{align}
It is important to note that the Bogoliubov quasiparticle operators are not independent and satisfy a conjugate relation 
\begin{align} \label{mutpfvvldl}
    \gamma^{\dagger}_{\vec{k}s\eta}&=  \gamma^{}_{-\vec{k}-s-\eta}.
\end{align}
Each spinful Bogoliubov quasiparticle always has its conjugate alternative with opposite spin and energy. Therefore, we can have $4^{N}$ combinations of complete and orthogonal quasiparticle bases, where $N$ describes the size of the normal metal or the number of kinetic momenta. In principle, we can choose any OBSs, such as spin-momentum OBSs $(\gamma^{}_{\vec{k}\uparrow+},\gamma^{}_{\vec{k}\downarrow+})$ and $(\gamma^{}_{\vec{k}\uparrow-},\gamma^{}_{\vec{k}\downarrow-})$,  or Nambu-momentum OBSs $(\gamma^{}_{\vec{k}\uparrow+},\gamma^{}_{\vec{k}\uparrow-})$ and $(\gamma^{}_{\vec{k}\downarrow+},\gamma^{}_{\vec{k}\downarrow-})$ for $\vec{k}$ in the entire Brillouin zone. 
Furthermore, by choosing the OBS $(\gamma^{}_{\vec{k}\uparrow+},\gamma^{}_{\vec{k}\downarrow-},\gamma^{}_{\vec{k}\downarrow+},\gamma^{}_{\vec{k}\downarrow-})$ for $\vec{k}$ in the half Brillouin zone, the semiconductor analogy of the superconductor produces dramatic features in various spectroscopies of the superconducting condensate \cite{coleman2015introduction}.

\subsubsection{Spin-momentum OBSs} \label{SMOBS}
STEP I): Let us first study the two spin-momentum OBSs conjugate with each other $(\gamma^{}_{\vec{k}\uparrow+},\gamma^{}_{\vec{k}\downarrow+})$  and $(\gamma^{}_{\vec{k}\uparrow-},\gamma^{}_{\vec{k}\downarrow-})$ for $\vec{k}$ in the whole Brillouin zone, whose elements  satisfy the anticommutation relation $\{\gamma^{}_{\vec{k}s\eta},\gamma^{}_{\vec{k}'s'\eta}\}=\delta_{\vec{k},\vec{k}'}\delta_{s,s'}$. We can derive the corresponding Hamiltonian from the BdG Hamiltonian \eqref{mgfeegblrgk} by transforming the other quasiparticles into the chosen OBSs using the conjugate relation \eqref{mutpfvvldl}
\begin{align} \label{dfvfgldv}
    H_L&=\sum_{\vec{k}s}E_{\vec{k}s+}\gamma^{\dagger}_{\vec{k}s+}\gamma^{}_{\vec{k}s+}+\mathcal{E}^{L}_{+}=\sum_{\vec{k}s}E_{\vec{k}s-}\gamma^{\dagger}_{\vec{k}s-}\gamma^{}_{\vec{k}s-}+\mathcal{E}^{L}_{-}.
\end{align}
The energy constants $\mathcal{E}^{L}_{+}$ and $\mathcal{E}^{L}_{-}$ arise from the mathematical transformation of the BdG Hamiltonian \eqref{mgfeegblrgk} into the Hamiltonian \eqref{dfvfgldv} using the OBSs $(\gamma_{\vec{k}\uparrow+},\gamma_{\vec{k}\downarrow+})$ and $(\gamma_{\vec{k}\uparrow-},\gamma_{\vec{k}\downarrow-})$
\begin{align} \label{fdamkvmdkf}
    \mathcal{E}^{L}_{+}=\mathcal{E}+\frac{1}{2}\sum_{\vec{k}s}E_{\vec{k}s-},
\end{align}
\begin{align} \label{amvkgdf}
    \mathcal{E}^{L}_{-}=\mathcal{E}+\frac{1}{2}\sum_{\vec{k}s}E_{\vec{k}s+}.
\end{align}

STEP II): The EVSs of the spin-momentum OBSs $(\gamma^{}_{\vec{k}\uparrow+},\gamma^{}_{\vec{k}\downarrow+})$ and $(\gamma^{}_{\vec{k}\uparrow-},\gamma^{}_{\vec{k}\downarrow-})$ are, respectively, defined by  
\begin{align}
    \gamma^{}_{\vec{k}s+}\vert V\rangle_{+}=0, \text{ for all $\vec{k}$ and $s$},
\end{align}
\begin{align}
    \gamma^{}_{\vec{k}s-}\vert V\rangle_{-}=0, \text{ for all $\vec{k}$ and $s$}.
\end{align}
Solving these equations, we obtain the corresponding EVSs
\begin{align} \label{pmfvdkdmfk1}
     \vert V\rangle_+=\prod_{\vec{k}}\left(u_{\vec{k}}-v^{*}_{\vec{k}} c_{\vec{k} \uparrow}^{\dagger} c_{-\vec{k} \downarrow}^{\dagger}\right)|0\rangle,
\end{align}
\begin{align} \label{pmfvdkdmfk2}
   \vert V\rangle_-=\prod_{\vec{k}}\left(v^{}_{\vec{k}}+u_{\vec{k}} c_{\vec{k} \uparrow}^{\dagger} c_{-\vec{k} \downarrow}^{\dagger}\right)|0\rangle.
\end{align}
These states exhibit entanglement between electrons with energies $E_{\vec{k} \uparrow}$ and $E_{-\vec{k} \downarrow}$, which form Cooper pairs.

STEP III): Let us calculate the superconducting ground state in term of the spin-momentum OBS  $(\gamma^{}_{\vec{k}\uparrow+},\gamma^{}_{\vec{k}\downarrow+})$. The superconductor has $4^{N}$ eigenstates. These eigenstates include $C(2N,0)$ EVS $\vert V \rangle_+$, $C(2N,1)$ one-Bogoliubon states $\gamma^{\dagger}_{\vec{k}_1s_1+}\vert V \rangle_+$, $C(2N,2)$ two-Bogoliubon states $\gamma^{\dagger}_{\vec{k}_1s_1+}\gamma^{\dagger}_{\vec{k}_2s_2+}\vert V \rangle_+$, and so on, up to  $C(2N,2N)$ $2N$-Bogoliubon states $(\prod_{\vec{k}s}\gamma^{\dagger}_{\vec{k}s+})\vert V \rangle_+$. The corresponding energies of these states are $\mathcal{E}^{L}_{+}$, $\mathcal{E}^{L}_{+}+E_{\vec{k}_1s_1+}$, $\mathcal{E}^{L}_{+}+E_{\vec{k}_1s_1+}+E_{\vec{k}_2s_2+}$, and so on, up to $\mathcal{E}^{L}_{+}+\sum_{\vec{k}s}E_{\vec{k}s+}$,  respectively, where $\vec{k}_i,s_i$ are required to satisfy the Pauli exclusion principle  for many Bogoliubov quasiparticles. The ground state is the state with the lowest energy among these $4^{N}$ states. In this case, the ground state is the Fermi sea, which corresponds to filling all the negative-energy Bogoliubov quasiparticles in the OBS $(\gamma_{\vec{k}\uparrow+},\gamma_{\vec{k}\downarrow+})$ starting from the EVS $\vert V \rangle_+$. Similarly, for the OBS $(\gamma_{\vec{k}\uparrow-},\gamma_{\vec{k}\downarrow-})$, the ground state is given by filling all the negative-energy Bogoliubov quasiparticles in the OBS $(\gamma_{\vec{k}\uparrow-},\gamma_{\vec{k}\downarrow-})$ starting from the EVS $\vert V \rangle_-$. Therefore, the superconducting ground state can be written as
\begin{align} \label{fvklal}
    \vert G \rangle=\left(\prod_{E^{}_{\vec{k}s+}<0}\gamma^{\dagger}_{\vec{k}s+}\right)\vert V\rangle_+=\left(\prod_{E^{}_{\vec{k}s-}<0}\gamma^{\dagger}_{\vec{k}s-}\right)\vert V\rangle_-,
\end{align}
with energy 
\begin{align} \label{fdvfmkdfmv}
\mathcal{E}_G=\mathcal{E}^{L}_{+}+\sum_{E^{}_{\vec{k}s+}<0}E^{}_{\vec{k}s+}=\mathcal{E}^{L}_{-}+\sum_{E^{}_{\vec{k}s-}<0}E^{}_{\vec{k}s-}.
\end{align}
When $h_L<\Delta$, we have $E_{\vec{k}s+}>0$ and $E_{\vec{k}s-}<0$. In this case, the ground state can be written as the EVS $\vert V \rangle_+$ with energy $\mathcal{E}_G = \mathcal{E}^{L}_{+} = \mathcal{E}+\frac{1}{2}\sum_{\vec{k}s}E_{\vec{k}s-}$.

PROPERTY I): The relation between the EVSs $\vert V\rangle_+$ and $\vert V\rangle_-$ can be derived as follows. For $h_L = 0$, the ground state can be expressed in terms of two spin-momentum OBSs as  $\vert G
\rangle=\vert V\rangle_{+}=\prod_{ \vec{k}} \gamma_{ \vec{k} \uparrow-\eta}^{\dagger}\gamma_{-\vec{k} \downarrow-\eta}^{\dagger} \vert V\rangle_{-}$.  Again, the EVSs $\vert V\rangle_{+}$ and $\vert V\rangle_{-}$ form a conjugate loop
\begin{align} \label{pmvfnjk1}
\vert V\rangle_{\eta}=\prod_{ \vec{k}} \gamma_{ \vec{k} \uparrow-\eta}^{\dagger}\gamma_{-\vec{k} \downarrow-\eta}^{\dagger} \vert V\rangle_{-\eta},
\end{align}
where $\eta=-$ of Eq. \eqref{pmvfnjk1} is derived from the $\eta=+$ of Eq. \eqref{pmvfnjk1} via the conjugate relation \eqref{mutpfvvldl}.

\subsubsection{Nambu-momentum OBSs} \label{NMOBS}

STEP I): Then, we study the Nambu-momentum OBSs conjugate with each other, denoted as $(\gamma^{}_{\vec{k}\uparrow+},\gamma^{}_{\vec{k}\uparrow-})$  and $(\gamma^{}_{\vec{k}\downarrow+},\gamma^{}_{\vec{k}\downarrow-})$ for $\vec{k}$ in the entire Brillouin zone, whose elements anticommutate with each other, i.e., $\{\gamma^{}_{\vec{k}s\eta},\gamma^{}_{\vec{k}'s\eta'}\}=\delta_{\vec{k},\vec{k}'}\delta_{\eta,\eta'}$.  The corresponding Hamiltonian can be derived from the BdG Hamiltonian $\eqref{mgfeegblrgk}$ by transforming the remaining quasiparticles into the chosen Nambu-momentum OBSs using the conjugate relation $\eqref{mutpfvvldl}$ 
\begin{align} \label{gbgklbml}
    H_L&=\sum_{\vec{k}\eta}E_{\vec{k}\uparrow\eta}\gamma^{\dagger}_{\vec{k}\uparrow\eta}\gamma^{}_{\vec{k}\uparrow\eta}+\mathcal{E}^{L}_{\uparrow}=\sum_{\vec{k}\eta}E_{\vec{k}\downarrow\eta}\gamma^{\dagger}_{\vec{k}\downarrow\eta}\gamma^{}_{\vec{k}\downarrow\eta}+\mathcal{E}^{L}_{\downarrow}, 
\end{align}
where $\mathcal{E}^{L}_{\uparrow}$ and $\mathcal{E}^{L}_{\downarrow}$ are energy constants resulting from the mathematical transformation of the BdG Hamiltonian $\eqref{mgfeegblrgk}$ into the form $\eqref{gbgklbml}$ using the Nambu-momentum operators $(\gamma_{\vec{k}\uparrow+},\gamma_{\vec{k}\uparrow-})$ and $(\gamma_{\vec{k}\downarrow+},\gamma_{\vec{k}\downarrow-})$
\begin{align} \label{erifdamkvmdkf}
    \mathcal{E}^{L}_{\uparrow}=\mathcal{E}+\frac{1}{2}\sum_{\vec{k}\eta}E_{\vec{k}\downarrow\eta},
\end{align}
\begin{align} \label{eriamvkgdf}
    \mathcal{E}^{L}_{\downarrow}=\mathcal{E}+\frac{1}{2}\sum_{\vec{k}\eta}E_{\vec{k}\uparrow\eta}.
\end{align}

STEP II): Next, we consider the EVSs of the Nambu-momentum OBSs $(\gamma^{}_{\vec{k}\uparrow+},\gamma^{}_{\vec{k}\uparrow-})$ and $(\gamma^{}_{\vec{k}\downarrow+},\gamma^{}_{\vec{k}\downarrow-})$, respectively, defined by  
\begin{align}
    \gamma^{}_{\vec{k}\uparrow \eta}\vert V\rangle_{\uparrow}=0, \text{ for all $\vec{k}$ and $\eta$},
\end{align}
\begin{align}
    \gamma^{}_{\vec{k}\downarrow \eta}\vert V\rangle_{\downarrow}=0, \text{ for all $\vec{k}$ and $\eta$}.
\end{align}
We can then obtain the corresponding EVSs \cite{datta1996scattering,datta1999can}
\begin{align} \label{rpmfvdkdmfk1}
     \vert V\rangle_{\uparrow}=\prod_{\vec{k}}c_{\vec{k} \downarrow}^{\dagger} |0\rangle,
\end{align}
\begin{align} \label{rpmfvdkdmfk2}
   \vert V\rangle_{\downarrow}=\prod_{\vec{k}}c_{\vec{k} \uparrow}^{\dagger}|0\rangle.
\end{align}
The EVS $\vert V\rangle_{\uparrow}$ ($\vert V\rangle_{\downarrow}$) contains all spin down (up) electrons.

STEP III): Let us express the superconducting ground state using the Nambu-momentum OBSs -- $(\gamma^{}_{\vec{k}\uparrow+},\gamma^{}_{\vec{k}\uparrow-})$ for all $\vec{k}$. In a superconducting system, there are $4^{N}$ eigenstates: $C(2N,0)$ EVS $\vert V \rangle_{\uparrow}$, $C(2N,1)$ one-Bogoliubon states $\gamma^{\dagger}_{\vec{k}_1\uparrow\eta_1}\vert V \rangle_{\uparrow}$, $C(2N,2)$ two-Bogoliubon states $\gamma^{\dagger}_{\vec{k}_1\uparrow\eta_1}\gamma^{\dagger}_{\vec{k}_2\uparrow\eta_2}\vert V \rangle_{\uparrow}$, and so on, up to $C(2N,2N)$ $2N$-Bogoliubon state $(\prod_{\vec{k}\eta}\gamma^{\dagger}_{\vec{k}\uparrow\eta})\vert V \rangle_{\uparrow}$. The corresponding energies are $\mathcal{E}^{L}_{\uparrow}$, $\mathcal{E}^{L}_{\uparrow}+E_{\vec{k}_1\uparrow\eta_1}$, $\mathcal{E}^{L}_{\uparrow}+E_{\vec{k}_1\uparrow\eta_1}+E_{\vec{k}_2\uparrow\eta_2}$, and so on, up to  $\mathcal{E}^{L}_{\uparrow}+\sum_{\vec{k}\eta}E_{\vec{k}\uparrow\eta}$,  respectively. Here, $\vec{k}_i,\eta_i$ are required to satisfy the Pauli exclusion principle  for many Bogoliubov quasiparticles. The ground state is the state with the lowest energy among these $4^{N}$ eigenstates. It corresponds to filling all negative-energy Bogoliubov quasiparticles of the Nambu-momentum OBS $(\gamma^{}_{\vec{k}\uparrow+},\gamma^{}_{\vec{k}\uparrow-})$ into the EVS $\vert V\rangle_{\uparrow}$. The same applies to the Nambu-momentum OBS $(\gamma^{}_{\vec{k}\downarrow+},\gamma^{}_{\vec{k}\downarrow-})$ for all $\vec{k}$, using the EVS \eqref{rpmfvdkdmfk2}. Thus, the superconducting ground state is given by
\begin{align} \label{fksdlfd}
    \vert G\rangle =\left(\prod_{E^{}_{\vec{k}\uparrow\eta}<0}\gamma^{\dagger}_{\vec{k}\uparrow\eta}\right)\vert V\rangle_{\uparrow}=\left(\prod_{E^{}_{\vec{k}\downarrow\eta}<0}\gamma^{\dagger}_{\vec{k}\downarrow\eta}\right)\vert V\rangle_{\downarrow},
\end{align}
with energy 
\begin{align} \label{dfvfdvfmkdfmv}
\mathcal{E}_G=\mathcal{E}^{L}_{\uparrow}+\sum_{E^{}_{\vec{k}\uparrow\eta}<0}E^{}_{\vec{k}\uparrow\eta}=\mathcal{E}^{L}_{\downarrow}+\sum_{E^{}_{\vec{k}\downarrow\eta}<0}E^{}_{\vec{k}\downarrow\eta}.
\end{align}
When $h_L<\Delta$, we have $E^{}_{\vec{k}s+}>0$ and $E^{}_{\vec{k}s-}<0$. By substituting Eqs. \eqref{rpmfvdkdmfk1} and \eqref{rpmfvdkdmfk2}, the ground state \eqref{fksdlfd} can be written as
$\vert G\rangle =\left(\prod_{\vec{k}}\gamma^{\dagger}_{\vec{k}\uparrow-}\right)\prod_{\vec{k}}c_{\vec{k} \downarrow}^{\dagger} |0\rangle=\left(\prod_{\vec{k}}\gamma^{\dagger}_{\vec{k}\downarrow-}\right)\prod_{\vec{k}}c_{\vec{k} \uparrow}^{\dagger} |0\rangle$, which is equivalent to Eq. \eqref{fvklal} after substituting the Bogoliubov quasiparticle operator \eqref{fdvmdfvdkmminu}.

PROPERTY I): To establish the relation between  EVSs \eqref{rpmfvdkdmfk1}  and \eqref{rpmfvdkdmfk2}, we consider the case when $h_L=0$. In this scenario, the ground state can be expressed in terms of the Nambu-momentum OBSs $\vert G
\rangle=\prod_{\vec{k}}\gamma^{\dagger}_{\vec{k}\uparrow-}\vert V\rangle_{\uparrow}=\prod_{\vec{k}}\gamma^{\dagger}_{\vec{k}\downarrow-}\vert V\rangle_{\downarrow}$. Using the conjugate relation \eqref{mutpfvvldl}, we find that the eigenstates $\vert V\rangle_{\uparrow}$ and $\vert V\rangle_{\downarrow}$ also form a conjugate loop:
\begin{align} \label{fdkfmvdk}
    \vert V\rangle_{s}=\prod_{ \vec{k}}\gamma^{\dagger}_{\vec{k}-s+}\gamma^{\dagger}_{\vec{k}-s-}\vert V\rangle_{-s}.
\end{align}
Here, the EVS $\vert V\rangle_{s}$ contains all the Bogoliubov quasiparticles with energies $E_{\vec{k}-s\eta}$. 

\subsubsection{Semiconductor analogy} \label{SAOBS}

STEP I): Next, the semiconductor analogy is revisited, considering a complete basis consisting of two spins and two Nambu operators for the half Brillouin zone.  Hereafter, we study two semiconductor-analogy OBSs conjugate with each other, i.e., $k_x>0$ OBS $(\gamma^{}_{\vec{k}\uparrow+},\gamma^{}_{\vec{k}\uparrow-},\gamma^{}_{\vec{k}\downarrow+},\gamma^{}_{\vec{k}\downarrow-})$ for $\vec{k}$ in the $k_x>0$ half Brillouin zone  and $k_x<0$ OBS $(\gamma^{}_{\vec{k}\uparrow+},\gamma^{}_{\vec{k}\uparrow-},\gamma^{}_{\vec{k}\downarrow+},\gamma^{}_{\vec{k}\downarrow-})$ for $\vec{k}$ in the $k_x<0$ half Brillouin zone. These quasiparticle operators satisfy the anticommutation relation $\{\gamma^{}_{\vec{k}s\eta},\gamma^{}_{\vec{k}'s'\eta'}\}=\delta_{\vec{k},\vec{k}'}\delta_{\eta,\eta'}\delta_{s,s'}$ when $\vec{k}$ is forced to be half Brillouin zone. The corresponding Hamiltonian can be derived from the BdG Hamiltonian \eqref{mgfeegblrgk} by transforming the remaining quasiparticles into the chosen OBSs using the conjugate relation \eqref{mutpfvvldl}. The Hamiltonian is given by
\begin{align} \label{pgbgklbml}
    H_L&=\sum_{k_x>0}\sum_{s\eta}E_{\vec{k}\uparrow\eta}\gamma^{\dagger}_{\vec{k}\uparrow\eta}\gamma^{}_{\vec{k}\uparrow\eta}+\mathcal{E}^{L}_{>}\\
    &=\sum_{k_x<0}\sum_{s\eta}E_{\vec{k}s\eta}\gamma^{\dagger}_{\vec{k}s\eta}\gamma^{}_{\vec{k}\downarrow\eta}+\mathcal{E}^{L}_{<}, \notag 
\end{align}
where the energy constants $\mathcal{E}^{L}_{>}$ and $\mathcal{E}^{L}_{<}$ arise from the mathematical transformation of the BdG Hamiltonian \eqref{mgfeegblrgk} into the Hamiltonian \eqref{pgbgklbml} using the chosen OBSs. The specific expressions for $\mathcal{E}^{L}_{>}$ and $\mathcal{E}^{L}_{<}$ are given as
\begin{align} \label{perifdamkvmdkf}
    \mathcal{E}^{L}_{>}=\mathcal{E}+\frac{1}{2}\sum_{k_x<0}\sum_{s\eta}E_{\vec{k}\downarrow\eta},
\end{align}
\begin{align} \label{periamvkgdf}
    \mathcal{E}^{L}_{<}=\mathcal{E}+\frac{1}{2}\sum_{k_x>0}\sum_{s\eta}E_{\vec{k}\uparrow\eta}.
\end{align}

STEP II): The EVSs of the two semiconductor-analogy OBSs, i.e., $k_x>0$ OBS $(\gamma^{}_{\vec{k}\uparrow+},\gamma^{}_{\vec{k}\uparrow-},\gamma^{}_{\vec{k}\downarrow+},\gamma^{}_{\vec{k}\downarrow-})$ for $\vec{k}$ in the $k_x>0$ half Brillouin zone  and $k_x>0$ OBS$(\gamma^{}_{\vec{k}\uparrow+},\gamma^{}_{\vec{k}\uparrow-},\gamma^{}_{\vec{k}\downarrow+},\gamma^{}_{\vec{k}\downarrow-})$ for $\vec{k}$ in the $k_x<0$ half Brillouin zone, are, respectively, defined by
\begin{align}
    \gamma^{}_{\vec{k}s\eta}\vert V\rangle_{>}=0, \text{ for $k_x>0$ and all $s$, $\eta$},
\end{align}
\begin{align}
    \gamma^{}_{\vec{k}s\eta}\vert V\rangle_{<}=0  \text{ for $k_x<0$ and all $s$, $\eta$}.
\end{align}
Then, we obtain the corresponding EVSs
\begin{align} \label{fdmkk1}
    \vert V\rangle_{>}=\prod_{ k_x<0}c^{\dagger}_{\vec{k}\uparrow}c^{\dagger}_{\vec{k}\downarrow}\vert 0\rangle,
\end{align}
\begin{align} \label{fdmkk2}
    \vert V\rangle_{<}=\prod_{ k_x>0}c^{\dagger}_{\vec{k}\uparrow}c^{\dagger}_{\vec{k}\downarrow}\vert 0\rangle.
\end{align}
The EVS $\vert V\rangle_{>}$ ($\vert V\rangle_{<}$) contains all $k_x<0$ ($k_x>0$) electrons.

STEP III): Let us first work on the semiconductor-analogy OBS  $(\gamma^{}_{\vec{k}\uparrow+},\gamma^{}_{\vec{k}\uparrow-},\gamma^{}_{\vec{k}\downarrow+},\gamma^{}_{\vec{k}\downarrow-})$ for $\vec{k}$ in the $k_x>0$ half Brillouin zone. The system has a total of  $4^{N}$  eigenstates. These eigenstates include $C(2N,0)$ EVS $\vert V \rangle_{>}$, $C(2N,1)$ one-Bogoliubon states $\gamma^{\dagger}_{\vec{k}_1s_1\eta_1}\vert V \rangle_{>}$, $C(2N,2)$ two-Bogoliubon states $\gamma^{\dagger}_{\vec{k}_1s_1\eta_1}\gamma^{\dagger}_{\vec{k}_2s_2\eta_2}\vert V \rangle_{>}$, and so on, up to $C(2N,2N)$ $2N$-Bogoliubon states $(\prod_{k_x>0}\prod_{s\eta}\gamma^{\dagger}_{\vec{k}s\eta})\vert V \rangle_{>}$. The corresponding energies of these states are given by $\mathcal{E}^{L}_{>}$, $\mathcal{E}^{L}_{>}+E_{\vec{k}_1s_1\eta_1}$, $\mathcal{E}^{L}_{>}+E_{\vec{k}_1s_1\eta_1}+E_{\vec{k}_2s_2\eta_2}$, and so on, up to  $\mathcal{E}^{L}_{>}+\sum_{k_x>0}\sum_{s\eta}E_{\vec{k}s\eta}$,  respectively, where $\vec{k}_i,s_i,\eta_i$ are required to satisfy the Pauli exclusion principle  for many Bogoliubov quasiparticles. The ground state corresponds to the lowest energy state among these $4^N$ states, and it represents a filled sea of negative Bogoliubov quasiparticles in this OBS. Similarly, for the OBS $(\gamma^{}_{\vec{k}\uparrow+},\gamma^{}_{\vec{k}\uparrow-},\gamma^{}_{\vec{k}\downarrow+},\gamma^{}_{\vec{k}\downarrow-})$ for $\vec{k}$ in the $k_x<0$ half Brillouin zone, we have a similar set of eigenstates and energies. The superconducting ground state is given by 
\begin{align} \label{dgfoog}
    \vert G\rangle_{}&=\left(\prod^{k_x>0}_{E^{}_{\vec{k}s\eta}<0}\gamma^{\dagger}_{\vec{k}s\eta}\right)\vert V\rangle_{>}=\left(\prod^{k_x<0}_{E^{}_{\vec{k}s\eta}<0}\gamma^{\dagger}_{\vec{k}s\eta}\right)\vert V\rangle_{<},
\end{align}
with energy 
\begin{align} \label{qdfvfdvfmkdfmv}
\mathcal{E}_G=\mathcal{E}^{L}_{>}+\sum^{k_x>0}_{E^{}_{\vec{k}s\eta}<0}E^{}_{\vec{k}s\eta}=\mathcal{E}^{L}_{<}+\sum^{k_x<0}_{E^{}_{\vec{k}s\eta}<0}E^{}_{\vec{k}s\eta}.
\end{align}
When $h_L<\Delta$, we have $E^{}_{\vec{k}s+}>0$ and $E^{}_{\vec{k}s-}<0$. By substituting the EVSs \eqref{fdmkk1} and \eqref{fdmkk2}, the ground state \eqref{dgfoog} can be represented as
$\vert G\rangle =\left(\prod^{k_x>0}_{E^{}_{\vec{k}s\eta}<0}\gamma^{\dagger}_{\vec{k}s\eta}\right)\left(\prod_{ k'_x<0}c^{\dagger}_{\vec{k}'\uparrow}c^{\dagger}_{\vec{k}'\downarrow}\right)\vert 0\rangle=\left(\prod^{k_x<0}_{E^{}_{\vec{k}s\eta}<0}\gamma^{\dagger}_{\vec{k}s\eta}\right)\left(\prod_{ k'_x>0}c^{\dagger}_{\vec{k}'\uparrow}c^{\dagger}_{\vec{k}'\downarrow}\right)\vert 0\rangle$, which is the same as Eq. \eqref{fvklal} after the substitution of the Bogoliubov quasiparticle operators \eqref{fdvmdfvdkmplusq} and \eqref{fdvmdfvdkmminuq}.

PROPERTY I): Then, we try to find the relation between EVSs \eqref{fdmkk1} and \eqref{fdmkk2}. For $h_L<\Delta$, the ground state can be expressed in terms of two semiconductor-analogy OBSs $ \vert G\rangle_{}=\left(\prod_{k_x>0}\gamma^{\dagger}_{\vec{k}s-}\right)\vert V\rangle_{>}=\left(\prod_{k_x<0}\gamma^{\dagger}_{\vec{k}s-}\right)\vert V\rangle_{<}$. By means of the conjugate relation \eqref{mutpfvvldl}, we find EVSs $\vert V\rangle_{>}$ and $\vert V\rangle_{<}$ also form a conjugate loop
\begin{align} 
    \vert V\rangle_{>}&=\prod_{ k_x<0}\gamma^{\dagger}_{\vec{k}\uparrow+}\gamma^{\dagger}_{\vec{k}\uparrow-}\gamma^{\dagger}_{\vec{k}\downarrow+}\gamma^{\dagger}_{\vec{k}\downarrow-}\vert V\rangle_{<},\label{dkfmvdk1}\\
    \vert V\rangle_{<}&=\prod_{ k_x>0}\gamma^{\dagger}_{\vec{k}\uparrow+}\gamma^{\dagger}_{\vec{k}\uparrow-}\gamma^{\dagger}_{\vec{k}\downarrow+}\gamma^{\dagger}_{\vec{k}\downarrow-}\vert V\rangle_{>}.\label{dkfmvdk2}
\end{align}

\section{Derivation of effective vacuum states} \label{effectivevacuumstates}

In this section, we derive the EVSs (or Fermi sea and sky), the Eq. \eqref{trufvdfmmvl} in the main text. 

For briefness, we combine $l$ ($n$) and $s$ as $a$ ($b$). Then, the quasiparticles [Eq. \eqref{fvvldl}] can be rewritten into 
\begin{align} \label{rtfvvldl}
    \gamma_{a\eta}&= \sum_{b}\left[(u_{\eta})_{ab}c^{}_{b}+(v_{\eta})_{ab}c^{\dagger}_{b}\right],
\end{align}
with
\begin{align} \label{fdvnjdfku}
    u_{\eta}=\begin{bmatrix}
      (u_{\eta})^{\Uparrow\uparrow} & (u_{\eta})^{\Uparrow\downarrow} \\
      (u_{\eta})^{\Downarrow\uparrow} &  (u_{\eta})^{\Downarrow\downarrow}
    \end{bmatrix},
\end{align}
\begin{align} \label{fdvnjdfkv}
    v_{\eta}=\begin{bmatrix}
      (v_{\eta})^{\Uparrow\downarrow} &-(v_{\eta})^{\Uparrow\uparrow} \\
       (v_{\eta})^{\Downarrow\downarrow} & -(v_{\eta})^{\Downarrow\uparrow}
    \end{bmatrix}.
\end{align}
Here, we assume the following ansatz for the Fermi sea and sky 
\begin{align} 
    \vert V\rangle_{\eta}=\frac{1}{N_\eta}e^{\sum_{bb'}\mathcal{Q}^{\eta}_{bb'}c^{\dagger}_{b}c^{\dagger}_{b'}}\vert 0\rangle.
\end{align}
It can be rewritten into 
\begin{align} \label{ufvkdklv}
    \vert V\rangle_{\eta}=\frac{1}{N^{1/2}_{\eta}}\sum_{n=0}^{\infty}\frac{1}{n!}\theta^n\vert 0\rangle,
\end{align}
with
\begin{align}
    \theta= \sum_{qq'}\mathcal{Q}^{\eta}_{qq'}c^{\dagger}_{q}c^{\dagger}_{q'}.
\end{align}
where $N_{\eta}$ is the renormalization coefficients to be given latter. 
We define the Bogoliubov vacuum states  as 
\begin{align}
    \gamma_{a\eta}\vert V\rangle_{\eta}=0, \text{for all a},
\end{align}
i.e.,
\begin{align} \label{rfvdkkkd}
   -\sum_{b}\mathcal{D}^{\eta}_{h,ab}c^{\dagger}_{b}\vert V\rangle_{\eta} =\sum_{b}\mathcal{D}^{\eta}_{e,ab}c^{}_{b}\vert V\rangle_{\eta} . 
\end{align}
The left- and right-hand sides of Eq. \eqref{rfvdkkkd} contain odd number of electrons. The term of $2n+1$ electrons comes from  $-\sum_{b}\mathcal{D}^{\eta}_{h,ab}c^{\dagger}_{b}$ working on the $n$th term and  $\sum_{b}\mathcal{D}^{\eta}_{e,ab}c^{}_{b}$ operating on the $(n+1)$th term. Then, Eq. \eqref{rfvdkkkd} becomes 
\begin{align} \label{rfdakvk}
     -\sum_{b}\mathcal{D}^{\eta}_{h,ab}c^{\dagger}_{b}\frac{1}{n!}\theta^n\vert 0\rangle=\sum_{b}\mathcal{D}^{\eta}_{e,ab}c^{}_{b}\frac{1}{(n+1)!}\theta^{n+1}\vert 0\rangle.
\end{align}

Let us first focus on the terms contain one electrons, i.e., $c^{\dagger}_{b}\vert 0\rangle$, which can be generated by $-\sum_{b}\mathcal{D}^{\eta}_{h,ab}c^{\dagger}_{b}$ working on the $n=0$ term of EVS \eqref{ufvkdklv} and $\sum_{b}\mathcal{D}^{\eta}_{e,ab}c^{}_{b}$ working on the $n=1$ term of EVS \eqref{ufvkdklv}. Thus, the superconducting cloud matrix, $\mathcal{Q}^{\eta}_{qq'}$ satisfy pair equation 
\begin{align}
    -\sum_{b}\mathcal{D}^{\eta}_{h,ab}c^{\dagger}_{b}\vert 0\rangle=\sum_{bqq'}\mathcal{D}^{\eta}_{e,ab}c^{}_{b}\mathcal{Q}^{\eta}_{qq'}c^{\dagger}_{q}c^{\dagger}_{q'}\vert 0\rangle,
\end{align}
i.e.,
\begin{align} \label{fvdkvmke}
    -\mathcal{D}^{\eta}_{h,ab}&=\sum_{c}\mathcal{D}^{\eta}_{e,ac}\mathcal{P}^{\eta}_{cb},
\end{align}
with
\begin{align}
     \mathcal{C}^{\eta}_{cb}=\mathcal{Q}^{\eta}_{cb}-\mathcal{Q}^{\eta}_{bc}.
\end{align}
Then, we reach
\begin{align} \label{fvdmdkfkv}
    \mathcal{C}^{\eta}=-u^{-1}_{\eta}v^{}_{\eta}. 
\end{align}
Knowing that 
\begin{align}
    \theta &= \frac{1}{2}\sum_{qq'} \left(\mathcal{Q}^{\eta}_{qq'}c^{\dagger}_{q}c^{\dagger}_{q'}+\mathcal{Q}^{\eta}_{q'q}c^{\dagger}_{q'}c^{\dagger}_{q}\right)\\
    &=\frac{1}{2}\sum_{qq'} (\mathcal{Q}^{\eta}_{qq'}-\mathcal{Q}^{\eta}_{q'q})c^{\dagger}_{q}c^{\dagger}_{q'},\notag 
\end{align}
Thus, we have 
\begin{align} \label{fdvmkfk}
    \theta= \frac{1}{2}\sum_{qq'}\mathcal{C}^{\eta}_{qq'}c^{\dagger}_{q}c^{\dagger}_{q'}.
\end{align}

So far we have shown that Eq. \eqref{fvdmdkfkv} solves Eq. \eqref{rfdakvk} for $n=0$. Next we show by mathematical induction that Eq. \eqref{fvdmdkfkv} solves Eq. \eqref{rfdakvk} also for all $n>0$. Then, we derive the $(n+1)$th term from the $n$th term of Eq. \eqref{rfdakvk}. First, we derive a useful relation
\begin{align} \label{rfdamfv}
    \sum_{b}\mathcal{D}^{\eta}_{e,ab}c^{}_{b}\theta &=\sum_{bqq'}\mathcal{D}^{\eta}_{e,ab}\mathcal{Q}^{\eta}_{qq'}c^{}_{b}c^{\dagger}_{q}c^{\dagger}_{q'}\\
    &= -\sum_{b}\mathcal{D}^{\eta}_{h,ab}c^{\dagger}_b+\sum_{b}\mathcal{D}^{\eta}_{e,ab}\theta c^{}_{b}, \notag
\end{align}
where we have used Eq. \eqref{fvdkvmke}. By means of Eq. \eqref{rfdamfv},
the $(n+1)$th term of the right-hand side of Eq. \eqref{rfdakvk} can be divided into two parts
\begin{align} \label{rfdmavl}
    R_{n+1}&=-\frac{1}{n+2}\sum_{b}\mathcal{D}^{\eta}_{h,ab}c^{\dagger}_{b}\frac{1}{(n+1)!}\theta^{n+1}\vert 0\rangle \\
    &+\frac{\theta}{n+2}\sum_{k}\mathcal{D}^{\eta}_{e,ab} c^{}_{b}\frac{1}{(n+1)!}\theta^{n+1}\vert 0\rangle. \notag 
\end{align}
By substitution of Eq. \eqref{rfdakvk}, Eq.  \eqref{rfdmavl} becomes
\begin{align} \label{rifdmavl}
    R_{n+1}&=-\sum_{b}\mathcal{D}^{\eta}_{h,ab}c^{\dagger}_{b}\frac{1}{(n+1)!}\theta^{n+1}\vert 0\rangle,
\end{align}
which is the $(n+1)$th term of the left-hand side of Eq. \eqref{rfdakvk}. Therefore, the superconducting cloud matrix \eqref{fvdmdkfkv} is our solution for EVSs.

In the presence of spin-flip,  Eq. \eqref{fvdmdkfkv} becomes
\begin{align} 
    \mathcal{C}=-\begin{bmatrix}
      (u_{\eta})^{\Uparrow\uparrow} & (u_{\eta})^{\Uparrow\downarrow} \\
      (u_{\eta})^{\Downarrow\uparrow} &  (u_{\eta})^{\Downarrow\downarrow}
    \end{bmatrix}^{-1} \begin{bmatrix}
      (v_{\eta})^{\Uparrow\downarrow} &-(v_{\eta})^{\Uparrow\uparrow} \\
       (v_{\eta})^{\Downarrow\downarrow} & -(v_{\eta})^{\Downarrow\uparrow}
    \end{bmatrix}.
\end{align}
Though $\mathcal{C}^{\eta}$ can be non-Hermitian, we can still diagonalize it by singular value decomposition
\begin{align} \label{yfdvkp}
 \mathcal{C}^{\eta}=U^{}_{\eta}\mathcal{A}^{\eta}V^{+}_{\eta},
\end{align}
where $U_{\eta}$ and $V_{\eta}$ are unitary matrix and $\mathcal{A}^{\eta}$ is diagonal matrix with non-negative real Bogoliubov coefficients $\mathcal{A}^{\eta}_{ll}$.  Note that Bogoliubov coefficients are double degenerate, which is donated by $\sigma$ subscript
\begin{align} \label{ffrdvmkfk}
    \theta= \frac{1}{2}\sum_{l}\tilde{a}^{\dagger}_{L,l\eta}\mathcal{A}_{ll}\tilde{a}^{\dagger}_{R,l\eta}=\frac{1}{2}\sum_{k\sigma}\tilde{a}^{\dagger}_{L,k\sigma\eta}\mathcal{A}_{k}\tilde{a}^{\dagger}_{R,k\sigma\eta},
\end{align}
with
\begin{align} \label{yvdfkvkp}
 \tilde{a}^{\dagger}_{L,k\sigma\eta}=\sum_{n}c^{\dagger}_{ns}(U_{\eta})^{s\sigma }_{nk},
\end{align}
\begin{align} \label{yvtfkvkp}
   \tilde{a}^{\dagger}_{R,k\sigma\eta}=\sum_{k}(V^{+}_{\eta})^{\sigma s}_{kn}c^{\dagger}_{ns}.
\end{align}
($\tilde{a}^{\dagger}_{L,k\sigma\eta},\tilde{a}^{\dagger}_{R,k\sigma\eta}$) for all $k$ and $\sigma$ is an overcomplete basis set and satisfy
\begin{align}
\tilde{a}^{\dagger}_{R,k\sigma\eta}=e^{i\phi^{\eta}_k}\sigma \tilde{a}^{\dagger}_{L,k-\sigma\eta}. 
\end{align}
Then Eq. \eqref{ffrdvmkfk} becomes 
\begin{align} \label{ffrsfdvmkfk}
    \theta=\sum_{k}a^{\dagger}_{k\Uparrow\eta}e^{i\phi_k}\mathcal{A}_{k} a^{\dagger}_{k\Downarrow\eta},
\end{align}
with
\begin{align}
a^{\dagger}_{k\sigma\eta}=\tilde{a}^{\dagger}_{L,k\sigma\eta}.
\end{align}
Thus, the EVS has Cooper-like pair form 
\begin{align} \label{yptrufvdfmmvl}
    \vert V\rangle_{\eta}=\frac{1}{N^{1/2}_\eta}\prod^{N}_{k=1}\left(1+e^{i\phi^{\eta}_k}\mathcal{A}^{\eta}_{k}a^{\dagger}_{k\Uparrow\eta} a^{\dagger}_{k\Downarrow\eta}\right)\vert 0\rangle,
\end{align}
with 
\begin{align}
    N_{\eta}=\prod^{N}_{k=1}(1+\vert\mathcal{A}^{\eta}_{k}\vert^2).
\end{align}
The complete and  orthogonal superconducting pseudospin cloud operators are given by
\begin{align} \label{yvdfkdvkp}
 a^{\dagger}_{k\sigma\eta}=\sum_{n}c^{\dagger}_{ns}(U_{\eta})^{ s\sigma}_{nk}.
\end{align}
Next, we express the quasiparticle operators \eqref{fvvldl} in terms of the superconducting pseudospin clouds  $\{a_{k\sigma\eta}\}$ for all $k$ and $\sigma$ -- an OBS
\begin{align} 
    \gamma_{l\sigma\eta}= \sum^{}_{k\sigma'}\left[(\mathcal{U}_{\eta})^{\sigma\sigma'}_{lk}a^{}_{k\sigma'\eta}+(\mathcal{V}_{\eta})^{\sigma\sigma'}_{lk}a^{\dagger}_{k\sigma'\eta}\right],
\end{align} 
with
\begin{align}
    (\mathcal{U}_{\eta})^{\sigma\sigma'}_{lk}=\sum^{}_{ns}(u_{\eta})^{\sigma s}_{ln}(U^{}_{\eta})^{  s\sigma'}_{nk},
\end{align}
\begin{align}
    (\mathcal{V}_{\eta})^{\sigma\sigma'}_{lk}=\sum^{}_{ns}-s(v_{\eta})^{\sigma s}_{ln}(U^{+}_{\eta})^{ \sigma' -s}_{kn},
\end{align}
where we have used the identities 
\begin{align} 
c^{\dagger}_{n-s}=\sum_{k\sigma'}a^{\dagger}_{k\sigma'\eta}(U^{+}_{\eta})^{ \sigma' -s}_{kn},
\end{align}
\begin{align} 
c^{}_{ns}=\sum_{k\sigma'}(U^{}_{\eta})^{  s\sigma'}_{nk}a^{}_{k\sigma'\eta}.
\end{align}

Hereafter, we consider a simple case without spin-flip, and hence spin becomes a good quantum number and Eqs. \eqref{fdvnjdfku} and \eqref{fdvnjdfkv} reduce to 
\begin{align} 
    u_{\eta}=\begin{bmatrix}
      (u^{}_{\eta})^{\Uparrow\uparrow} & 0 \\
      0 &  (u_{\eta})^{\Downarrow\downarrow}
    \end{bmatrix},
\end{align}
\begin{align} 
    v_{\eta}=\begin{bmatrix}
      0 &-(v_{\eta})^{\Uparrow\uparrow} \\
       (v_{\eta})^{\Downarrow\downarrow} & 0
    \end{bmatrix}.
\end{align}
Then, Eq. \eqref{fvdmdkfkv} becomes
\begin{align} \label{fvdmkfkv}
    \mathcal{C}^{\eta}=\begin{bmatrix}
     0 & +(u^{-1}_{\eta})^{\Uparrow\uparrow}(v_{\eta})^{\Uparrow\uparrow} \\
     -(u^{-1}_{\eta})^{\Downarrow\downarrow}(v_{\eta})^{\Downarrow\downarrow} & 0
    \end{bmatrix}. 
\end{align}
Notably,  the cloud matrix \eqref{fvdmkfkv} is off-diagonal in spin space. Thus, the Cooper-like pair consists of two superconducting clouds with different spins. By substitution of Eq. \eqref{fvdmkfkv}, Eq. \eqref{fdvmkfk} becomes 
\begin{align} \label{yfdvmkfk}
    \theta&= \frac{1}{2}\sum_{nn'}\left\{[(u^{-1}_{\eta})^{\Uparrow\uparrow}(v_{\eta})^{\Uparrow\uparrow}]_{nn'} c^{\dagger}_{n\uparrow}c^{\dagger}_{n'\downarrow}\right.\\
    &-\left.[(u^{-1}_{\eta})^{\Downarrow\downarrow}(v_{\eta})^{\Downarrow\downarrow}]_{n'n}c^{\dagger}_{n'\downarrow}c^{\dagger}_{n\uparrow}\right\},\notag 
\end{align}
i.e.,
\begin{align} \label{eq:theta_solution}
\theta=\sum_{nn'}\tilde{\mathcal{C}}^{\eta}_{nn'}c^{\dagger}_{n\uparrow}c^{\dagger}_{n'\downarrow},
\end{align}
with
\begin{align} \label{fvdvkk}
    \tilde{\mathcal{C}}^{\eta}=\frac{1}{2}\left[(u^{-1}_{\eta})^{\Uparrow\uparrow}(v_{\eta})^{\Uparrow\uparrow}+(u^{-1}_{\eta})^{\Downarrow\downarrow}(v_{\eta})^{\Downarrow\downarrow}\right].
\end{align}
Here, we specialize the spin subscript of field operators $c^{\dagger}_{ks}$ for better description of the Cooper-like pair. Then, we obtain 
\begin{align}
    \vert V\rangle_{\eta}=\frac{1}{N^{1/2}_\eta}e^{\sum_{nn'}\tilde{\mathcal{C}}^{\eta}_{nn'}c^{\dagger}_{n\uparrow}c^{\dagger}_{n'\downarrow}}\vert 0\rangle. 
\end{align}
Alternatively, the Fermi sea and sky can be rewritten into a compact form 
\begin{align} \label{fvdkvkd}
    \vert V\rangle_{\eta}=\frac{1}{N^{1/2}_\eta}\prod_{nn'}\left(1+\tilde{\mathcal{C}}^{\eta}_{nn'}c^{\dagger}_{n\uparrow}c^{\dagger}_{n'\downarrow}\right)\vert 0\rangle.
\end{align}
Though $\tilde{\mathcal{C}}^{\eta}$ can be non-Hermitian, we can still diagonalize it by singular value decomposition
\begin{align} \label{fdvkp}
    \tilde{\mathcal{C}}^{\eta}=U^{}_{\uparrow\eta}\mathcal{A}^{\eta}U^{+}_{\downarrow\eta},
\end{align}
where $U_{s\eta}$ is unitary matrix and $\mathcal{A}^{\eta}$ is diagonal matrix with non-negative real Bogoliubov coefficients $\mathcal{A}^{\eta}_{nn}$.  Thus,  EVS \eqref{fvdkvkd} has Cooper-like pair form 
\begin{align} \label{ptrufvdfmmvl}
    \vert V\rangle_{\eta}=\frac{1}{N^{1/2}_\eta}\prod^{N}_{k=1}\left(1+a^{\dagger}_{k\uparrow\eta}\mathcal{A}^{\eta}_{kk} a^{\dagger}_{k\downarrow\eta}\right)\vert 0\rangle,
\end{align}
with 
\begin{align}
    N_{\eta}=\prod^{N}_{k=1}(1+\vert\mathcal{A}^{\eta}_{kk}\vert^2).
\end{align}
The complete and  orthogonal superconducting spin cloud operators are given by
\begin{align} \label{vdfkvkp} 
    a^{\dagger}_{k\uparrow\eta}=\sum_{n}c^{\dagger}_{n\uparrow}(U_{\uparrow\eta})_{nk},
\end{align}
\begin{align} \label{vtfkvkp}
    a^{\dagger}_{k\downarrow\eta}=\sum_{n}(U^{+}_{\downarrow\eta})_{kn}c^{\dagger}_{n\downarrow}.
\end{align}
The superconducting spin cloud operators \eqref{vdfkvkp} and \eqref{vtfkvkp} show how the electrons in the nonuniform superconductor pair with each other. The superconducting spin cloud picture of the ground state describes the quantum entanglement of the hybrid system. This is a widely entangled system as shown by the superconducting cloud matrix \eqref{fvdvkk}, which might be interesting for future quantum computation.

\section{Microscopic derivation of the tunneling current}\label{tunnelingcurrent}

In this subsection, we derive the tunneling current between tip and  quantum-dot Josephson junction from a microscopic perspective, i.e., Eq. \eqref{fvnjfvn} in the main text.

The tip electrons are described by the non-interacting Hamiltonian 
\begin{align}
    H_t=\sum_{\vec{p}s}E_{\vec{p}s}c^{\dagger}_{\vec{p}s}c^{}_{\vec{p}s},
\end{align}
where $c^{}_{\vec{p}s}$ is the field operator of electron in the tip with momentum $\vec{p}$, spin $s$, and energy $E_{\vec{p}}$. 
The electron tunneling Hamiltonian between the quantum dot and the tip  can be expressed as follows
\begin{align} \label{glfdd}
  V_{ep}=\sum_{l\vec{p}s} \left[t_d (u^{*}_{+})^{s s}_{l1}c^{\dagger}_{\vec{p}s}\gamma_{ls+}+t_ds(v_{+})^{-s -s}_{l1}c^{\dagger}_{\vec{p}s}\gamma_{ls-}+h.c.\right], 
\end{align}
where $t_d$ is the tunnel coupling strength between dot and tip. We begin with the Liouville-von Neumann equation in the interaction representation of $H_0=H_{e}+H_{}$, 
\begin{equation}
\dfrac{\partial}{\partial t}\check{\varrho}(t)=\dfrac{i}{\hbar}[\check{\varrho}(t), \check{\mathcal{V}}_{ep}(t)],
\end{equation}
where 
\begin{eqnarray}
 \check{\mathcal{O}}(t)=e^{+i\check{H}_0t/\hbar}O(t)e^{-i\check{H}_0t/\hbar}.
\end{eqnarray}
This equation can be solved by time-convolutionless (TCL) projection operator method, and we reach TCL equation, in \textit{Born-Markov approximation}, 
\begin{align} \label{NZE}
\dfrac{\partial}{\partial t} \hat{\varrho}(t)&=& \frac{i}{\hbar} [\hat{\varrho}(t), \check{\mathcal{V}}_{\mathrm{ep}}(t)]- \int_{t_0}^{t} \frac{d\tau}{\hbar^2}  [[ \hat{\varrho}(t),\check{\mathcal{V}}_{\mathrm{ep}}(\tau)],\check{\mathcal{V}}_{\mathrm{ep}}(t)].
\end{align}
The transition rate between  $\vert E_{ls\eta}  \rangle $  and $\vert E_{\vec{p}s} \rangle $ is given by  $\hat{\mathcal{V}}_{ep}^{\vec{p}s,ls\eta}=t_dD_{ls\eta}$, where $D_{ls+}=(u^{*}_{+})^{ss}_{l1}$ and $D_{ls-}=s(v_{+})^{-s -s}_{l1}$. Then, the electron tunneling Hamiltonian \eqref{glfdd} becomes 
\begin{align} \label{glfafdd}
   V_{ep}=\sum_{l\vec{p}s\eta} \left(\hat{\mathcal{V}}_{ep}^{\vec{p}s,ls\eta}c^{\dagger}_{\vec{p}s}\gamma^{}_{ls\eta}+\hat{\mathcal{V}}_{ep}^{ls\eta,\vec{p}s}\gamma^{\dagger}_{ls\eta}c^{}_{\vec{p}s}\right). 
\end{align}
The second term, i.e., collision term, reads 
\begin{eqnarray}
\hat{ \mathcal{J}}(\hat{\varrho}\vert t)&=&-\sum_{\vec{q}_i}  \int_{t_0}^{t}  d\tau
 \left\lbrace   \left[ \hat{\mathcal{V}}_{ep}(t),\hat{\mathcal{V}}_{ep}(\tau) \hat{\varrho}(t) \right] \right.\notag \\
 &-& \left.  \left[\hat{\mathcal{V}}_{ep}(t),\hat{\varrho}(t)\hat{\mathcal{V}}_{ep}(\tau) \right] \right\rbrace.
\end{eqnarray}
Converting into Schrodinger representation, we reach  
\begin{equation}  \label{SREPRE}
\dfrac{\partial}{\partial t} \hat{\rho}(t) - \hat{B}_0(t)\hat{\rho}(t)- \hat{B}_1(t)\hat{\rho}(t)=\hat{ J}(\hat{\rho}\vert t),
\end{equation} 
with 
\begin{eqnarray}
 \hat{B}_0(t)\hat{\rho}(t)=\frac{i}{\hbar} \left[\hat{\rho}(t), \hat{H}_0  \right],
\end{eqnarray}
\begin{eqnarray}  \label{djanv}
      \hat{B}_1(t)\hat{\rho}(t) =\frac{i}{\hbar}  [\hat{\rho}(t), \check{V}_{\mathrm{ep}}],
\end{eqnarray}
\begin{align} \label{fsanfvfnv}
 \hat{ J}(\hat{\rho})&=\sum_{\vec{q}_i}  \int_{0}^{t-t_0} d\tau  \left\lbrace  \left[ \hat{V}_{eq},\hat{\rho}(t)\hat{\mathcal{V}}_{eq}(-\tau) \right]\right.\notag \\
 &-\left.   \left[ \hat{V}_{eq},\hat{\mathcal{V}}_{eq}(-\tau) \hat{\rho}(t)\right]   \right\rbrace. 
\end{align}
The collision term $\hat{ J}(\hat{\rho})$ describes the scattering induced by dot-tip tunneling coupling.

We here define the tunneling current between tip and quantum-dot Josephson junction as the rate of change of the electron number in the tip 
\begin{align} \label{dfvadfm}
    I_T=e\frac{d}{dt} \sum_{\vec{p}s} \left\langle\hat{c}^+_{\vec{p}s}\hat{c}^{}_{\vec{p}s}\right\rangle(t),
\end{align}
with
\begin{align}
    \left\langle\hat{c}^+_{\vec{p}s}\hat{c}^{}_{\vec{p}s}\right\rangle(t)=\text{Tr} \{\hat{c}^+_{\vec{p}s}\hat{c}^{}_{\vec{p}s}\hat{\rho}(t)\}.
\end{align}
Then, we define one-body density matrix from the many-body density matrix  as follows
\begin{equation} \label{jgsufgnn}
    \varrho_{\kappa_2\kappa_1}(t)= \langle \kappa_2\vert \hat{\varrho}(t) \vert \kappa_1\rangle\equiv\mathrm{Tr}\{\hat{f}^+_{\kappa_1}\hat{f}^{}_{\kappa_2}\hat{\rho}(t)\}.
\end{equation}
For tip electrons, we have $\kappa_i=\alpha_i=(\vec{p}_i,s_i)$, corresponding to $\hat{f}^{}_{\kappa_i}=f_{\alpha_i}=c_{\vec{p}_i,s_i}$. For the quantum-dot Josephson junction, we have $\kappa_i=\lambda_i=(l_i,s_i,\eta_i)$, corresponding to $\hat{f}^{}_{\kappa_i}=f_{\lambda_i}=\gamma_{l_i,s_i,\eta_i}$.  As shown in Eq. \eqref{dfvadfm}, we are only interested in the time evolution of the diagonal and tip components of the one-body density matrix, i.e., $\varrho_{\alpha\alpha}(t)=\varrho_{\alpha}(t)$, given by 
\begin{equation} \label{RFE} 
\dfrac{\partial}{\partial t} \varrho_{\alpha}(t)-\frac{i}{\hbar}[\varrho(t), \hat{\mathcal{V}}_{ep}]_{\alpha\alpha}=\mathcal{ J}_{\alpha}(\hat{\varrho}).
\end{equation} 
The collision integral in the right hand side of Eq. (\ref{RFE}) is given by 
\begin{align} \label{CIRF}
 \mathcal{ J}_{\alpha}(\hat{\varrho})&=  \int_{0}^{t-t_0} d\tau  \text{Tr}\left\lbrace   \left[ \hat{V}_{eq},\hat{f}^+_{\alpha}\hat{f}^{}_{\alpha} \right] \left[ \hat{\mathcal{V}}_{eq}(-\tau), \hat{\rho}(t)\right]  \right\rbrace. 
\end{align}
It is required to calculate commutator 
\begin{align} \label{fhanfu}
    [V_{eq},\hat{f}^+_{\alpha}\hat{f}^{}_{\alpha}]
    &=\sum_{\lambda_1}  \left[\hat{\mathcal{V}}_{ep}^{\lambda_1\alpha} \hat{f}^{+}_{\lambda_1} \hat{f}^{}_{\alpha}-\hat{\mathcal{V}}_{ep}^{\alpha\lambda_1} \hat{f}^{+}_{\alpha} \hat{f}^{}_{\lambda_1}\right].
\end{align}
Substituting Eq. \eqref{fhanfu} into \eqref{CIRF}, we reach
\begin{align} \label{CIRFfanjgra}
\mathcal{ J}_{\alpha}(\hat{\varrho})&= \sum_{\lambda_1}  \int_{0}^{t-t_0} d\tau  \text{Tr}\left\lbrace   \left[\hat{\mathcal{V}}_{ep}^{\lambda_1\alpha} \hat{f}^{+}_{\lambda_1} \hat{f}^{}_{\alpha}\right.\right.\\
 & -\left.\left. \hat{\mathcal{V}}_{ep}^{\alpha\lambda_1} \hat{f}^{+}_{\alpha} \hat{f}^{}_{\lambda_1}\right] \left[  \hat{\mathcal{V}}_{eq}(-\tau), \hat{\rho}(t)\right]  \right\rbrace. \notag
\end{align}
By substitution of 
\begin{align} 
   \hat{\mathcal{V}}_{eq}(-\tau)=\sum_{\kappa_5,\kappa_6} e^{+i\omega^{\kappa_6}_{\kappa_5}\tau}\hat{\mathcal{V}}_{ep}^{\kappa_5,\kappa_6}f^{\dagger}_{\kappa_5}f^{}_{\kappa_6}.  
\end{align}
Eq. \eqref{CIRFfanjgra} becomes 
\begin{align} \label{fagayotar1}
 \mathcal{ J}_{\alpha}(\hat{\varrho})&= \sum_{\lambda_i\kappa_i} \text{Tr}\left\lbrace   \left[\hat{\mathcal{V}}_{ep}^{\lambda_3\alpha} \hat{f}^{+}_{\lambda_3} \hat{f}^{}_{\alpha}-\hat{\mathcal{V}}_{ep}^{\alpha\lambda_3} \hat{f}^{+}_{\alpha} \hat{f}^{}_{\lambda_3}\right]\right.\\
 &\times \left. \left[ \delta^+_{}(\omega^{\kappa_6}_{\kappa_5}) \hat{\mathcal{V}}_{ep}^{\kappa_5\kappa_6}\hat{f}^+_{\kappa_5}\hat{f}^{}_{\kappa_6}, \hat{\rho}(t)\right]  \right\rbrace, \notag
\end{align}
with
\begin{align} \label{fdkaknp}
    \delta^{+}_{}(\omega) =\int^{+\infty}_{0} d\tau e^{+i\omega\tau-\eta \tau},
\end{align}
where $t_0\rightarrow -\infty$, and  $\eta\rightarrow 0^+$ was introduced to remove the divergence of the infinity integral.  
Eq. \eqref{fagayotar1} can be rewritten into  
\begin{align} \label{CIRFfanjgr}
\hat{\mathcal{J}}_{\alpha}(\hat{\varrho})&= \sum_{\kappa_i\lambda_i}\delta^+_{}(\omega^{\kappa_6}_{\kappa_5}) \hat{\mathcal{V}}_{ep}^{\kappa_5\kappa_6} \left \{  \hat{\mathcal{V}}_{ep}^{\alpha\lambda_3}[ R^{\kappa_5\alpha}_{\kappa_6\lambda_3}(t)-R^{\alpha\kappa_5}_{\lambda_3\kappa_6}(t)]\right. \notag \\
&+\left.\hat{\mathcal{V}}_{ep}^{\lambda_3\alpha}  [R^{\lambda_3\kappa_5}_{\alpha\kappa_6}(t)-R^{\kappa_5\lambda_3}_{\kappa_6\alpha}(t)] \right\}.
\end{align}
The many-electron correlation functions, are defined as
\begin{align} \label{lbneruyt}
    R^{\kappa_1\kappa_3}_{\kappa_2\kappa_4}(t)=\text{Tr}\left\{\hat{f}^+_{\kappa_1}\hat{f}^{}_{\kappa_2}\hat{f}^{+}_{\kappa_3} \hat{f}^{}_{\kappa_4}\hat{\rho}(t)\right\},
\end{align}
To calculate Eq. \eqref{CIRFfanjgr}, it is required to calculate  many-electron correlation function \eqref{lbneruyt}. To reach a normal order inside the trace $\text{Tr}$, we reach 
\begin{align} \label{kkvpgbw1}
    R^{\kappa_1\kappa_3}_{\kappa_2\kappa_4}(t)=\delta_{\kappa_2\kappa_3}\varrho_{\kappa_4\kappa_1}-\text{Tr}_e\left\{\hat{f}^+_{\kappa_1}\hat{f}^{+}_{\kappa_3}\hat{f}^{}_{\kappa_2} \hat{f}^{}_{\kappa_4}\hat{\rho}(t)\right\}.
\end{align} 
The second term in normal order, can be approximated into  
\begin{align} \label{kkvpgbw2}
    \text{Tr}_e\left\{\hat{f}^+_{\kappa_1}\hat{f}^{+}_{\kappa_3}\hat{f}^{}_{\kappa_2} \hat{f}^{}_{\kappa_4}\check{\rho}_e(t)\right\}\simeq \varrho_{\kappa_4\kappa_1}\varrho_{\kappa_2\kappa_3}-\varrho_{\kappa_4\kappa_3}\varrho_{\kappa_2\kappa_1}.
\end{align}
With the help from  Eqs. \eqref{kkvpgbw1} and \eqref{kkvpgbw2}, Eq. \eqref{CIRFfanjgr} can be divided into two terms
\begin{align}
    \mathcal{ J}_{\alpha}(\hat{\varrho})&= \mathcal{ J}^e_{\alpha}(\hat{\varrho})+\mathcal{ J}^a_{\alpha}(\hat{\varrho})
\end{align}
with 
\begin{widetext}
\begin{align} \label{farnyay1}
   \hat{\mathcal{J}}^{e}_{\alpha}(\hat{\varrho})&= \sum_{\lambda_3\kappa_i}  \delta^+_{}(\omega^{\kappa_6}_{\kappa_5}) [\hat{\mathcal{V}}_{ep}^{\alpha\lambda_3}\varrho_{\lambda_3\kappa_5}\hat{\mathcal{V}}_{ep}^{\kappa_5\kappa_6} \bar{\varrho}_{\kappa_6\alpha}-\varrho_{\alpha\kappa_5}\hat{\mathcal{V}}_{ep}^{\kappa_5\kappa_6}\bar{\varrho}_{\kappa_6\lambda_3}\hat{\mathcal{V}}_{ep}^{\lambda_3\alpha} ]\\
   &+\sum_{\alpha_i}  \delta^+_{}(\omega^{\kappa_6}_{\kappa_5}) [\bar{\varrho}_{\alpha\kappa_5} \hat{\mathcal{V}}_{ep}^{\kappa_5\kappa_6} \varrho_{\kappa_6\lambda_3}  \hat{\mathcal{V}}_{ep}^{\lambda_3\alpha}
   -\hat{\mathcal{V}}_{ep}^{\alpha\lambda_3} \bar{\varrho}_{\lambda_3\kappa_5}\hat{\mathcal{V}}_{ep}^{\kappa_5\kappa_6}\varrho_{\kappa_6\alpha} ] , \notag
\end{align}
\begin{align}  \label{farnyay4}
   \hat{\mathcal{J}}^{a}_{\alpha}(\hat{\varrho})&=  \sum_{\lambda_3\kappa_i}   \delta^+_{}(\omega^{\kappa_6}_{\kappa_5})[\hat{\mathcal{V}}_{ep}^{\alpha\lambda_3}\varrho_{\lambda_3\alpha} \varrho_{\kappa_6\kappa_5} \hat{\mathcal{V}}_{ep}^{\kappa_5\kappa_6}-\varrho_{\alpha\lambda_3}\hat{\mathcal{V}}_{ep}^{\lambda_3\alpha} \hat{\mathcal{V}}_{ep}^{\kappa_5\kappa_6}\varrho_{\kappa_6\kappa_5} ]   \\
   &+ \sum_{\alpha_i}  \delta^+_{}(\omega^{\kappa_6}_{\kappa_5})[\varrho_{\alpha\lambda_3}\hat{\mathcal{V}}_{ep}^{\lambda_3\alpha}  \varrho_{\kappa_6\kappa_5} \hat{\mathcal{V}}_{ep}^{\kappa_5\kappa_6}
   - \hat{\mathcal{V}}_{ep}^{\alpha\lambda_3}  \varrho_{\lambda_3\alpha}  \hat{\mathcal{V}}_{ep}^{\kappa_5\kappa_6}\varrho_{\kappa_6\kappa_5}],\notag
\end{align}
\end{widetext}
with 
\begin{align}
    \bar{\varrho}_{\kappa_1\kappa_2}=\delta_{\kappa_1\kappa_2}-\varrho_{\kappa_1\kappa_2}.
\end{align}

By means of Eq. \eqref{RFE}, the tunneling current is given by 
\begin{align} \label{fvfldvfl}
   I_T= e\sum_{\alpha} \mathcal{J}_{\alpha}(\check{\varrho}),
\end{align}
with
\begin{align} \label{farynnqdllaap}
   \mathcal{J}_{\alpha}(\check{\varrho})&=\sum_{\lambda} \left\lbrace   [\delta^+_{}(\omega^{\alpha}_{\lambda})\hat{\mathcal{V}}_{ep}^{\alpha,\lambda}\hat{\mathcal{V}}_{ep}^{\lambda,\alpha} \varrho_{\lambda}\bar{\varrho}_{\alpha}-\delta^+_{}(\omega^{\lambda}_{\alpha})\hat{\mathcal{V}}_{ep}^{\alpha,\lambda}\hat{\mathcal{V}}_{ep}^{\lambda,\alpha}\varrho_{\alpha}\bar{\varrho}_{\lambda} ] \right. \notag \\
  &+\left.  [\delta^+_{}(\omega^{\lambda}_{\alpha}) \hat{\mathcal{V}}_{ep}^{\alpha,\lambda} \hat{\mathcal{V}}_{ep}^{\lambda,\alpha} \varrho_{\lambda}  \bar{\varrho}_{\alpha}
   -\delta^+_{}(\omega^{\alpha}_{\lambda})\hat{\mathcal{V}}_{ep}^{\alpha,\lambda} \hat{\mathcal{V}}_{ep}^{\lambda,\alpha}\varrho_{\alpha} \bar{\varrho}_{\lambda}]\right\}, 
\end{align}
where we have used the diagonal approximation 
\begin{align} \label{fanfsgunba}
    \varrho_{\kappa_1\kappa_2}\simeq \delta_{\kappa_1\kappa_2}\varrho_{\kappa_1\kappa_1}=\delta_{\kappa_1\kappa_2}\varrho_{\kappa_1},
\end{align}
which also results in  
\begin{align} 
\sum_{\alpha}[\varrho(t), \hat{\mathcal{V}}_{ep}]_{\alpha\alpha}=0.
\end{align} 
Noting that  
\begin{align}
    \delta^+_{}(\omega_{\lambda}^{\alpha})+\delta^+_{}(\omega^{\lambda}_{\alpha})=\delta (\omega_{\lambda}^{\alpha}),
\end{align}
and hence Eq. \eqref{farynnqdllaap} reduces to 
\begin{align}  \label{fvdfgav}
   \mathcal{J}^{s}_{\vec{p}}(\check{\varrho})&=\sum_{l\eta} \left\lbrace   [\delta(\omega^{\vec{p}s}_{ls\eta})\hat{\mathcal{V}}_{ep}^{\vec{p}s,ls\eta}\hat{\mathcal{V}}_{ep}^{ls\eta,\vec{p}s} \varrho_{ls\eta}\bar{\varrho}_{\vec{p}s}\right.\\
   &-\left.\delta(\omega^{ls\eta}_{\vec{p}s})\hat{\mathcal{V}}_{ep}^{\vec{p}s,ls\eta}\hat{\mathcal{V}}_{ep}^{ls\eta,\vec{p}s}\varrho_{\vec{p}s}\bar{\varrho}_{ls\eta} ] \right\}. \notag 
\end{align}
In the presence of the applied voltage, $V$, we assume
\begin{align} \label{favmfdmv}
    \varrho_{\vec{p}s}=f(E_{\vec{p}}-V), \varrho_{ls\eta}=f(E_{ls\eta}),
\end{align}
where $f(E)=1/(1+e^{E/k_BT})$ is the Fermi-Dirac distribution for temperature $\beta=1/k_BT$. 
Then, Eq. \eqref{fvdfgav} reduces to 
\begin{align} \label{fvnjdfvn}
    \mathcal{J}^{s}_{\vec{p}}(\check{\varrho})&=\frac{2\pi}{h}  \sum_{l\eta} f(E_{ls\eta})[1-f(E_{\vec{p}}-V)]\vert T^{\vec{p}s}_{ls\eta}\vert^2 \delta(E_{\vec{p}}-E_{ls\eta})\notag\\
    &-\frac{2\pi}{h} \sum_{l\eta} f(E_{\vec{p}}-V)[1-f(E_{ls\eta})]\vert T^{\vec{p}s}_{ls\eta} \vert^2\delta(E_{\vec{p}}-E_{ls\eta}).
\end{align}
with $\vert T^{\vec{p}s}_{ls\eta}\vert^2=\hat{\mathcal{V}}_{ep}^{\vec{p}s+,ls\eta}\hat{\mathcal{V}}_{ep}^{ls\eta,\vec{p}s+}$. 
Finally, we obtain the tunneling current as follows 
\begin{align} \label{fvdanjfvn}
    I_T&=\frac{2\pi e}{h}  \sum_{\vec{p} ls\eta} f(E_{ls\eta})[1-f(E_{\vec{p}}-V)]\vert T^{\vec{p}s}_{ls\eta}\vert^2 \delta(E_{\vec{p}}-E_{ls\eta})\notag\\
    &-\frac{2\pi e}{h} \sum_{\vec{p} ls\eta} f(E_{\vec{p}}-V)[1-f(E_{ls\eta})]\vert T^{\vec{p}s}_{ls\eta} \vert^2\delta(E_{\vec{p}}-E_{ls\eta}).
\end{align}
This is the tunneling current between tip and  quantum-dot Josephson junction from a microscopic perspective, given by Eq. \eqref{fvnjfvn} in the main text.

\end{document}